\documentclass[a4paper,12pt]{article}

\pdfoutput=1
\usepackage{jheppub}
\usepackage[T1]{fontenc}

\usepackage{graphicx}
\usepackage{cancel}
\usepackage{amssymb}
\usepackage{textcomp}
\usepackage{amsmath}
\usepackage{bm}
\usepackage{times}
\usepackage{epsfig}
\usepackage{epstopdf}
\usepackage{color}
\usepackage{multirow}

\def\lsim{\mathrel{\raise.3ex\hbox{$<$\kern-.75em\lower1ex\hbox{$\sim$}}}}
\def\gsim{\mathrel{\raise.3ex\hbox{$>$\kern-.75em\lower1ex\hbox{$\sim$}}}}

\def\beq{\begin{equation}}
\def\eeq{\end{equation}}
\def\be{\begin{equation}}
\def\ee{\end{equation}}
\def\bea{\begin{eqnarray}}
\def\eea{\end{eqnarray}}

\def\ev{\,{\rm eV}}
\def\to{\rightarrow}

\usepackage[normalem]{ulem}

\newcommand{\minigraph}[5][0.25in]{\begin{minipage}{#2}\begin{center}\includegraphics[width=#2]{#5}\\\vspace{#3}\hspace{#1}{\footnotesize #4}\end{center}\end{minipage}}

\title{Revealing the origin of neutrino masses through the Type II Seesaw mechanism at high-energy muon colliders}

\author[a]{Tong Li}
\emailAdd{litong@nankai.edu.cn}
\affiliation[a]{
School of Physics, Nankai University, Tianjin 300071, China}

\author[a,b]{Chang-Yuan Yao}
\emailAdd{yaocy@nankai.edu.cn}
\affiliation[b]{
Deutsches Elektronen-Synchrotron DESY, Notkestr. 85, 22607 Hamburg, Germany}

\author[a]{Man Yuan}
\emailAdd{yuanman@mail.nankai.edu.cn}

\abstract{
The future muon collider can play as an ideal machine to search for new physics at high energies. In this work, we study the search potential of the heavy Higgs triplet in the Type II Seesaw mechanism at muon colliders with high collision energy and high luminosity. The latest neutrino oscillation data are taken into account for realizing the leptonic decay modes of the charged Higgs bosons $(H^{\pm\pm},~H^{\pm})$ in the Type II Seesaw.
We show the impact of neutrino mass and mixing parameters on the purely leptonic decays. The pair production of doubly charged Higgs $H^{++}H^{--}$ is through direct $\mu^+\mu^-$ annihilation and vector boson fusion (VBF) processes at muon collider. The associated production $H^{\pm\pm}H^{\mp}$ can only be induced by VBF processes. We simulate both the purely leptonic and bosonic signal channels of charged Higgs bosons in Type II Seesaw, together with the Standard Model backgrounds. We show the required luminosity for the discovery of the charged Higgses and the reachable limits on the leptonic decay branching fractions.
}

\begin{document}

\begin{flushright}
DESY-23-006
\end{flushright}

\makeatletter\renewcommand{\@fpheader}{\ }\makeatother

\maketitle
\flushbottom
\newpage

\section{Introduction}

It is well-known that, in the context of the Standard Model (SM), the small but non-zero neutrino masses can be realized at leading order through a dimension-5 operator~\cite{Weinberg:1979sa}
\begin{eqnarray}
{\kappa\over \Lambda}\ell_L \ell_LHH\;,
\label{weinberg}
\end{eqnarray}
where $\ell_L$ and $H$ stand for the SM lepton doublet and the Higgs doublet, respectively. There are only three ultraviolet (UV) completions of this ``Weinberg operator'' at tree level~\cite{Ma:1998dn}, known as the Type I~\cite{Minkowski:1977sc,Yanagida:1979as,GellMann:1980vs,Glashow:1979nm,Mohapatra:1979ia,Shrock:1980ct,Schechter:1980gr}, Type II~\cite{Konetschny:1977bn,Cheng:1980qt,Lazarides:1980nt,Schechter:1980gr,Mohapatra:1980yp} and Type III~\cite{Foot:1988aq} Seesaw mechanisms. These mechanisms extend the SM by introducing singlet right-handed neutrinos $N_R$, SU(2)$_L$ scalar triplet $\Delta$, and SU(2)$_L$ fermionic triplet, respectively.
Among them, the Type II Seesaw model has an extended scalar sector with rich phenomena. The leptonic Yuakawa interaction of the scalar triplet provides the tiny Majorana neutrino masses after the neutral component of the triplet acquires the vacuum expectation value (vev) $v_{\Delta}$. There are seven physical mass eigenstates of Higgs boson in the Type II Seesaw, including two singly-charged Higgs ($H^{\pm}$), two doubly-charged Higgs ($H^{\pm\pm}$), two CP-even ($H_1\,, H_2$) and one CP-odd ($A$) neutral Higgs~\cite{FileviezPerez:2008jbu,Arhrib:2011uy,BhupalDev:2013xol}.

The low-scale Type II Seesaw can be experimentally accessible at high-energy colliders when the triplet Higgs mass $M_\Delta$ is at TeV level.
The searches of the triplet Higgs have been studied extensively at the Large Hadron Collider (LHC)~\cite{Huitu:1996su,Gunion:1996pq,Chun:2003ej,Han:2005nk,Han:2007bk,Garayoa:2007fw,Kadastik:2007yd,Akeroyd:2007zv,Chao:2007mz,Chao:2008mq,delAguila:2008cj,FileviezPerez:2008wbg,FileviezPerez:2008jbu,Akeroyd:2009hb,Akeroyd:2010ip,Akeroyd:2011zza,Melfo:2011nx,Aoki:2011pz,Chiang:2012dk,Chun:2012zu,Chun:2013vma,delAguila:2013mia,Kanemura:2013vxa,Kanemura:2014ipa,Kang:2014lwn,Godunov:2014waa,Chen:2014qda,Han:2015hba,Han:2015sca,Bonilla:2015jdf,Mitra:2016wpr,Babu:2016rcr,Ghosh:2017pxl,Antusch:2018svb,BhupalDev:2018tox,Primulando:2019evb,deMelo:2019asm,Chun:2019hce,Padhan:2019jlc,Ashanujjaman:2021txz,Ghosh:2021khk,Ashanujjaman:2022ofg,Butterworth:2022dkt}, its upgrades~\cite{Li:2018jns,Du:2018eaw,deMelo:2019asm,Arhrib:2019ywg,Padhan:2019jlc,Fuks:2019clu,Aoki:2020til}, leptonic colliders~\cite{Rodejohann:2010jh,Shen:2015pih,Blunier:2016peh,Sui:2017qra,Nomura:2017abh,Agrawal:2018pci,Dev:2018sel,Rahili:2019ixf,Li:2019xvv,Gluza:2020icp,Costantini:2020stv,Bandyopadhyay:2020mnp,Li:2021lnz,Bai:2021ony,Ashanujjaman:2022tdn,Mandal:2022ysp,Mandal:2022zmy} and $ep$ collider~\cite{Dev:2019hev,Yang:2021skb}, in terms of lepton number violation (LNV) signatures for $H^{\pm\pm}$ and $H^\pm$. The decay modes of charged Higgs bosons can be categorized into leptonic decays and the decays to vector bosons. The comprehensive reviews of search for LNV at colliders can be found in Refs.~\cite{Deppisch:2015qwa,Cai:2017mow}. Up to now, no significant excess beyond the SM expectations was observed at the LHC with a centre-of-mass (c.m.) energy of 13 TeV. The lower limit on the mass of a doubly charged Higgs boson is $900-1080$ GeV at the 95\% confidence level (CL) from the same-sign lepton channel, depending on specific models~\cite{ATLAS:2022pbd}. The vector boson channels exclude the charged Higgs boson lighter than $230-350$ GeV at 95\% CL~\cite{ATLAS:2021jol}. However, due to the limitation of the current collision energy, it will be difficult to observe the charged Higgs bosons with mass larger than 1 TeV. To search for new electroweak physics, there is an urgent need of the future hadron or lepton colliders with much higher collision energy and integrated luminosity.

Recently, due to the technological breakthrough of the ionization cooling by the Muon Ionization Cooling Experiment (MICE)~\cite{MICE:2019jkl}, the establishment of the muon collider has rekindled hope and again received much attention in the community.
The key factor limiting the collision energy is the energy loss during the acceleration, i.e., the synchrotron radiation. For a circular machine with a radius of $R$, the energy loss per revolution is given by $\Delta E\approx {1\over R}\Big({E\over m}\Big)^4$
with $E$ being the beam energy and $m$ the mass of colliding particles. Thus, an accelerator is more efficient for a more massive particle. The muon mass is about 200 times greater than that of electron. Then, the relatively large mass of muon suppresses the synchrotron radiation. Thus, a muon collider can exceed the energy reach of the $e^+e^-$ colliders, and achieve both high energies and high
luminosities~\cite{Delahaye:2019omf, Bartosik:2020xwr}. It would offer a great opportunity to provide an unprecedented new energy threshold for new physics search and a clean environment of leptonic collisions for precision measurements~\cite{Han:2020uid,AlAli:2021let,Bose:2022obr,Maltoni:2022bqs,Narain:2022qud}.
For the $\mu^+\mu^-$ annihilation well above the $Z$ pole, the cross section falls off as $1/s$ with $\sqrt{s}$ being the c.m. energy. Thus, the annihilation cross sections decrease rapidly with colliding energy and meanwhile these cross sections are insensitive to the
mass of the new physics particles in final states when producing above threshold. By contrast, the cross sections of vector boson fusion (VBF) processes typically scale with c.m. energy as ${\rm ln}(s)$ above threshold. Thus, from logarithmic enhancement, VBF becomes an important channel to search for new physics particles at high energies. Moreover, the cross sections decrease with larger mass of new particles due to the suppression of EW PDF threshold and thus the VBF mechanism is sensitive to the mass of the new physics particles~\cite{AlAli:2021let,Han:2021udl}.

In this work, we study typical electroweak production channels of triplet Higgs bosons in Type II Seesaw at future high-energy muon collider. The major pair production channels of $H^{++}H^{--}$ include direct $\mu^+\mu^-$ annihilation and the VBF processes~\cite{Han:2020uid,Costantini:2020stv,Ruiz:2021tdt}. The associated production of $H^{\pm\pm}H^{\mp}$ can only be generated by the VBF processes at muon collider.
The decay modes of the charged Higgs bosons are closely related to the value of triplet Higgs vev. For $v_{\Delta}\lesssim 10^{-4}$ GeV, depending on their charges, the charged Higgs bosons mainly decay to a charged lepton plus a neutrino or a pair of same-sign leptons. For $v_{\Delta} \gtrsim 10^{-4}$ GeV, they decay into SM gauge bosons.
More details can be found in Ref.~\cite{FileviezPerez:2008jbu}.
Four possible signatures of the triplet Higgs will be discussed in detail in this paper, including both leptonic decay modes and gauge boson decay modes for $H^{++}H^{--}$ or $H^{\pm\pm}H^{\mp}$ production. We choose the muon collider with multi-TeV c.m. energies $\sqrt{s}$ and the integrated luminosity scaling with energy quadratically~\cite{Delahaye:2019omf}
\begin{equation}
\label{eq:lumi}
    \mathcal{L}=\left(\frac{\sqrt{s}}{10\ {\rm TeV}}\right)^2 \times 10~\textrm{ab}^{-1}\;.
\end{equation}
In particular, the benchmark choices of the collider energies and the corresponding integrated luminosities are
\begin{equation}
    \sqrt{s}=3,~10~{\rm and}~30~{\rm TeV},~~~ \mathcal{L}=1,~10~{\rm and}~90~{\rm ab}^{-1}.
\end{equation}

On the other hand, in recent years, the precision of neutrino mixing parameters has been significantly improved in neutrino oscillation experiments. For instance, Double Chooz~\cite{Abe:2011fz}, RENO~\cite{Ahn:2012nd} and in particular Daya Bay~\cite{An:2012eh}, have reported non-zero experimental results of $\theta_{13}$ by looking for the disappearance of anti-electron neutrinos. In addition, the long-baseline neutrino experiments T2K~\cite{Abe:2013hdq,Abe:2017uxa} and NOvA~\cite{Adamson:2016tbq} suggest a non-zero leptonic CP phase. These experiments provide us with up-to-date neutrino oscillation experimental results to investigate the impact on neutrino mass models. As the sign of the neutrino mass-squared difference $\Delta m^2_{3\ell} \equiv m^2_3-m^2_\ell$ ($\ell=1~\text{or}~2$) has not been determined, two possible scenarios are considered separately in data analysis, which are commonly called normal hierarchy~(NH) and inverted hierarchy~(IH) correspond to positive and negative sign respectively. The future neutrino oscillation experiments, such as T2HK~\cite{Hyper-KamiokandeProto-:2015xww}, JUNO~\cite{JUNO:2015sjr} and DUNE~\cite{DUNE:2015lol} will measure some oscillation parameters with better than $0.5\%$ precision, and the neutrino mass hierarchy will also be revealed. With the development of low-energy experiments, the theoretical and phenomenological studies of neutrino models will benefit from the constrained properties of neutrinos. We will investigate the constraints of the latest neutrino oscillation data on the decay patterns of triplet Higgs bosons. Moreover, the phenomenological study at high-energy colliders will provide us with more physical information about a specific neutrino model, and reveal the neutrino properties in a full UV theory. These studies also work as a supplementary strategy to explore the properties of neutrinos. We will emphasize the flavor structure of the lepton number violating decays of the charged Higgs bosons and the implications for neutrino properties.

The rest of this paper is organized as follows. In Sec.~\ref{sec:typeIImodel}, we first outline the Type II Seesaw model and discuss the constraints from neutrino oscillation experiments. We show the flavor structure of charged Higgs decay and the impact of neutrino mass and mixing parameters on the decay branching fractions. In Sec.~\ref{sec:simulation}, we perform the collider simulation of charged Higgs production and decay with different benchmark collision energies and luminosities. The promising signatures will be considered separately. The projected discovery bounds are presented in terms of required luminosities and branching ratios. Finally, we summarize the main results and draw our conclusions in Sec.~\ref{sec:Concl}.

\section{Type II Seesaw and the impact of neutrino data}
\label{sec:typeIImodel}

In this section, we first briefly review the Type II Seesaw model and then discuss the impact of neutrino oscillation data on charged Higgs decay.

\subsection{Type II Seesaw mechanism}

In the Type II Seesaw model, the Higgs sector of the SM is extended by adding an SU$(2)_L$ scalar triplet $\Delta\sim (1,3,1)$, which can be decomposed as
\begin{eqnarray}
\Delta= \left(
  \begin{array}{cc}
    \delta^+/\sqrt{2} & \delta^{++} \\
    \delta^0 & -\delta^+/\sqrt{2} \\
  \end{array}
\right).
\end{eqnarray}
In the lepton sector, the scalar triplet $\Delta$ interacts with the SM lepton doublet $\ell_L$ through the Yukawa interaction
\begin{eqnarray}
Y_\nu \ \ell^T_L \ C\ i\sigma_2 \ \Delta \ \ell_L+h.c. \,,
\end{eqnarray}
where $C$ represents the charge conjugation operator, $\sigma_2$ is the Pauli matrix and the Yukawa coupling $Y_{\nu}$ is a $3\times3$ symmetric complex matrix. In the Higgs sector, the scalar triplet $\Delta$ also couples with the SM Higgs doublet $H$ via the mixing term
\begin{eqnarray}
\mu H^T \ i\sigma_2 \ \Delta^\dagger H+h.c. \ .
\end{eqnarray}
Then the mass matrix of the neutrino is given by the following relations
\begin{eqnarray}
m_\nu=\sqrt{2}Y_\nu v_\Delta, \ \ \ v_\Delta={\mu v_0^2\over \sqrt{2} M_\Delta^2}  \ ,
\label{eqn:M_v}
\end{eqnarray}
where the $v_0$ and $v_\Delta$ satisfying $\sqrt{v_0^2+v_\Delta^2}\approx 246 \ {\rm GeV}$ are the vevs of the neutral components of the Higgs doublet and triplet, respectively. $M_\Delta$ is the mass of the heavy triplet Higgs. Due to the presence of the triplet $\Delta$, the lepton number is explicitly broken. After the electroweak symmetry breaking, there are seven physical massive Higgses, including a SM-like Higgs $H_1$, a $\Delta$-like Higgs $H_2$, a CP-odd scalar $A$ , two singly charged Higgs $H^\pm\approx \delta^\pm$ and two doubly charged Higgs $H^{\pm\pm}=\delta^{\pm\pm}$ with $M_{H_2}\simeq M_{A}\simeq M_{H^\pm}\simeq M_{H^{\pm\pm}}= M_\Delta$.~\footnote{We assume degenerate triplet Higgs spectrum in the following analysis. For the phenomenological studies of non-degenerate case, see Refs.~\cite{Akeroyd:2011zza,Melfo:2011nx,Aoki:2011pz,Chun:2013vma,Chen:2014qda,Han:2015hba,Han:2015sca,Shen:2015pih,Primulando:2019evb,Cheng:2022hbo,Butterworth:2022dkt}.} See Ref.~\cite{FileviezPerez:2008jbu} for detailed discussions. In the physical basis for the leptons, the Yukawa interaction of the singly  and doubly charged Higgs can be respectively written as
\begin{align}
\nu_L^T \ C \ Y^+_{\nu} \ H^+ \ \ell_L \quad {\rm and } \quad \ell_L^T \ C \ Y^{++}_\nu \ H^{++} \ \ell_L \; ,
\end{align}
where
\begin{align}
\label{eq:yukawa}
Y^+_{\nu}  =  \cos \theta_+ \ \frac{m_\nu^{diag}}{v_{\Delta}} \ U_{PMNS}^\dagger  \quad {\rm and } \quad Y^{++}_\nu = {M_\nu\over \sqrt{2} v_\Delta} = U_{PMNS}^* \ \frac{m_{\nu}^{diag}}{\sqrt{2} \ v_{\Delta}} \ U_{PMNS}^{\dagger}\;,
\end{align}
with $\theta_+ \thickapprox \sqrt{2} v_{\Delta}/v_0 $ being the singly charged Higgs mixing angle and $U_{PMNS}$ being the Pontecorvo-Maki-Nakagawa-Sakata (PMNS) neutrino mixing matrix. The partial width of doubly charged Higgs decay into same-sign leptons is thus given by
\begin{eqnarray}
\Gamma(H^{++}\to \ell^+_i \ell^+_j)={1\over 4\pi(1+\delta_{ij})}|(Y^{++}_{\nu})_{ij}|^2 M_{H^{++}},
\label{width-Hpp}
\end{eqnarray}
and the partial width of singly charged Higgs decay into the charged lepton and neutrino can be written as
\begin{eqnarray}
\Gamma(H^{+}\to \ell^+_i \bar{\nu}_j)={1\over 16\pi}|(Y^{+}_{\nu})_{ij}|^2 M_{H^{+}},
\label{width-Hp}
\end{eqnarray}
with $i,j=e, \mu, \tau$. According to Eq.~\eqref{eq:yukawa}, the Yukawa matrices that characterize the strength of the coupling between charged Higgs and leptons are inversely proportional to $v_\Delta$. When $v_{\Delta}$ is below $10^{-4}$ GeV, the decay modes of charged Higgs are dominated by the leptonic channels as given in Eq.~\eqref{width-Hpp} and Eq.~\eqref{width-Hp}. Their branching ratios will only depend on the PMNS matrix and neutrino masses.
If $v_{\Delta}$ is larger than $10^{-4}$ GeV, the main decay modes of charged Higgs will be the gauge boson channels~\cite{FileviezPerez:2008jbu}, i.e. $H^\pm\to W^\pm h/Z$~\footnote{The singly charged Higgs $H^+$ can also decay into $t\bar{b}$ if kinematically allowed. This channel is suppressed compared with $W^+ Z$ and $W^+ H_1$ when $m_{H^+}>400$ GeV~\cite{FileviezPerez:2008jbu}.} and $H^{\pm\pm}\to W^\pm W^\pm$. In this paper, we will discuss both the lepton and gauge boson decay channels in detail. For the purely leptonic decays, in particular, the branching ratios of the charged Higgs decay to different lepton flavors can be determined by the experimental data of neutrino oscillation. In the following subsection, we will focus on this impact and give a numerical analysis of the branching fractions of triplet Higgs leptonic decay.

In 2021, Fermilab released the result of the muon $g-2$ measurement and its combination with the Brookhaven experimental measurement led to a $4.2\sigma$ tension~\cite{Muong-2:2021ovs,Muong-2:2021vma,Muong-2:2021ojo}
\begin{eqnarray}
\Delta a_\mu = a_\mu^{\rm exp} - a_\mu^{\rm SM}=(2.51\pm 0.59)\times 10^{-9}\;.
\end{eqnarray}
The Yukawa interactions of scalar triplet can induce muon magnetic dipole moment at one loop level. It is given by~\cite{Li:2019xvv}
\begin{eqnarray}
\Delta a_\mu = {(Y_\nu^\dagger Y_\nu)^{\mu\mu}\over 6\pi^2} \Big({m_\mu^2\over m_{H^{++}}^2}+{m_\mu^2\over 8m_{H^{+}}^2}\Big)={3(Y_\nu^\dagger Y_\nu)^{\mu\mu} m_\mu^2\over 16\pi^2 m_{\Delta}^2}\;,
\end{eqnarray}
where the second equal sign is obtained by degenerate assumption. The muon $g-2$ anomaly would constrain
the $\mu\mu$ component of the coupling combination $(Y_\nu^\dagger Y_\nu)^{\mu\mu}$ as
\begin{eqnarray}
(Y_\nu^\dagger Y_\nu)^{\mu\mu}/m_{\Delta}^2 = (0.0000118\pm 0.0000028)~{\rm GeV}^{-2} \;.
\end{eqnarray}

\subsection{Impact of neutrino data on charged Higgs decay}
\label{sec:flavor}

The neutrino oscillation experiments provide us with the most precise data of neutrino masses and mixing so far. If neutrinos are Majorana particles, the neutrino mixing is characterized by the PMNS matrix
\begin{eqnarray}
\label{eq:PMNS}
U_{PMNS}&=& \left(\begin{array}{ccc}
c_{12} c_{13} & s_{12} c_{13} & s_{13} e^{-i \delta_{\mathrm{CP}}} \\
-s_{12} c_{23}-c_{12} s_{23} s_{13} e^{i \delta_{\mathrm{CP}}} & c_{12} c_{23}-s_{12} s_{23} s_{13} e^{i \delta_{\mathrm{CP}}} & s_{23} c_{13} \\
s_{12} s_{23}-c_{12} c_{23} s_{13} e^{i \delta_{\mathrm{CP}}} & -c_{12} s_{23}-s_{12} c_{23} s_{13} e^{i \delta_{\mathrm{CP}}} & c_{23} c_{13}
\end{array}\right) \nonumber \\
&&\times \text{diag} (e^{i \Phi_1/2}, 1, e^{i \Phi_2/2})\,,
\end{eqnarray}
where $s_{ij}\equiv\sin{\theta_{ij}}$, $c_{ij}\equiv\cos{\theta_{ij}}$, $0 \le
\theta_{ij} \le \pi/2$ and $0 \le \delta_{\mathrm{CP}}, \Phi_i < 2\pi$ with $\delta_{\mathrm{CP}}$ being the Dirac CP phase and $\Phi_i$ being the Majorana phases.
For the neutrino masses, the oscillation experiments can determine the mass-squared splitting between three neutrino mass eigenstates up to normal and inverted neutrino mass hierarchies. With the inclusion of the data on atmospheric neutrinos provided by the Super-Kamiokande collaboration, the latest best global fit results of the neutrino masses and mixing parameters are shown in Table~\ref{table:mixing}~\cite{Esteban:2020cvm,nufit2021} for both NH and IH.
We also adopt the tightest constraint on the sum of neutrino masses
by combining the TT, TE, EE+lowE+lensing+BAO data~\cite{Vagnozzi:2017ovm,Planck:2018vyg} at $95\%$ CL,
\beq
\sum_{i=1}^3 m_i < \ 0.12 \ \ev .
\label{cosmo}
\eeq
The neutrino mixing parameters can be varied according to a normal distribution with central values and $1\sigma$ ranges given in Table~\ref{table:mixing} for NH and IH. The constraints on neutrino masses are applicable together with Eq.~\eqref{cosmo}. We are then able to perform a numerical analysis of the leptonic decay branching ratios of the charged Higgs bosons based on Eq.~\eqref{width-Hpp} and Eq.~\eqref{width-Hp}.

\begin{table}[t!]
\centering
\begin{tabular}{|c|c|c|}
\hline
\hline
    Parameter & Normal Hierarchy & Inverted Hierarchy   \\ \hline
    $\sin^2{\theta_{12}}$ & $0.304^{+0.012}_{-0.012}$ & $0.304^{+0.013}_{-0.012}$ \\
    $\sin^2{\theta_{23}}$ & $0.450^{+0.019}_{-0.016}$ & $0.570^{+0.016}_{-0.022}$ \\
    $\sin^2{\theta_{13}}$ & $0.02246^{+0.00062}_{-0.00062}$ & $0.02241^{+0.00074}_{-0.00062}$  \\
    $\delta_{\mathrm{CP}}[^{\circ}]$ & $230^{ +36}_{-25}$ & $278^{ +22}_{-30}$  \\
    $\Delta m_{21}^2[10^{-5} \ev^2]$ & $7.42^{+0.21}_{-0.20}$ & $7.42^{+0.21}_{-0.20}$  \\
    $\Delta m_{3\ell}^2[10^{-3} \ev^2]$ & $+2.510^{+0.027}_{-0.027}$ & $-2.490^{+0.026}_{-0.028}$  \\ \hline
\hline
\end{tabular}
\caption{The best fit of the neutrino oscillation parameters with $1\sigma$ range~\cite{Esteban:2020cvm,nufit2021}. $\Delta m_{3\ell}^2 \equiv \Delta m_{31}^2$ for NH and $\Delta m_{3\ell}^2 \equiv \Delta m_{32}^2$ for IH.
}
\label{table:mixing}
\end{table}

Nowadays, we still can not tell the nature of the neutrinos from the neutrino oscillation experiments or the neutrinoless double beta decay experiments. It means neutrinos can be either Dirac or Majorana particles. If neutrino has a Dirac nature, the two Majorana phases in Eq.~\eqref{eq:PMNS} can be further absorbed by rephrasing the fields, and we can simply set them to be zeros in a physical basis. If neutrinos are Majorana particles, the Majorana phases could have non-vanishing values.

\begin{figure}[htb!]
\begin{center}
\includegraphics[height=0.23\textheight]{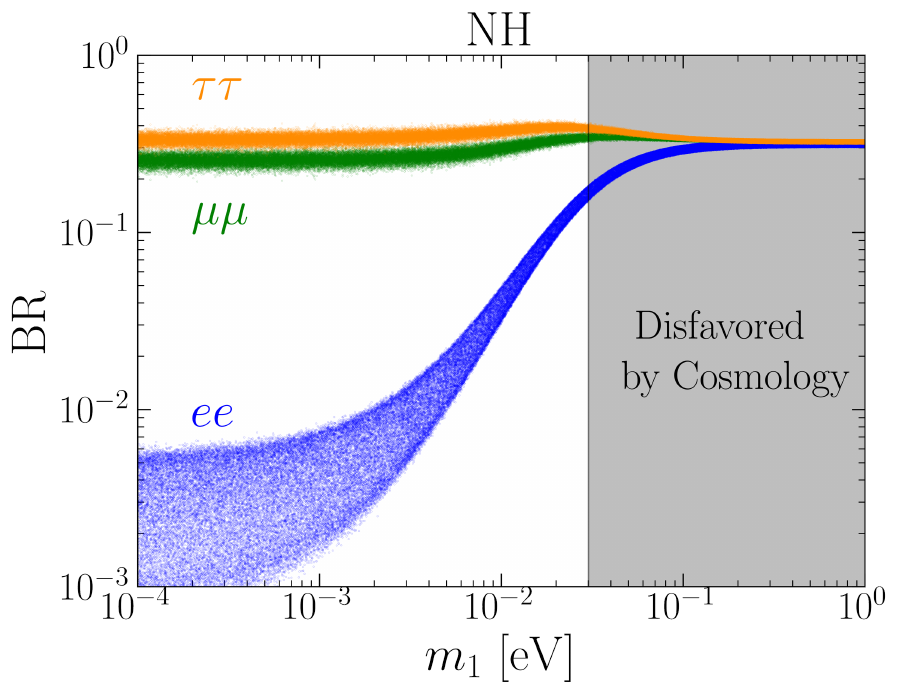}
\includegraphics[height=0.23\textheight]{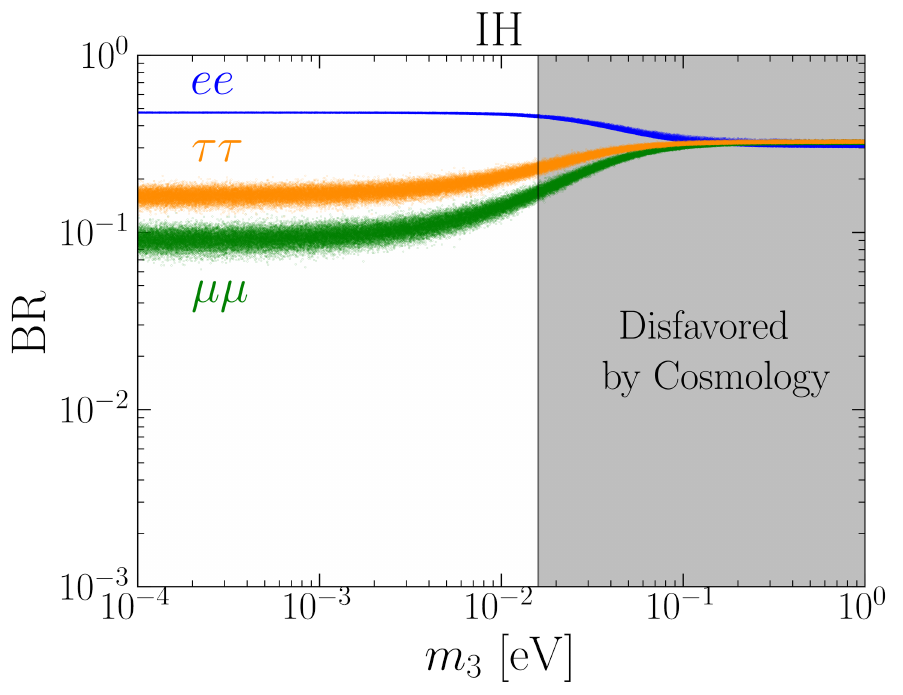}
\includegraphics[height=0.23\textheight]{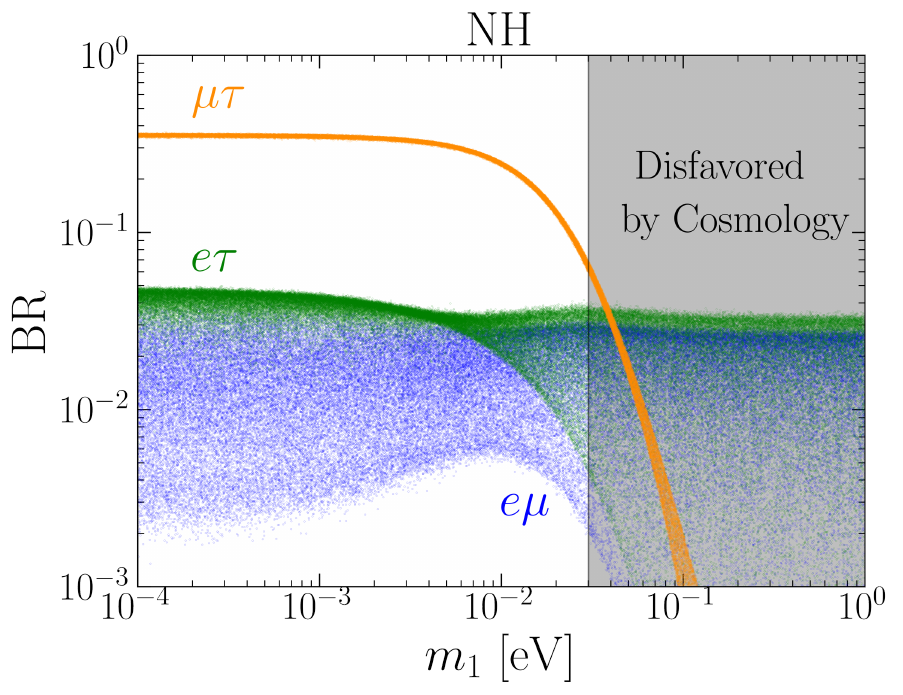}
\includegraphics[height=0.23\textheight]{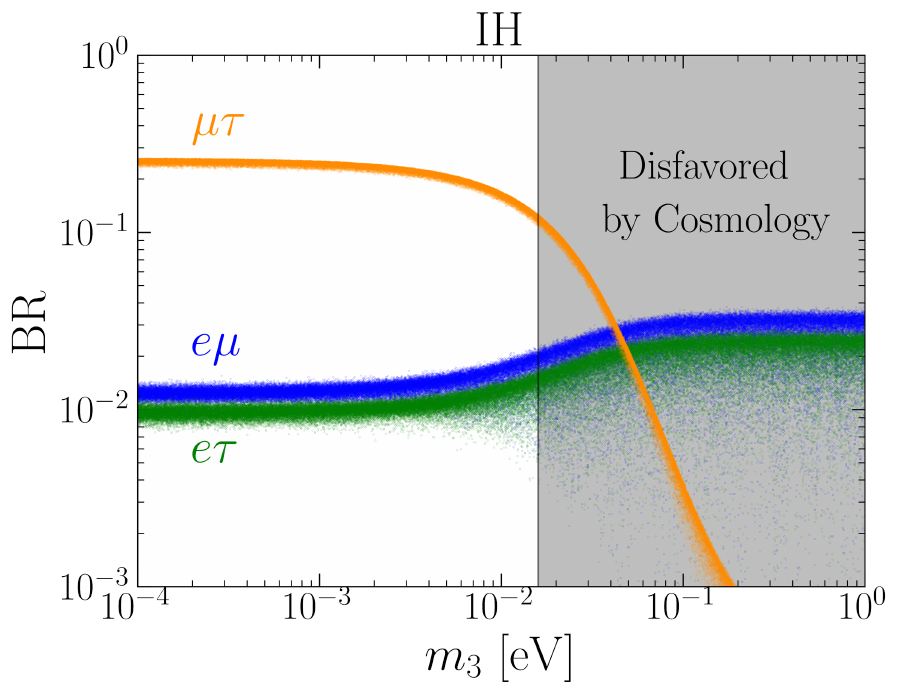}
\end{center}
\caption{Scatter plots for the branching ratios of $H^{++}$ decay into the same flavor~(two upper panels) and different flavor~(two lower panels) like-sign leptons versus the lowest neutrino mass for NH (two left panels) and IH (two right panels) with $\Phi_1 = \Phi_2 = 0$. The grey bands correspond to the current limit on the lightest neutrino mass $m_{1(3)}$ from the cosmological data of $\sum m_i<0.12$~eV by the Planck collaboration~\cite{Planck:2018vyg}.
}
\label{scatterBR-Hpp}
\end{figure}

\begin{figure}[htb!]
\begin{center}
\includegraphics[height=0.23\textheight]{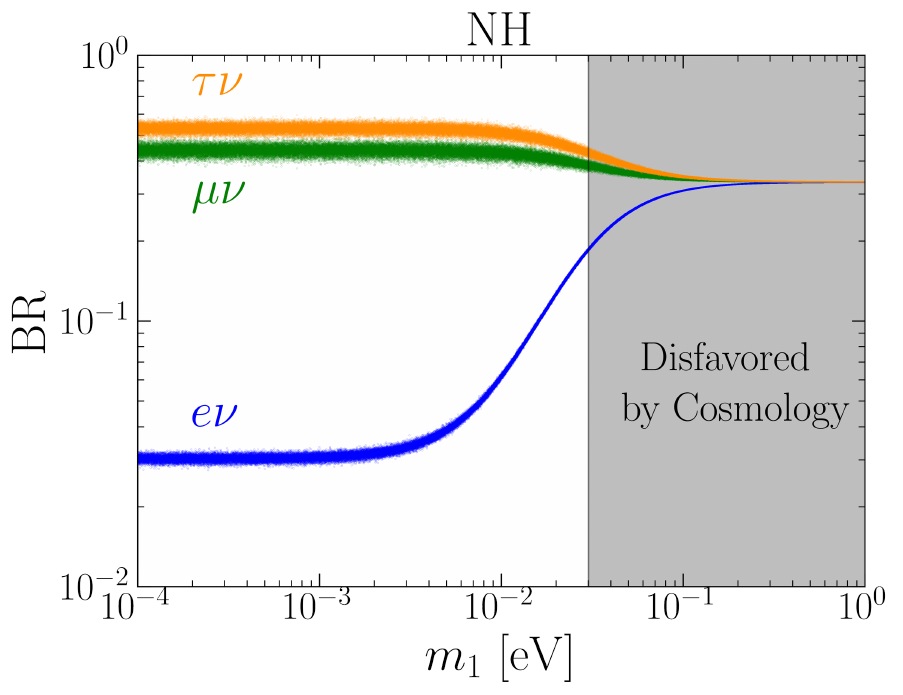}
\includegraphics[height=0.23\textheight]{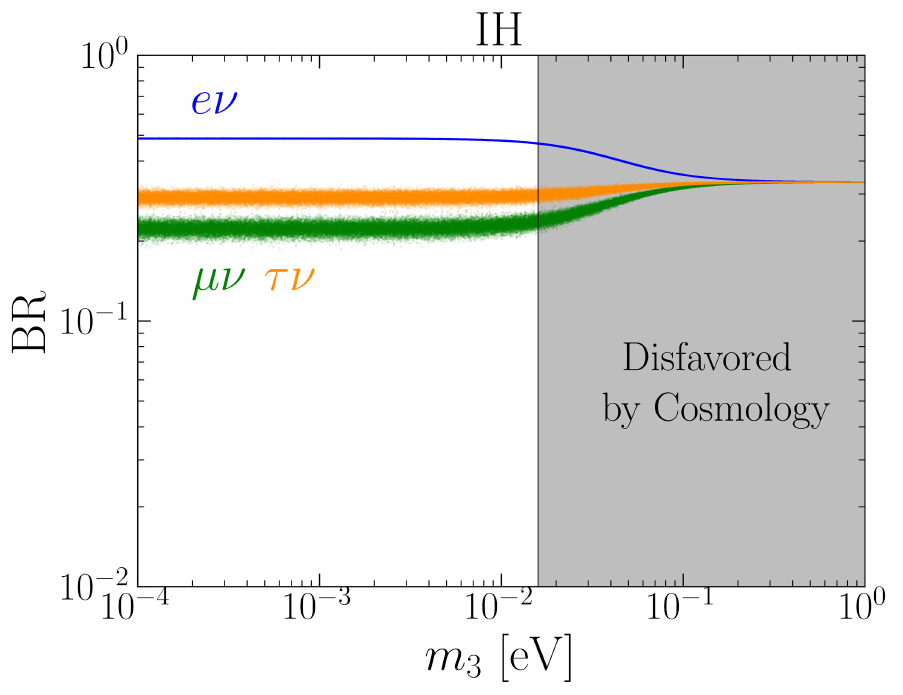}
\end{center}
\caption{Scatter plots for the $H^{+}$ decay branching ratios to leptons versus the lowest neutrino mass for NH~(left) and IH~(right) with $\Phi_1 = \Phi_2 = 0$. The grey bands correspond to the current limit on the lightest neutrino mass $m_{1(3)}$ from the cosmological data of $\sum m_i<0.12$~eV by the Planck collaboration~\cite{Planck:2018vyg}.
 }
\label{scatterBR-Hp}
\end{figure}

We first consider the case with $\Phi_1 = \Phi_2 = 0$. We show the scatter plots of the decay branching ratios of doubly charged Higgs ($H^{++}\to \ell_i^+\ell_j^+$) versus the lightest neutrino mass in Fig.~\ref{scatterBR-Hpp}. The lightest neutrino with $m_{1(3)}\gtrsim 10^{-2}$~eV would be excluded by the cosmological data of $\sum m_i < 0.12$~eV by the Planck collaboration~\cite{Planck:2018vyg}, as indicated by the grey region in the scatter plots.
The branching ratios of $H^{++}$ decay into the same flavor ($i=j$) and different flavor ($i\neq j$) like-sign leptons are shown in the upper and lower panels of Fig.~\ref{scatterBR-Hpp}, respectively.
The results for the NH and IH scenarios are also shown explicitly in the left and right panels of Fig.~\ref{scatterBR-Hpp}. In the NH case, the leading decay channels are given by the second and third lepton flavors, i.e., the $\mu\mu, \tau\tau$ and $\mu\tau$ channels. Each of their branching ratios approximately reaches $\sim30\%$.
The remaining flavor combinations relate to the electron, i.e., the $ee, e\tau$ and $e\mu$ channels. Their branching fractions are at least one order of magnitude smaller than the leading channels. For the IH case, by contrast, the $ee$ channel dominates the branching ratios with the value close to $50\%$. The $\mu\tau, \tau\tau$ and $\mu\mu$ channels have lower but same order branching ratios. The $e\mu$ and $e\tau$ channels are suppressed by one order of magnitude. It is also worth mentioning that some branching ratios of the decay channels are constrained within very narrow bands, for example the $\mu\tau$ channel in both NH and IH cases, and the $ee$ channel in IH case. The accuracy is considerably improved compared with the results from previous neutrino oscillation data~\cite{FileviezPerez:2008jbu}.
For the singly charged Higgs $H^+$, the scatter plots of the branching ratios versus the lightest neutrino mass are shown in Fig.~\ref{scatterBR-Hp} for both NH and IH spectra. In the NH case, the $\mu\bar{\nu}$ and $\tau\bar{\nu}$ channels are dominant and comparable. The $e\bar{\nu}$ channel is one order of magnitude smaller. By contrast, the $e\bar{\nu}$ channel is dominant with the branching fraction being $\sim 50\%$ in IH case. The $\mu\bar{\nu}$ and $\tau\bar{\nu}$ channels are at the same order of magnitude with $e\bar{\nu}$.

Next, we will only focus on some intuitive benchmarks with specific decay branching ratios and perform a phenomenological study at a muon collider. The benchmarks of doubly (singly) charged Higgs decay are shown in Table~\ref{BR-Hpp} (Table~\ref{BR-Hp}). The lightest neutrino mass is assumed to have a small value $m_{1(3)}=10^{-4}$ for NH (IH) spectrum. They are obtained by fixing the neutrino mass and mixing parameters to their best-fit values.

\begin{table}[htb!]
\begin{center}
\begin{tabular}{|c|c|c|c|c|c|c|}
\hline
${\rm BR}(H^{++})$  & $ee$ & $e\mu$ & $e\tau$ & $\mu\mu$ & $\mu\tau$ & $\tau\tau$
\\ \hline
NH & 0.28\% & 1.25\% & 4.27\% & 25.57\%  & 35.43\% & 33.20\%
\\ \hline
IH & 47.49\% & 1.23\% & 0.96\% & 8.84\%  & 25.63\% & 15.85\%
\\ \hline
\end{tabular}
\end{center}
\caption{Benchmark decay branching ratios of doubly charged Higgs for NH and IH spectra. They are obtained by fixing the neutrino mass and mixing parameters at their best-fit values, and the lightest neutrino mass is assumed to be $m_{1(3)}=10^{-4}$~eV for NH (IH). We assume vanishing Majorana phases.
}
\label{BR-Hpp}
\end{table}

\begin{table}[htb!]
\begin{center}
\begin{tabular}{|c|c|c|c|}
\hline
${\rm BR}(H^{+})$  & $e\bar{\nu}$ & $\mu\bar{\nu}$ & $\tau\bar{\nu}$
\\ \hline
NH & 3.04\% & 43.91\% & 53.05\%
\\ \hline
IH & 48.59\% & 22.27\% & 29.14\%
\\ \hline
\end{tabular}
\end{center}
\caption{Benchmark decay branching ratios of singly charged Higgs for NH and IH spectra. They are obtained by fixing the neutrino mass and mixing parameters at their best-fit values, and the lightest neutrino mass is assumed to be $m_{1(3)}=10^{-4}$~eV for NH (IH). We assume vanishing Majorana phases. The light neutrinos in final states are summed over.
}
\label{BR-Hp}
\end{table}

\begin{figure}[htb!]
\begin{center}
\includegraphics[height=0.23\textheight]{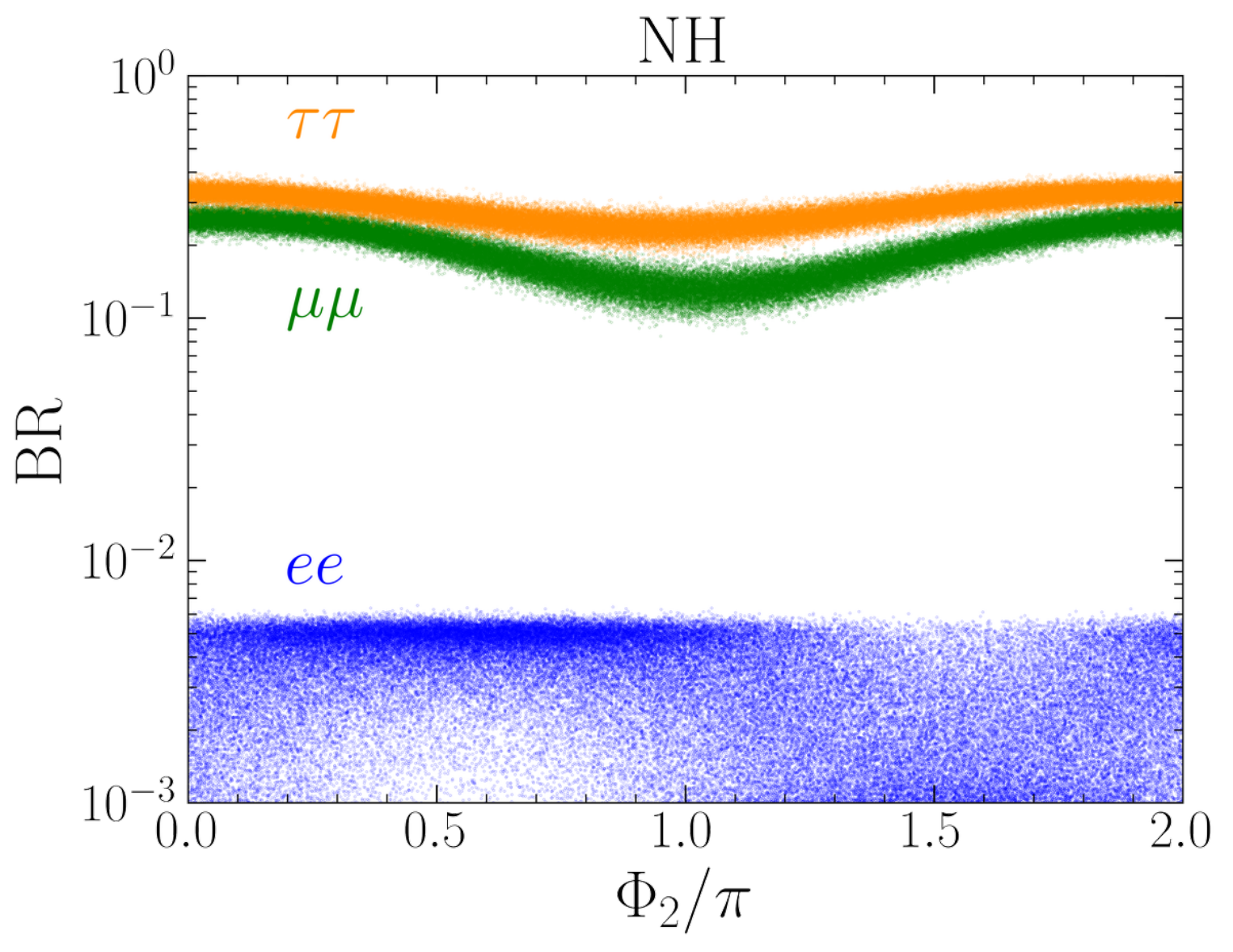}
\includegraphics[height=0.23\textheight]{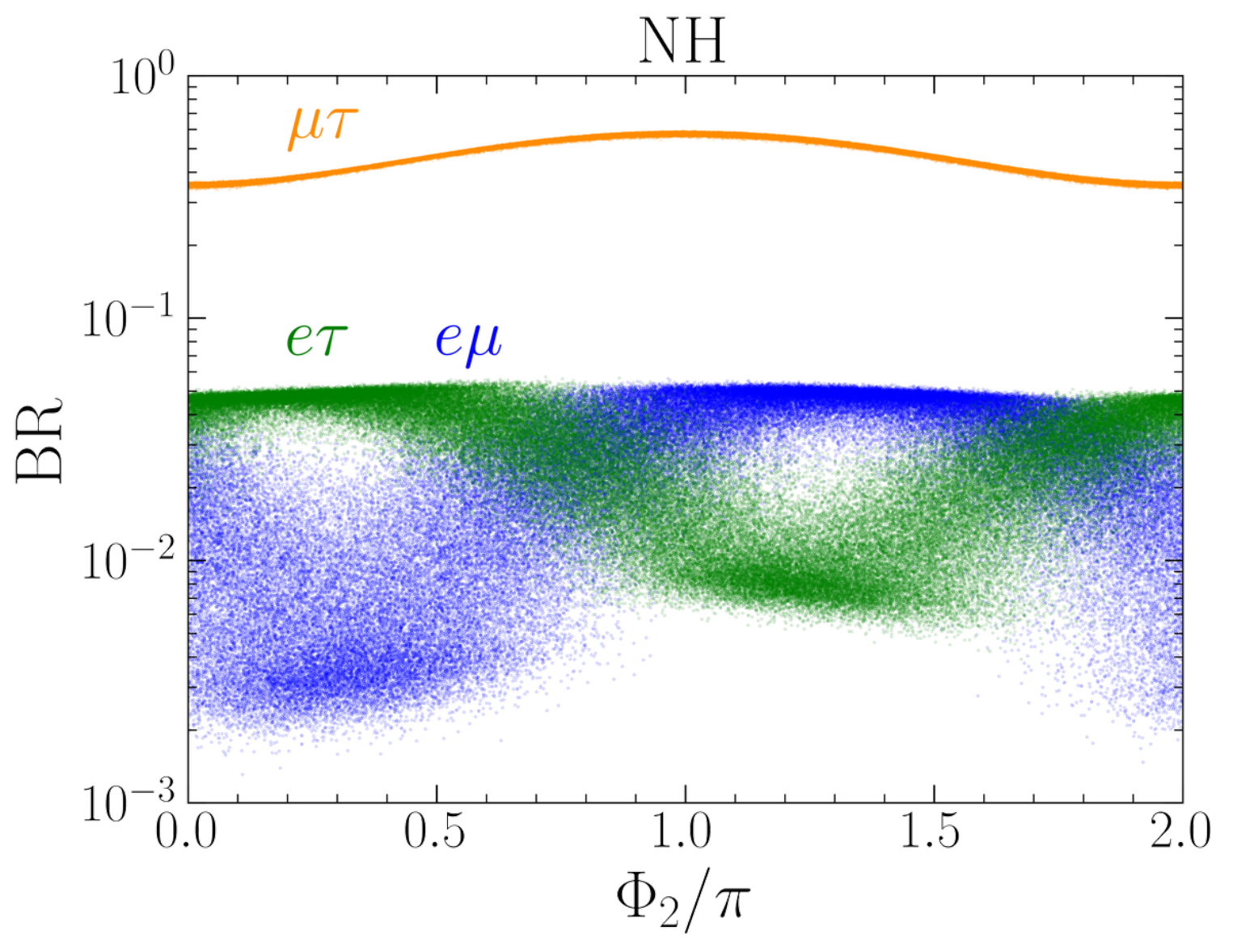}
\end{center}
\caption{Scatter plots for the same flavor (left) and different flavor (right) leptonic branching ratios for the $H^{++}$ decay versus Majorana phase $\Phi_2$ for the NH spectrum with $m_1 = 0$. The other Majorana phase $\Phi_1$ is uniformly sampled within $[0,2\pi)$.}
\label{scatterBR-phi2-NH}
\end{figure}

\begin{figure}[htb!]
\begin{center}
\includegraphics[height=0.23\textheight]{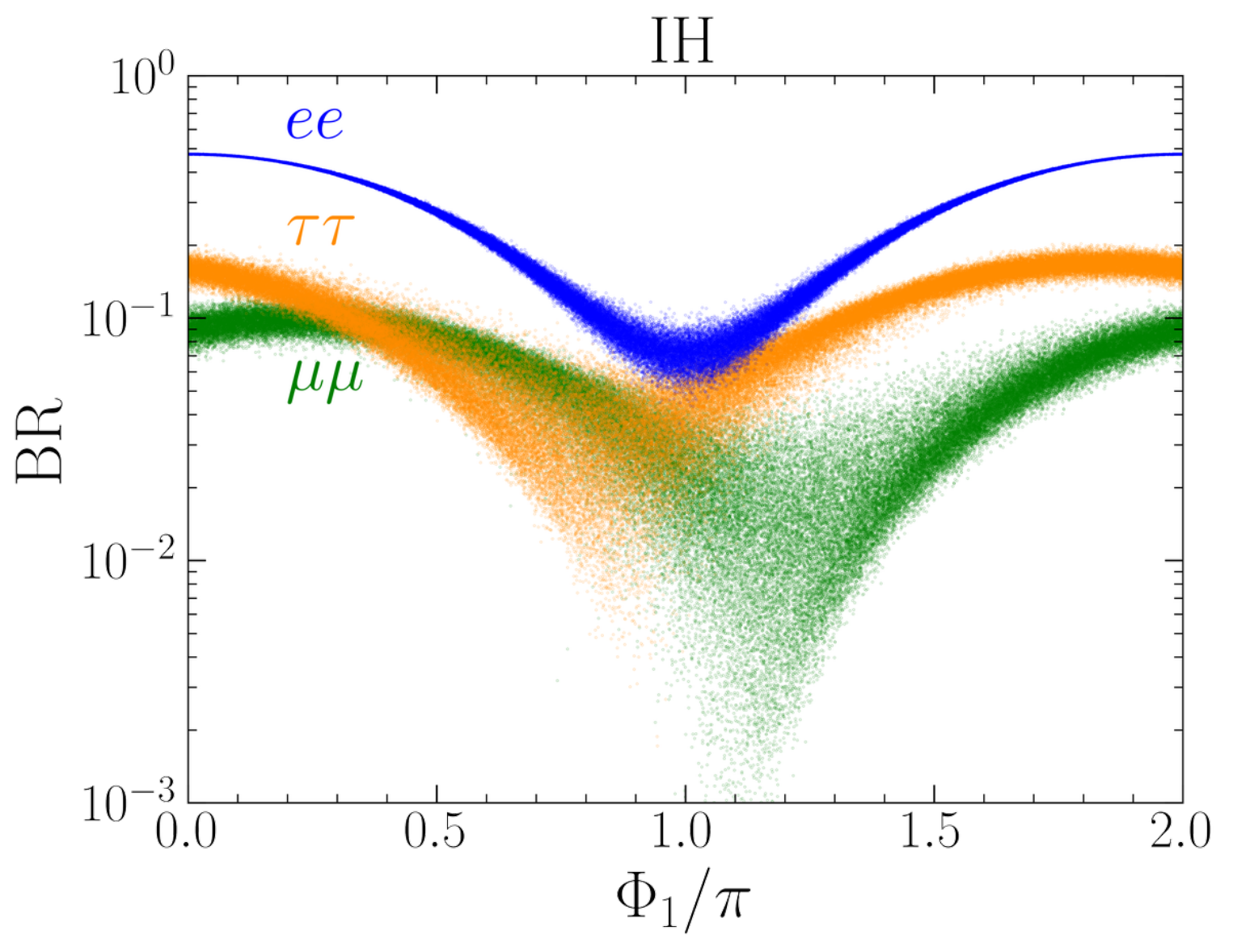}
\includegraphics[height=0.23\textheight]{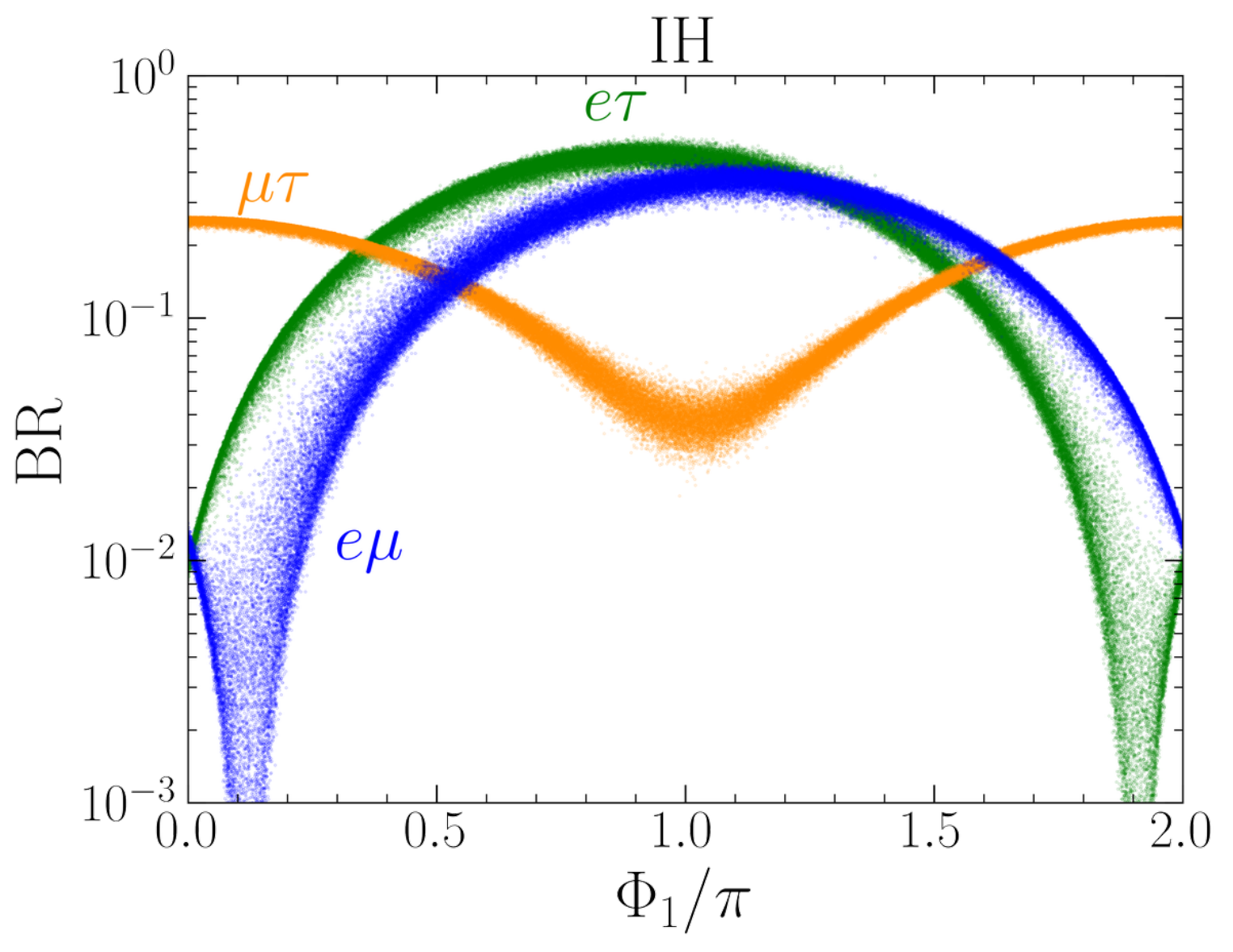}
\end{center}
\caption{Scatter plots for the same flavor~(left) and different flavor~(right) leptonic branching ratios for the $H^{++}$ decay versus Majorana phase $\Phi_1$ for the IH spectrum with $m_3 = 0$. The other Majorana phase $\Phi_1$ is uniformly sampled within $[0,2\pi)$.
}
\label{scatterBR-phi1-IH}
\end{figure}

Then we consider the case with nonzero Majorana phases ($\Phi_1,\Phi_2 \neq 0$). The effects of the Majorana phases on the Higgs decays have been investigated in Refs.~\cite{FileviezPerez:2008jbu,Garayoa:2007fw,Kadastik:2007yd,Akeroyd:2007zv}. The main conclusion is that the decay branching ratios of doubly charge Higgs slightly depend on $\Phi_2$ in NH case, whereas the dependence on $\Phi_1$ is strong in IH case. The singly charged Higgs decays do not depend on the Majorana phases. The $H^{++}$ decay branching ratios for all channels are shown in Fig.~\ref{scatterBR-phi2-NH} for the NH case. We can see that the branching ratios indeed have a rather weak dependence on the Majorana phase $\Phi_2$. The maximal suppression (enhancement) happens when $\Phi_2=\pi$ for the $\tau\tau$ and $\mu\mu$ ($\mu\tau$) channels. It changes the branching ratios by a factor of two at most. For the case of IH spectrum, we show the scatter plots of the doubly charged Higgs decay branching ratios versus $\Phi_1$ in Fig.~\ref{scatterBR-phi1-IH}. There is a strong dependence on the Majorana phase $\Phi_1$ for all decay branching ratios. The leading decay channels swap from $ee, \mu\tau$ at $\Phi_1=0$ to $e\mu, e\tau$ at $\Phi_1=\pi$. Such a significant oscillation signal can be used to determine the value of the Majorana phase $\Phi_1$.

\begin{figure}[htb!]
\begin{center}
\includegraphics[height=0.23\textheight]{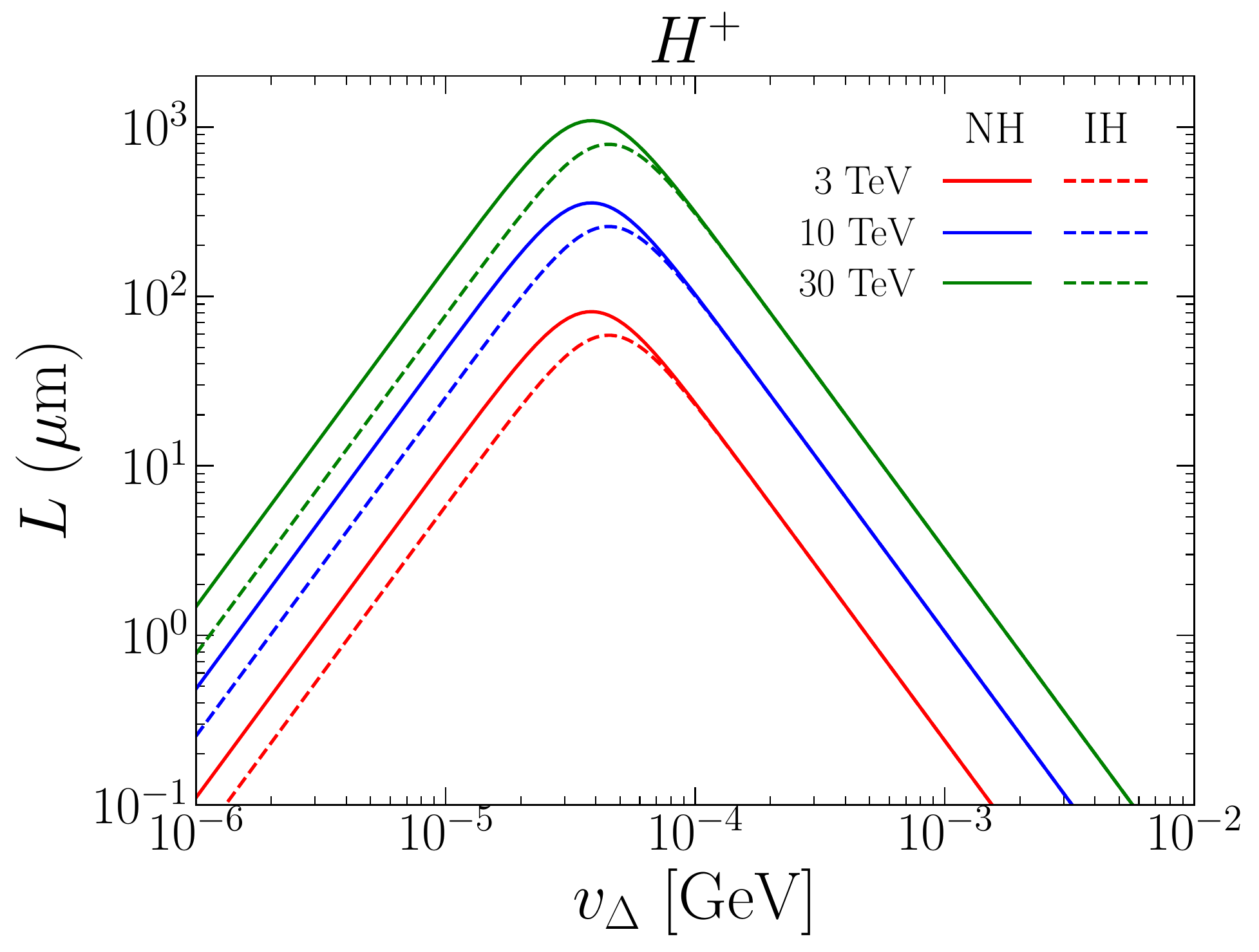}
\includegraphics[height=0.23\textheight]{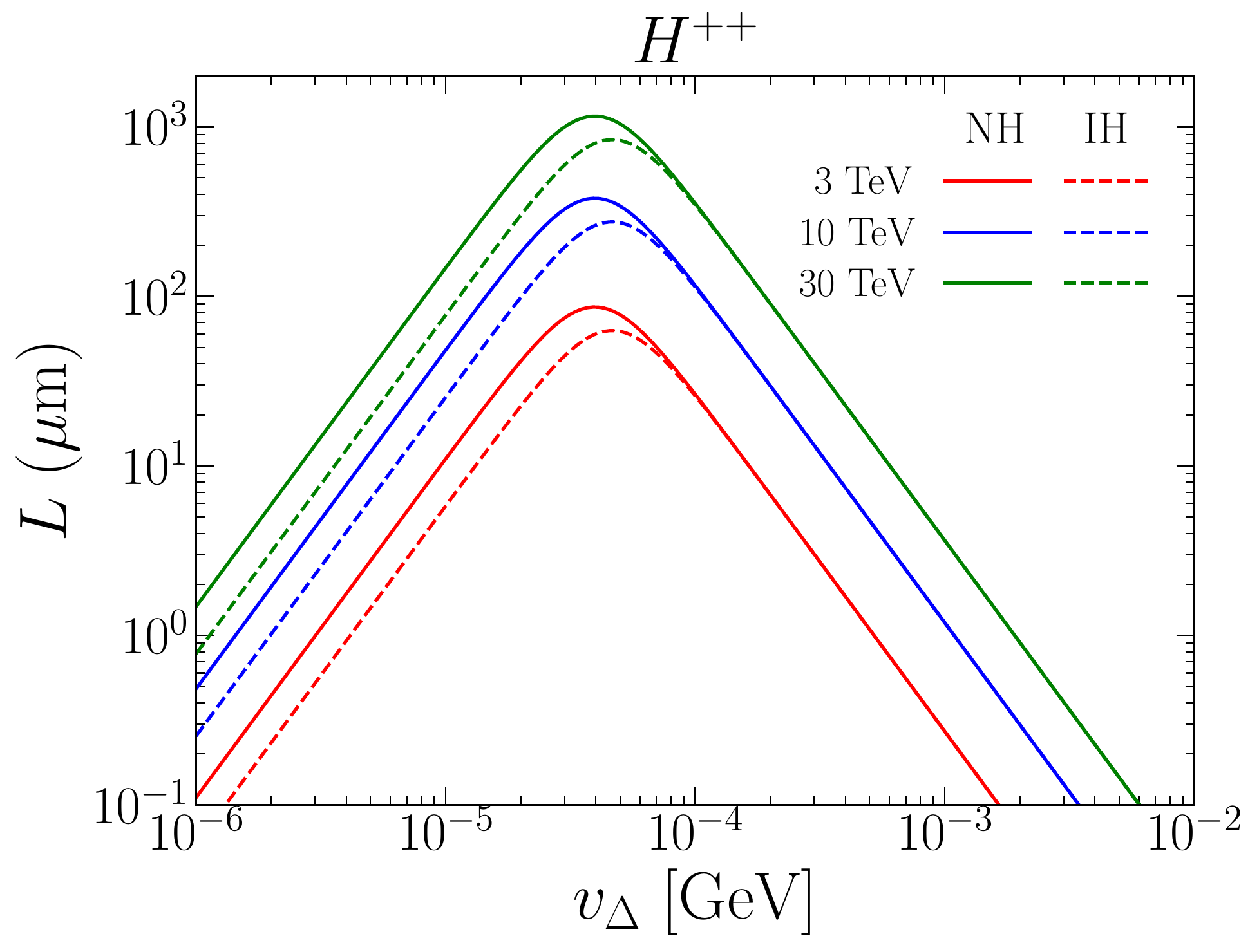}
\end{center}
\caption{The decay length of singly charged Higgs $H^+$ (left) and doubly charged Higgs $H^{++}$ (right). The c.m. energy is taken to be $\sqrt{s}=3,~10,~30$ TeV and $m_{H^{++}}=m_{H^{+}}=1$ TeV.
}
\label{lifetime}
\end{figure}

The leptonic Yukawa couplings of charged Higgs bosons are proportional to light neutrino masses. As a result, their decays may lead to long decay length and displaced vertex in the detector. The decay length of the charged Higgses is given by $L= \gamma\beta c\tau$, where $\tau=1/\Gamma$ with $\Gamma$ being the total width, $\gamma$ is the boost factor and $\beta$ is the ratio of Higgs velocity to the speed of light $c$. The product $\gamma\beta$ is $\sqrt{E_\Delta^2/M_\Delta^2-1}$ and $E_\Delta$ is taken to be $\sqrt{s}/2$ in the c.m. frame. We take into account the above best-fitted neutrino mass and mixing parameters, and show the decay lengths as a function of the triplet vev in Fig.~\ref{lifetime}. The c.m. energy is taken to be $\sqrt{s}=3,~10,~30$ TeV and we assume $m_{H^{++}(H^{+})}=1$ TeV. The decay widths of the $H^{+}$ and $H^{++}$ are approximately the same for large triplet Higgs mass. We find that the decay length could be as large as 1 mm and there is a clear distinction between NH and IH when $v_\Delta\lesssim 10^{-4}$ GeV.
If the displaced vertex searches can be performed in future, it could serve as an indication to distinguish the neutrino mass patterns.

\section{Testing Type II Seesaw at muon collider}
\label{sec:simulation}

In this section we discuss the productions and signatures of doubly and singly charged Higgs bosons in Type II Seesaw mechanism at high-energy muon collider.

\subsection{Production channels}

\begin{figure}[h!]
\begin{center}
\minigraph{5cm}{-0.05in}{(a)}{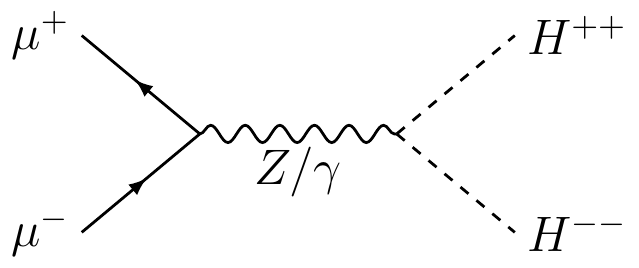}
\minigraph{5cm}{-0.05in}{(b)}{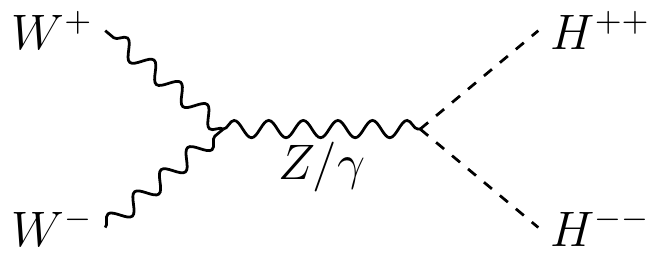}\\
\minigraph{4.7cm}{-0.05in}{(c)}{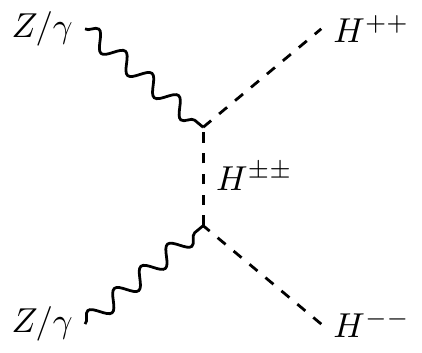}
\minigraph{4.7cm}{-0.05in}{(d)}{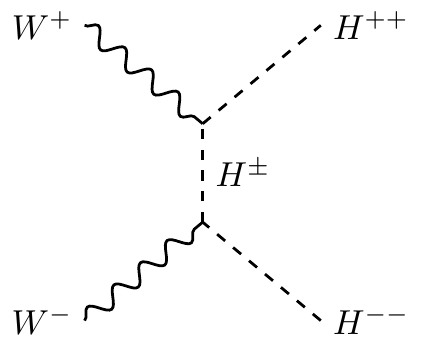}\\
\minigraph{4.7cm}{-0.05in}{(e)}{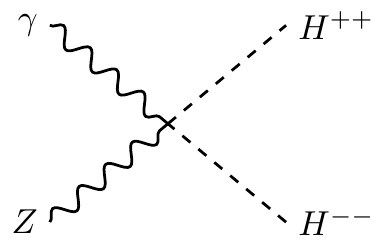}
\minigraph{4.7cm}{-0.05in}{(f)}{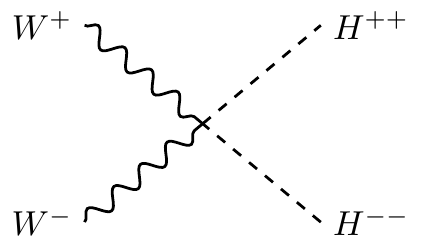}
\end{center}
\caption{Feynman diagrams of heavy doubly charged Higgs pair productions in Type II Seesaw. The diagram (a) is the $\mu\mu$ annihilation process ($\mu^+\mu^-\to H^{++}H^{--}$) and the others are though VBF processes ($ VV \to H^{++}H^{--}$).
}
\label{diagram:HppHmm}
\end{figure}

\begin{figure}[ht!]
\centering
\includegraphics[height=0.23\textheight]{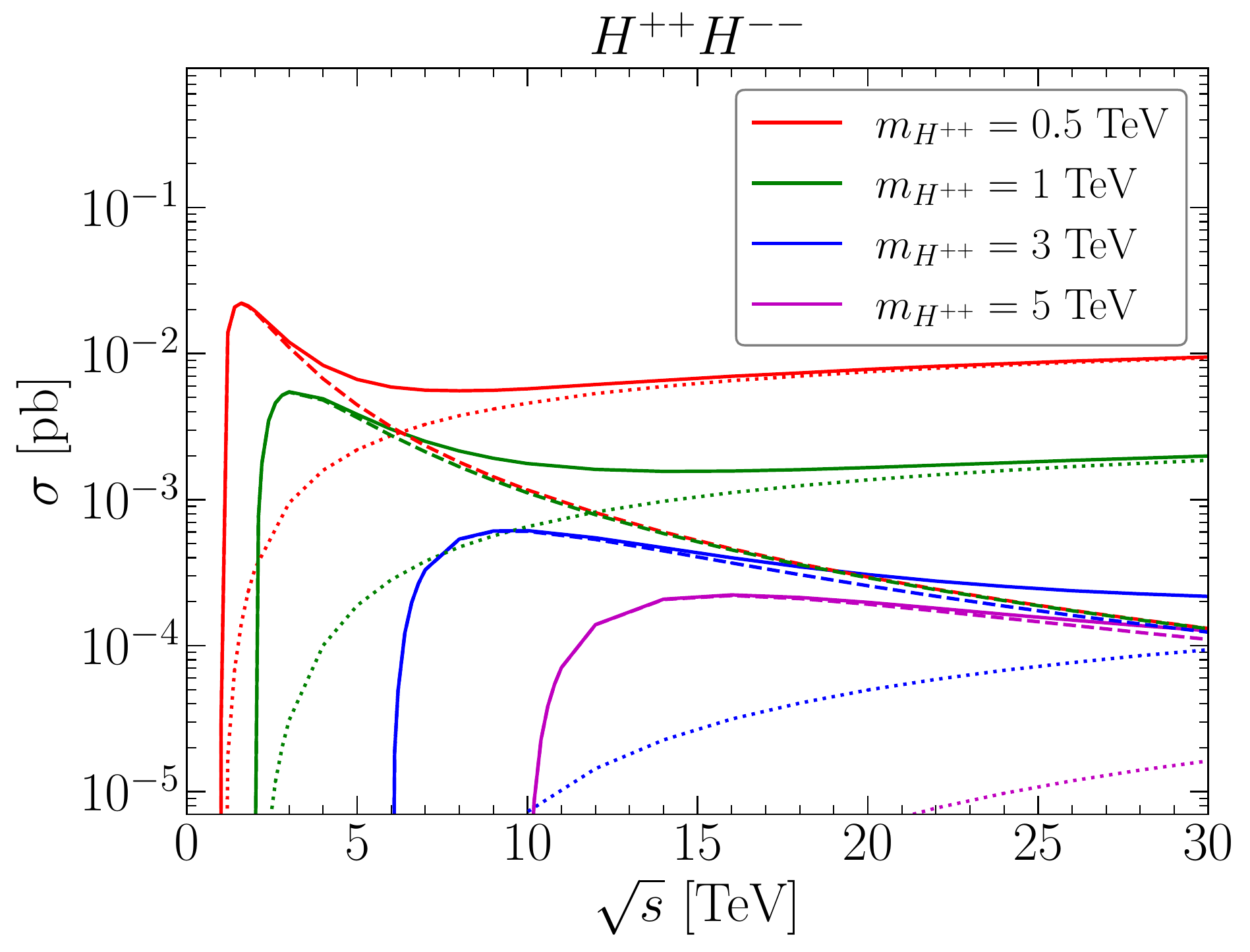}
\includegraphics[height=0.23\textheight]{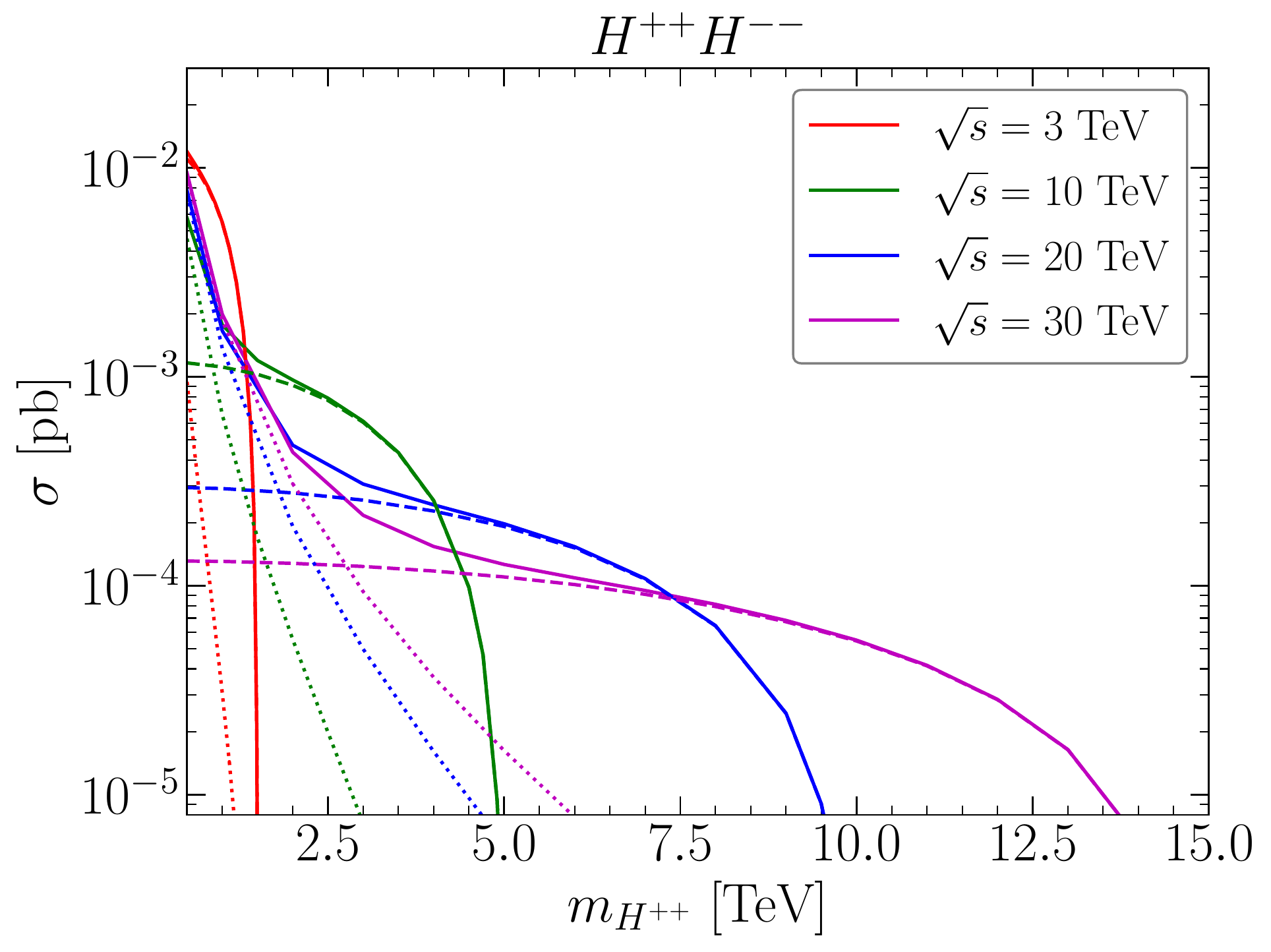}\\
\caption{Cross sections of $H^{++}H^{--}$ pair production as a function of the collider energy $\sqrt{s}$ (left) and heavy Higgs masses $m_{H^{\pm\pm}}$ (right) at muon colliders, through $\mu^+\mu^-$ annihilation (dashed lines) and VBF (dotted lines) processes. Their sum is shown as solid lines. We have chosen four benchmark masses of the $H^{\pm\pm}$, that are $0.5, 1, 3$ and $5$ TeV in the left panel. In the right panel, the c.m. energy is assumed to be $\sqrt{s}=3, 10, 20$ or 30 TeV. }
\label{fig:xsection_hpphmm}
\end{figure}

First of all, the doubly charged Higgs can be produced in pairs by either direct $\mu^+\mu^-$ annihilation or the fusion of electroweak vector bosons (generally denoted by $V$)
\begin{eqnarray}
\mu^+\mu^-, VV \to H^{++}H^{--}\;.
\label{eqn_HppHmm}
\end{eqnarray}
For the VBF processes, the vector bosons are taken as initial partons and produce ``inclusive'' processes.
The Feynman diagrams of these production processes in Type II Seesaw are collected in Fig.~\ref{diagram:HppHmm}.
We use FeynRules UFO file of the Type II Seesaw~\cite{Fuks:2019clu} and MadGraph5\_aMC@NLO~\cite{Alwall:2014hca} to calculate the cross sections and the results are shown in Fig.~\ref{fig:xsection_hpphmm}, as a function of c.m. energy $\sqrt{s}$ (left panel) and doubly charged Higgs mass $m_{H^{\pm\pm}}$ (right panel). The productions are categorized into $\mu^{+}\mu^{-}$ annihilation (dashed lines), VBF (dotted lines) and their sum (solid lines). In the new version of MadGraph5, the leading-order framework of electroweak parton distribution functions (EW PDFs) were embedded to calculate the VBF processes~\cite{Ruiz:2021tdt}.

In Fig.~\ref{fig:xsection_hpphmm}, one can see that the doubly charged Higgs can be produced in pairs via the direct $\mu^{+}\mu^{-}$ annihilation above the kinematic threshold. The cross sections of $\mu^{+}\mu^{-}$ annihilation ($\sigma^{\rm Ann}$) behave like $\sigma^{\rm Ann}\sim \beta^3/s$ with the velocity as $\beta = \sqrt{1-4m_{H^{\pm\pm}}^2/s}$. As a result, the cross sections of $\mu^{+}\mu^{-}$ annihilation decrease with increasing $\sqrt{s}$ well above threshold, and doubly charged Higgs bosons with different masses tend to be indistinguishable at high energies.
The cross sections of VBF processes ($\sigma^{\rm VBF}$) are enhanced at high beam energies by collinear logarithm ln($\hat{s}/m_{\mu}^2$) for photon or ln($\hat{s}/m_{V}^2$) for massive gauge boson $V$ with $\sqrt{\hat{s}}$ being the parton-level c.m. energy. It turns out that the VBF processes dominate at high energies compared with the $\mu^+\mu^-$ annihilation channel. Moreover, the VBF processes are more sensitive to the Higgs mass and the cross sections decrease along with increasing $m_{H^{\pm\pm}}$.

\begin{figure}[h!]
\begin{center}
\minigraph{4.7cm}{-0.05in}{(a)}{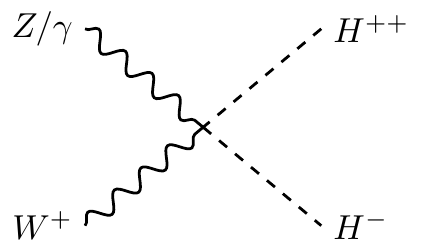}
\minigraph{6.5cm}{-0.05in}{(b)}{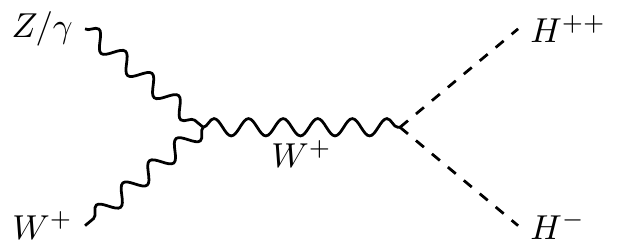}\\
\minigraph{4.7cm}{-0.05in}{(c)}{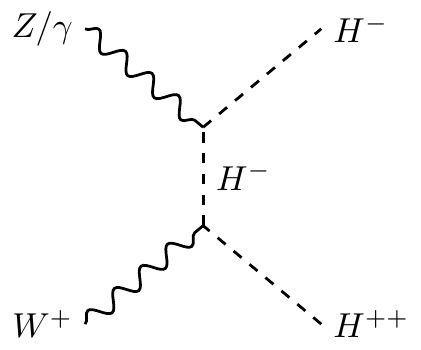}
\minigraph{4.7cm}{-0.05in}{(d)}{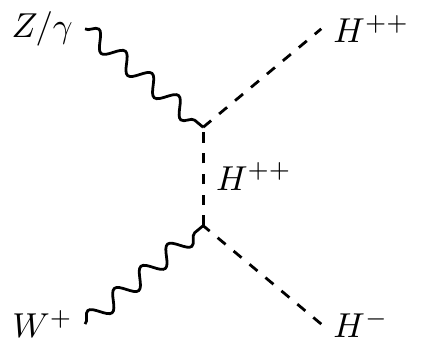}
\end{center}
\caption{Feynman diagrams of doubly and singly charged Higgs associated productions in Type II Seesaw. The diagrams are all produced by VBF processes (only $VV\to H^{++}H^{-}$ for illustration). }
\label{diagram:HppHm}
\end{figure}

\begin{figure}[ht!]
\centering
\includegraphics[height=0.23\textheight]{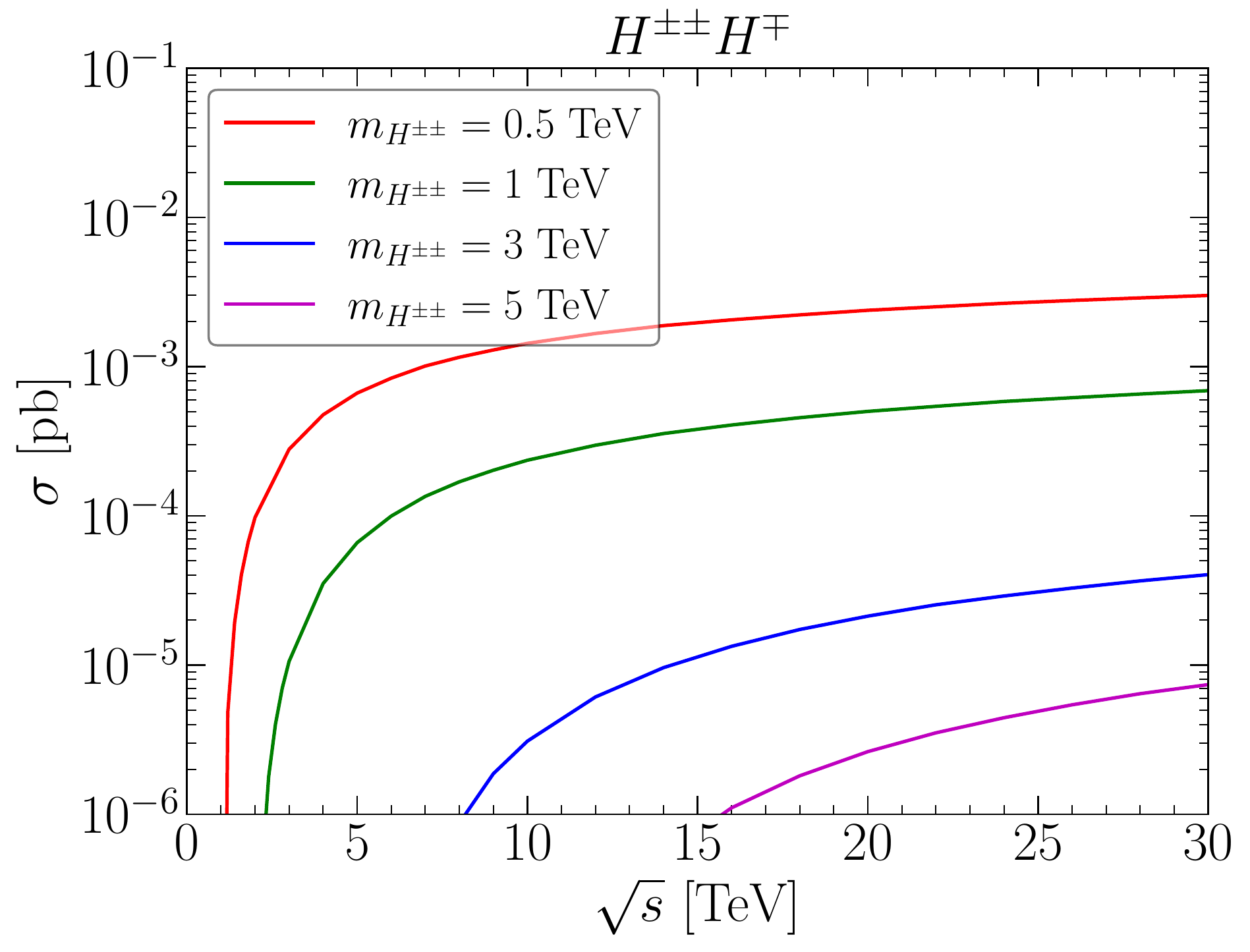}
\includegraphics[height=0.23\textheight]{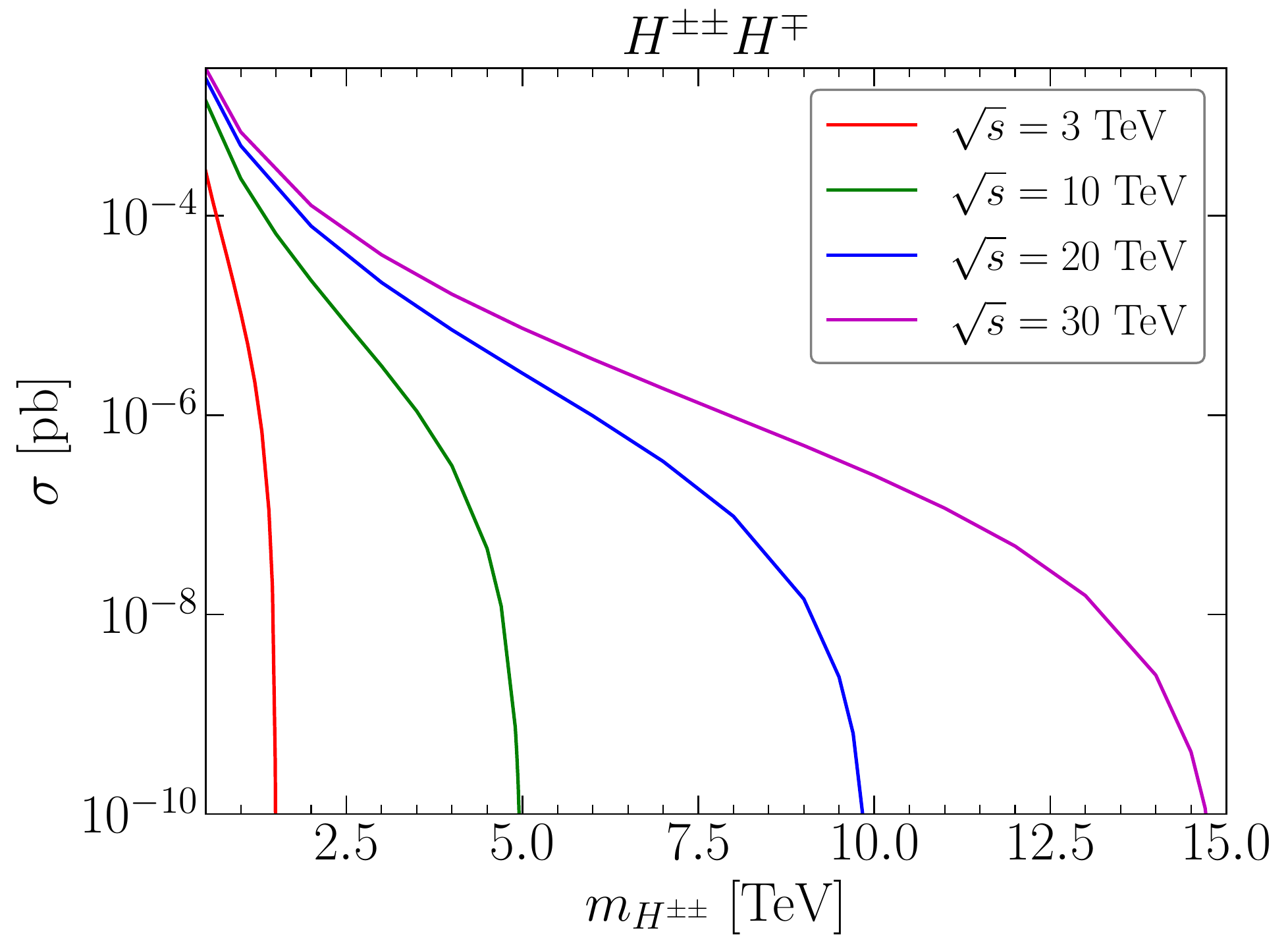}
\caption{Cross sections of $H^{\pm\pm}H^{\mp}$ associated production as a function of the collider energy $\sqrt{s}$ (left) and heavy Higgs masses $m_{H^{\pm\pm}}$ (equal to $m_{H^\pm}$) (right) at muon colliders only by VBF processes. The charged Higgs masses and the c.m. energies are assumed to be the same values as those in Fig.~\ref{fig:xsection_hpphmm}.}
\label{fig:xsection_hpphm}
\end{figure}

The associated production of doubly and singly charged Higgs can only be induced by the VBF processes
\begin{eqnarray}
VV\to H^{\pm\pm}H^{\mp}\;.
\end{eqnarray}
The Feynman diagrams are shown in Fig.~\ref{diagram:HppHm}.
The cross sections of these processes are shown in Fig.~\ref{fig:xsection_hpphm}, as a function of c.m. energy $\sqrt{s}$ (left) and heavy Higgs masses $m_{H^{\pm\pm}}$ (equal to $m_{H^{\pm}}$) (right).~\footnote{The cross sections of $H^{++}H^-$ and its charge conjugation are summed over.}
One can see that the VBF cross sections of $H^{\pm\pm}H^{\mp}$ are a few times smaller than those of $H^{++}H^{--}$.
This is due to the absence of dominant $\gamma\gamma$ fusion in the associated production.

In the following subsections, we simulate the productions of charged Higgs by using MadGraph5\_aMC@NLO~\cite{Alwall:2014hca}. The simulation of VBF process at muon collider is carried out according to the descriptions in Ref.~\cite{Ruiz:2021tdt}. The decays of the charged Higgs bosons are implemented using MadSpin~\cite{Artoisenet:2012st}. In the simulation, we independently generate the events of $\mu^+\mu^-$ annihilation and VBF processes. Then, we pass the parton-level events to PYTHIA 8~\cite{Sjostrand:2014zea} for performing parton shower. For the simulation of detector effects, we choose the card of muon collider in Delphes 3~\cite{deFavereau:2013fsa}. Finally, we combine the results of $\mu^+\mu^-$ annihilation and VBF processes weighted by their cross sections in local significance analysis.

\subsection{$H^{++}H^{--}\to \ell^+\ell^+\ell^-\ell^-$}

The decays of doubly charged Higgs $H^{++}$ are dominated by the leptonic channels when the vev of $\Delta$ is below $v_{\Delta}\approx 10^{-4}$~GeV~\cite{FileviezPerez:2008jbu}. In this subsection, we take one of the decay channels $H^{\pm\pm}\to\mu^{\pm}\mu^{\pm}$ as a benchmark to evaluate the detection effect of the Type II Seesaw at muon collider.
There are four muons in the final states of our signal. The corresponding SM backgrounds are simply chosen to have four-muon final states. They can also be categorized into $\mu^+\mu^-$ annihilation and VBF processes
\begin{equation}
  \label{eq:sm_bkg_4mu}
  \mu^+\mu^-,~VV \to \mu^+\mu^+\mu^-\mu^-\,.
\end{equation}
We have adopted the following basic cuts for the muons in final states
\begin{equation}
  \label{eq:basic_cut_4mu}
  p_T(\mu) > 50~\mathrm{GeV}\,,\quad
   \left| \eta(\mu) \right| < 2.5\,,\quad
  \Delta R_{\mu\mu} > 0.4\,.
\end{equation}
The first two cuts are essential to avoid the possible collinear divergence in the background VBF processes.

\begin{figure}[t!]
\centering
\includegraphics[width=0.6\textwidth]{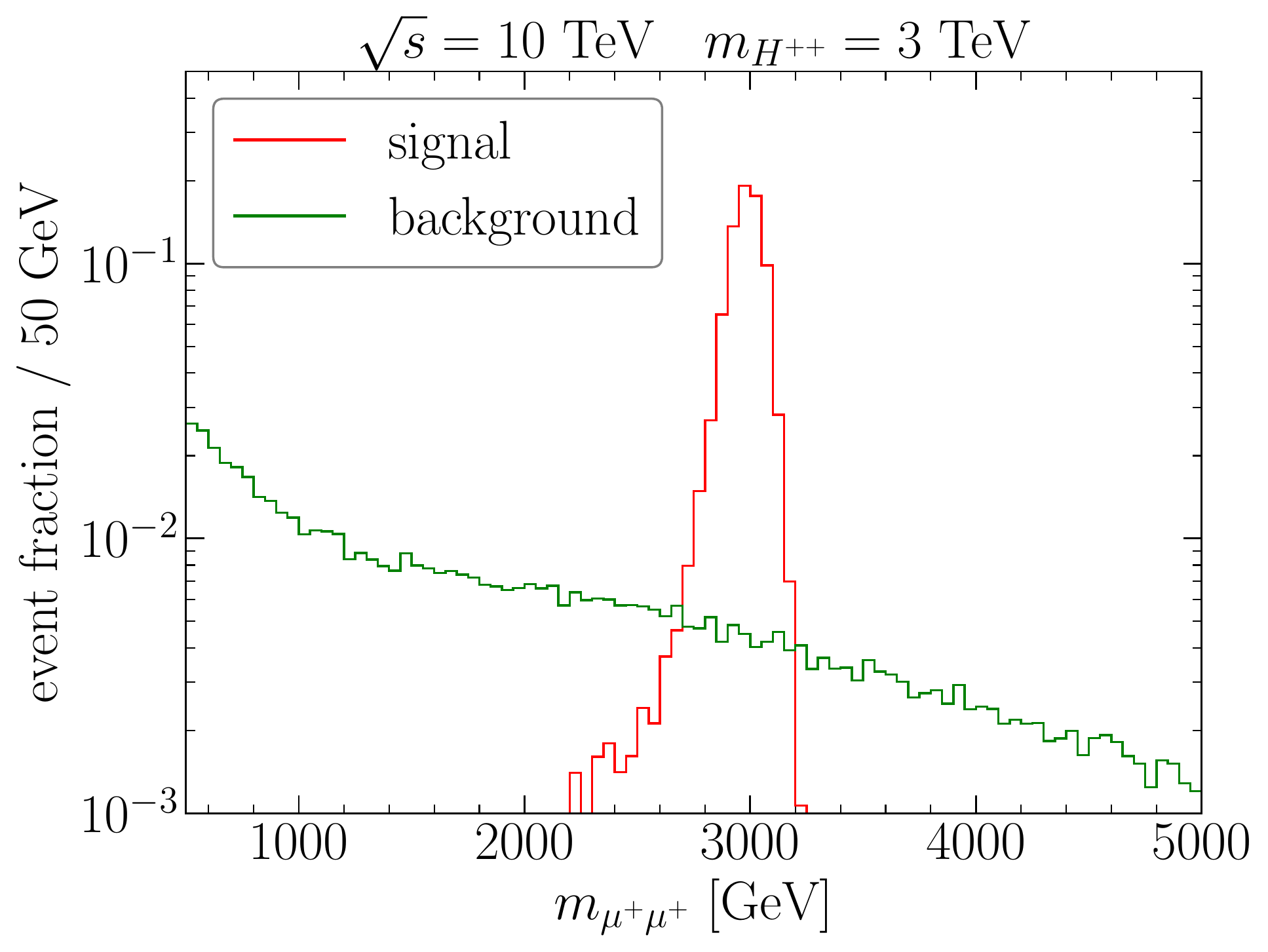}
\caption{The invariant mass of the same-sign muons $\mu^\pm\mu^\pm$ in $H^{++}H^{--}$ pair production for $\sqrt{s}=10$ TeV and $m_{H^{\pm\pm}}=3$ TeV. The signal and SM background are both shown for comparison. The contributions of $\mu^+\mu^+$ annihilation and VBF processes are weighted and summed over. The individual backgrounds are also summed over.
}
\label{fig:recon_m_mumu}
\end{figure}

We then select the generated events by requiring the number of muons $n_{\mu^{+}}\geq 2$ and $n_{\mu^{-}}\geq 2$. In our signal events, two same-sign muon pairs in the final states can form two heavy resonances of doubly charged Higgs bosons. As an illustration, we take $\sqrt{s}=10$~TeV for muon collider and $m_{H^{++}}=3$~TeV to show the invariant mass of leading same-sign muons in Fig.~\ref{fig:recon_m_mumu}. For this choice, the $\mu^+\mu^+$ annihilation cross section is dominant. Here we weight and combine the contributions of $\mu^+\mu^+$ annihilation and VBF processes. We can see that the invariant mass plot shows a resonance peak around the expected Higgs mass of 3 TeV in our signal histogram.
As a result, we can apply the following invariant mass window to suppress the background and enhance the signal-to-background ratio
\begin{equation}
  \label{eq:4mu_m_mumu_cut}
  |m_{\mu^\pm \mu^\pm}-m_{H^{\pm\pm}}| < m_{H^{\pm\pm}}/5\,.
\end{equation}
The background events can then be efficiently suppressed.

We use the following formula to evaluate the significance
\begin{equation}
  \mathcal{S} = \frac{N_{\rm S}}{\sqrt{N_{\rm S}+N_{\rm B}}}\;,
\label{eqn:significance}
\end{equation}
where $N_S$ and $N_B$ are the event numbers of signal and background, respectively. In this purely leptonic channel, we have
\begin{eqnarray}
N_{\rm S}&=&(\sigma_{\rm S}^{\rm Ann} \epsilon_{\rm S}^{\rm Ann}+\sigma_{\rm S}^{\rm{VBF}} \epsilon_{\rm S}^{\rm{VBF}})\times {\rm BR}^2(H^{++}\to \mu^+ \mu^+) \times \mathcal{L}\;, \nonumber \\
N_{\rm B}&=&(\sigma_{\rm B}^{\rm Ann} \epsilon_{\rm B}^{\rm Ann}+\sigma_{\rm B}^{\rm{VBF}} \epsilon_{\rm B}^{\rm{VBF}})\times \mathcal{L}\;,
\end{eqnarray}
where the subscript ``S'' (``B'') stands for the signal (SM background), $\epsilon_{\rm S,B}$ represent the efficiencies of the above cuts and $\mathcal{L}$ denotes the integrated luminosity.

\begin{figure}[ht!]
\centering
\includegraphics[height=0.23\textheight]{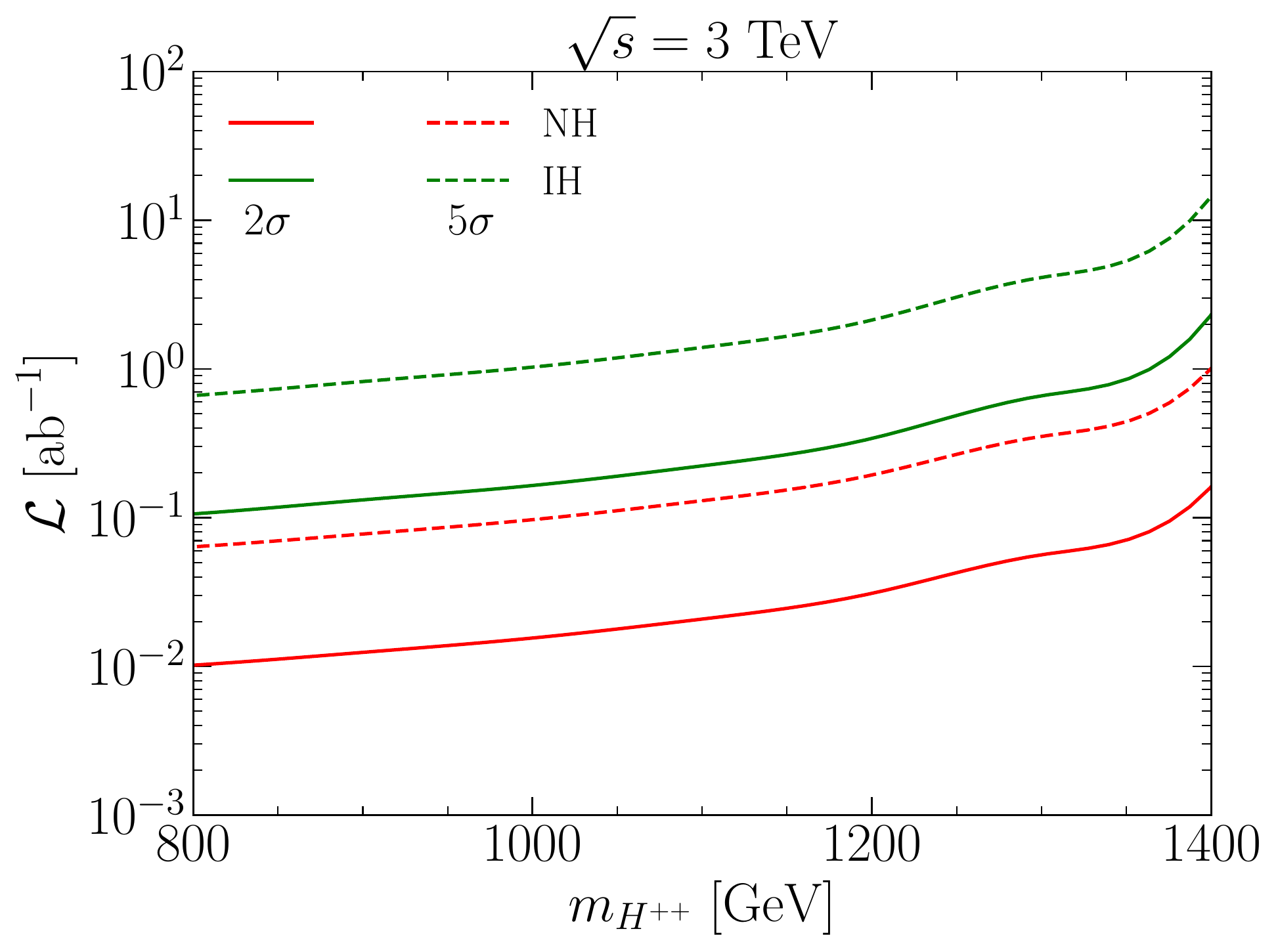}
\includegraphics[height=0.23\textheight]{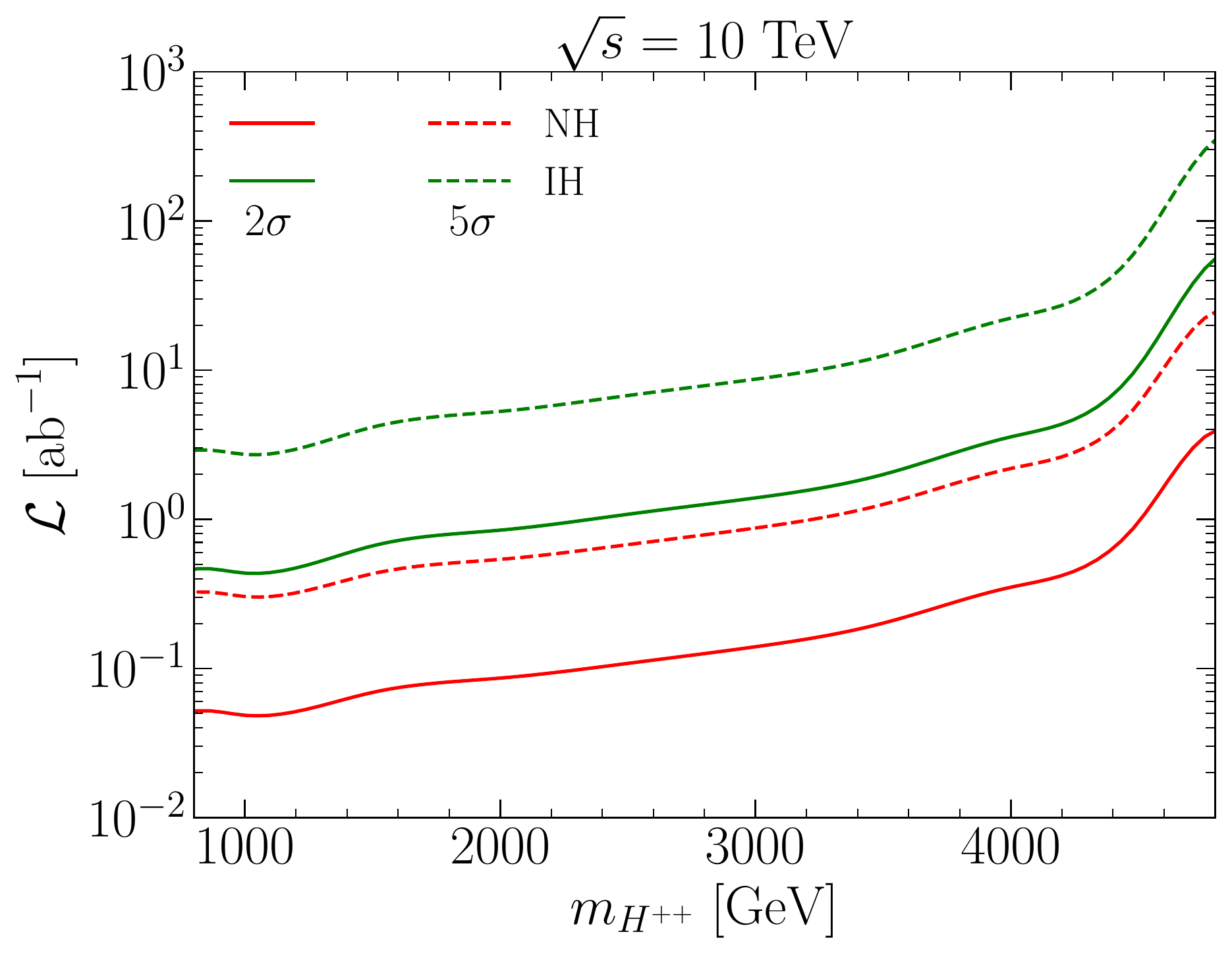}
\includegraphics[height=0.23\textheight]{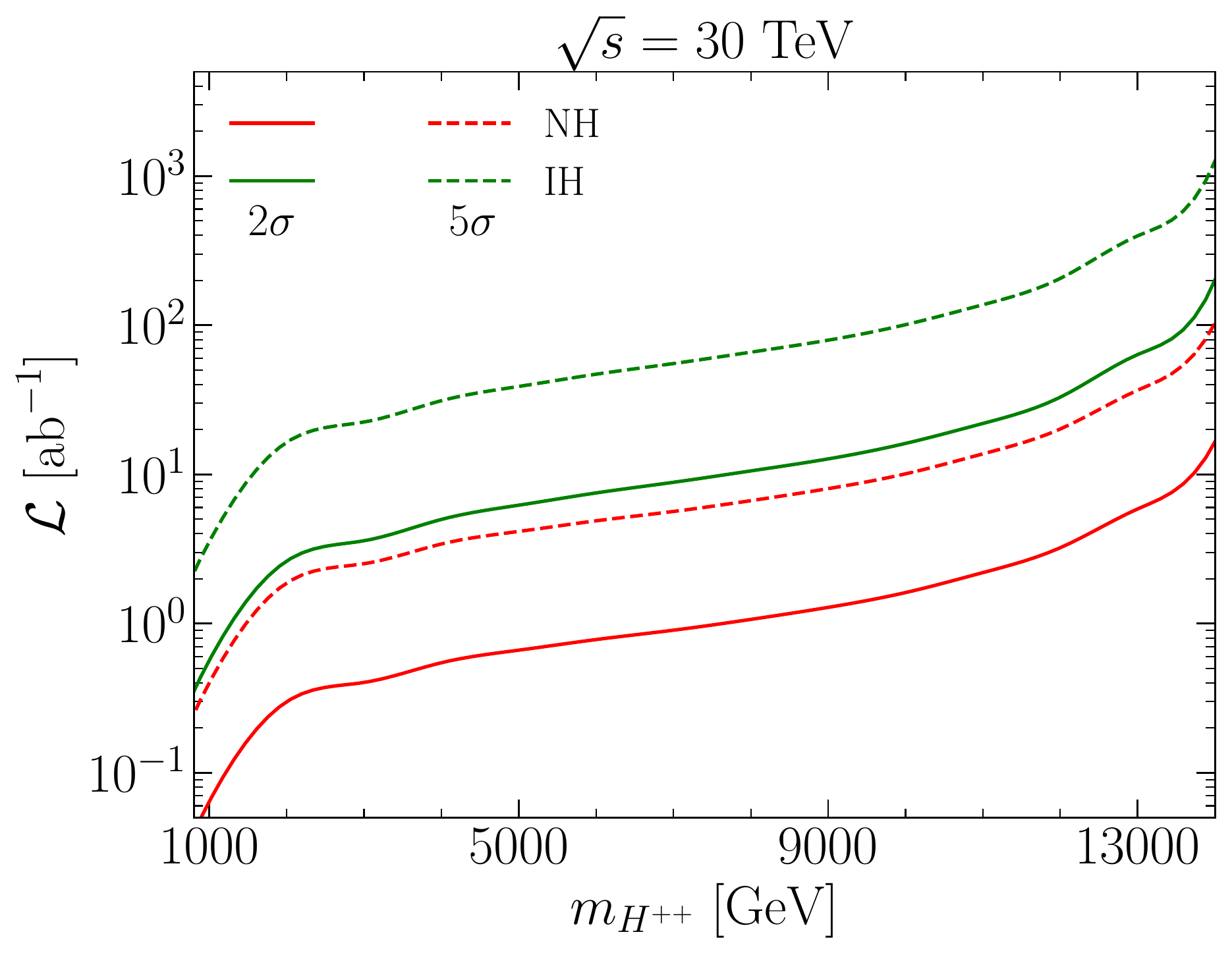}
\caption{The integrated luminosities required for 2$\sigma$ (solid lines) and 5$\sigma$ (dashed lines) significance versus $m_{H^{\pm\pm}}$ for $H^{++}H^{--}\to \mu^+\mu^+\mu^-\mu^-$ at muon colliders with $\sqrt{s}=3$ TeV, 10 TeV and 30 TeV, in neutrino mass pattern NH (red) or IH (green).
}
\label{fig:L-4mu}
\end{figure}

In Fig.~\ref{fig:L-4mu} we show the reach of doubly charged Higgs in purely leptonic channel.
For $\sqrt{s}=3$~TeV, we scan the value of $m_{H^{++}}$ and derive the required luminosities to achieve a $2\sigma$ or $5\sigma$ significance.
The benchmark decay branching fractions of $H^{++}\to \mu^+\mu^+$ in Table~\ref{BR-Hpp} are taken for the two different neutrino mass hierarchies, i.e., NH and IH.
The similar procedures of analysis are performed for $\sqrt{s}=10$~TeV and $\sqrt{s}=30$~TeV, and the results are also shown in Fig.~\ref{fig:L-4mu}. Except
for the near threshold regime, the doubly charged Higgs in NH mass pattern will be discovered with the optimistic integrated luminosity of 1, 10 and 90 ab$^{-1}$ for $\sqrt{s}=$ 3, 10 and 30 TeV, respectively. The 5$\sigma$ significance can be reached in IH for $m_{H^{\pm\pm}}$ below 1.0 TeV, 3.5 TeV and 10 TeV with the optimistic integrated luminosity and $\sqrt{s}=3$, 10 and 30 TeV, respectively.

\begin{figure}[ht!]
\centering
\includegraphics[height=0.23\textheight]{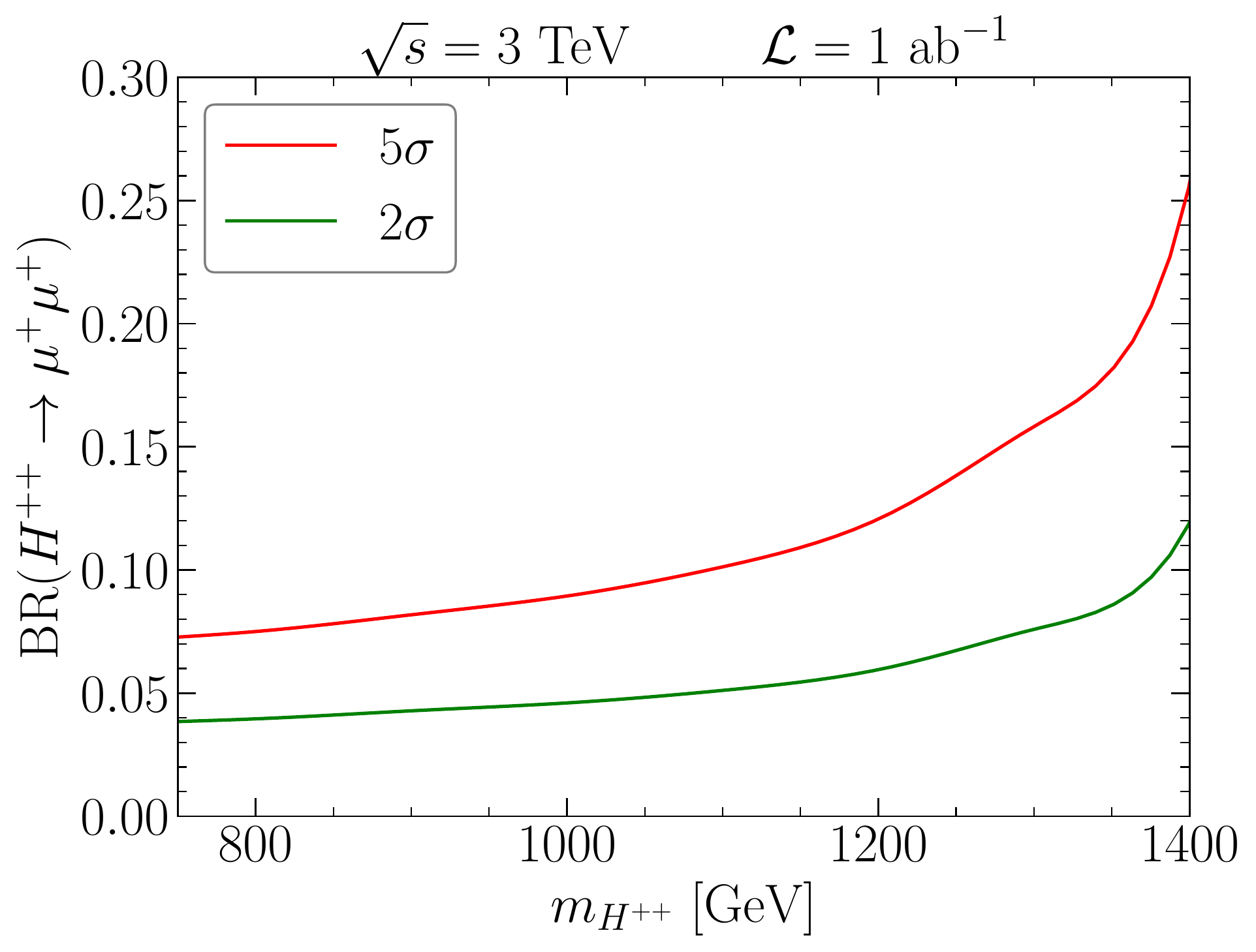}
\includegraphics[height=0.23\textheight]{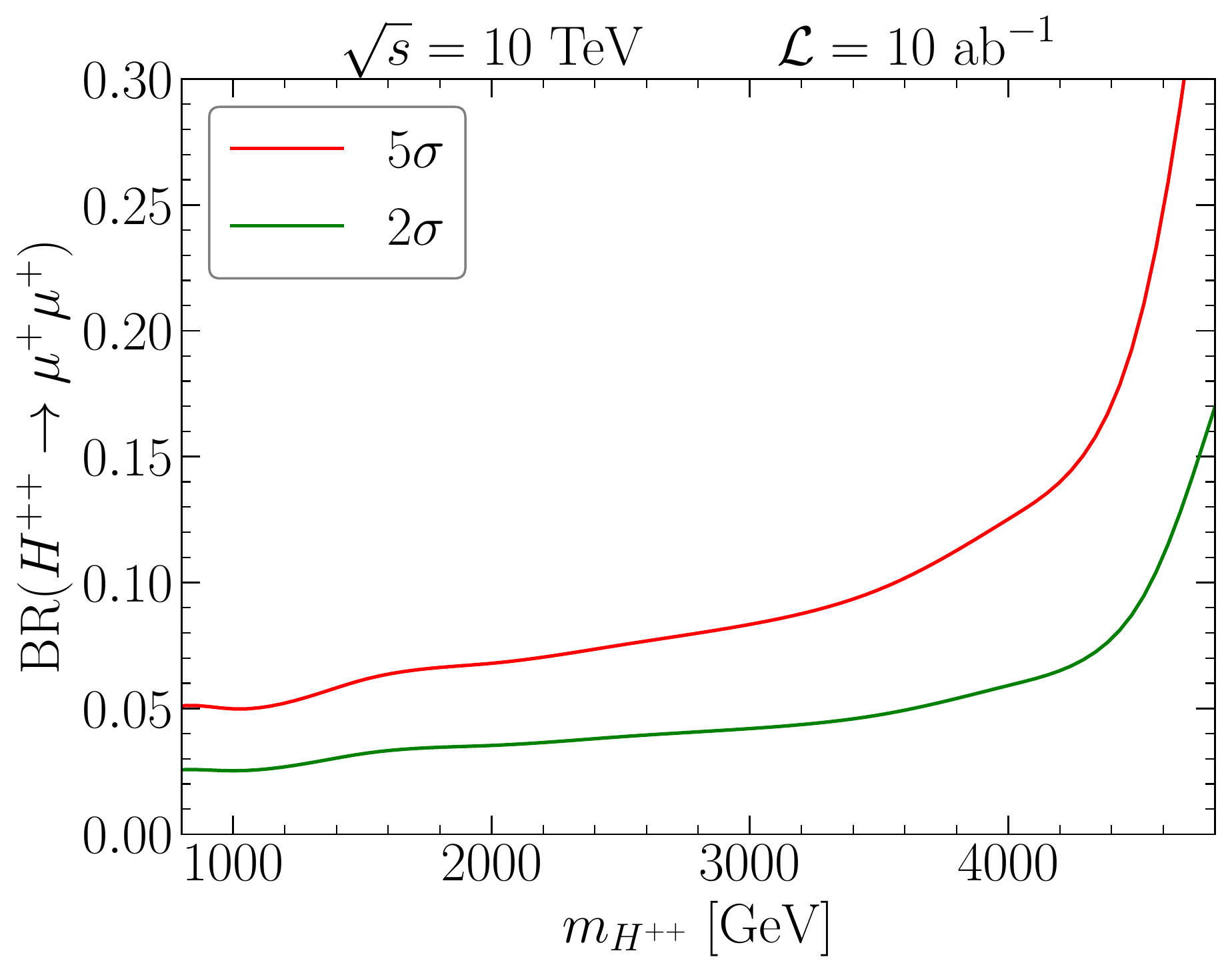}
\includegraphics[height=0.23\textheight]{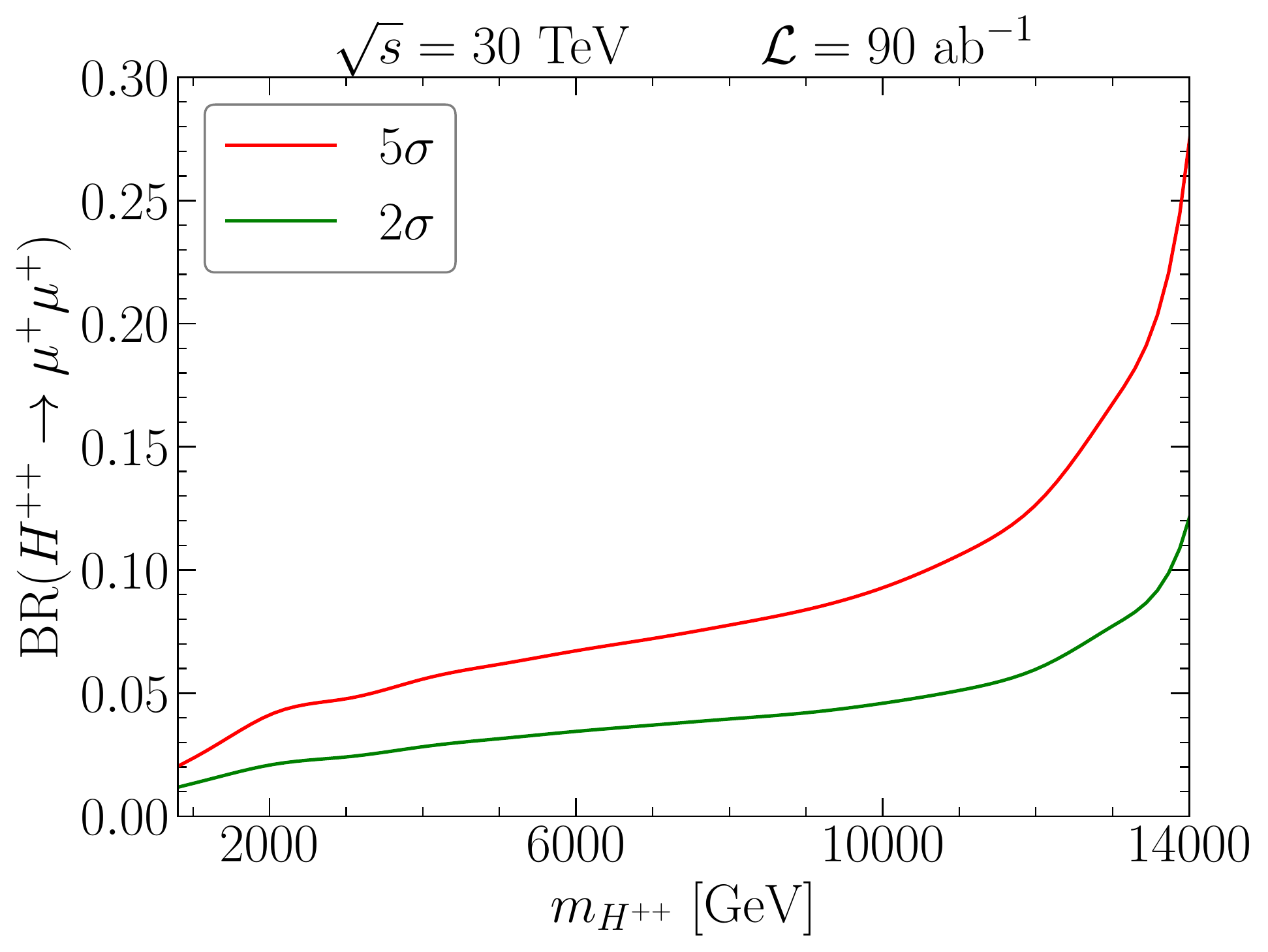}
\caption{The reachable branching ratios of $H^{++}\to \mu^+\mu^+$ corresponding to 2$\sigma$ (green) and 5$\sigma$ (red) significance versus $m_{H^{++}}$ for $H^{++}H^{--} \to \mu^+\mu^+\mu^-\mu^- $ at muon colliders with $\sqrt{s}=3$ TeV ($\mathcal{L} = 1~{\rm ab}^{-1}$), 10 TeV ($\mathcal{L} = 10~{\rm ab}^{-1}$) and 30 TeV ($\mathcal{L} = 90~{\rm ab}^{-1}$).
}
\label{fig:BR-4mu}
\end{figure}

In our analysis, we only consider the decay channel of $H^{\pm\pm}\to\mu^\pm \mu^\pm$ for illustration. For other decay channels, the main difference comes from the branching fractions of doubly charged Higgs decay into different lepton flavor combinations.
As discussed in Sec.~\ref{sec:flavor}, different neutrino mass patterns and mixing parameters affect the branching fraction of decay $H^{\pm\pm}\to\mu^\pm \mu^\pm$.
In order to show the sensitivity of muon colliders to charged Higgs decay, we next fix the integrated luminosity and show the bound on BR($H^{++}\to\mu^+\mu^+$). The reachable limits of BR($H^{++}\to\mu^+\mu^+$) corresponding to 2$\sigma$ or 5$\sigma$ significance are shown in Fig.~\ref{fig:BR-4mu} with different collision energies.
Taking $m_{H^{\pm\pm}} = 1.3$ TeV for illustration, one can see that BR($H^{++}\to\mu^+\mu^+$) can approach 15.8\% (7.6\%) for 5$\sigma$ (2$\sigma$) significance for $\sqrt{s}=3$ TeV and $\mathcal{L}=1~{\rm ab}^{-1}$. For $m_{H^{\pm\pm}}=3$ TeV at $\sqrt{s}=10$ TeV and $m_{H^{\pm\pm}}=10$ TeV at $\sqrt{s}=30$ TeV, the reachable limits of the branching fraction are 8.3\% (4.2\%) and 9.3\% (4.6\%) for 5$\sigma$ (2$\sigma$) significance, respectively.

\subsection{$H^{++}H^{--}\to W^+ W^+ W^- W^-$}
\label{sec:4w}

In this subsection, we analyze the decay of doubly charged Higgs $H^{\pm\pm}$ into same-sign $W^\pm$ bosons with subsequent decay to leptonic or hadronic products.
Here we consider the leptonic decay of two same-sign $W$ bosons into muon and neutrino ($W^{\pm} \to \mu^{\pm} \mathop{\nu_{\mu}}\limits^{(-)}$), and the hadronic decay of the other two $W$ bosons with opposite sign into di-jet ($W^{\mp} \to j~j$). For heavy Higgs boson decay, the $W$ bosons would be highly boosted and the produced di-jet can be regarded as a single fat jet $J$ in the detector of muon collider.

The LNV signal then becomes
\begin{equation}
    \mu^+\mu^-, VV \to H^{++}H^{--} \to W^{+}W^{+}W^{-}W^{-} \to  \mu^{\pm} \mu^{\pm} \mathop{\nu_{\mu}}\limits^{(-)} \mathop{\nu_{\mu}}\limits^{(-)} J~J.
\end{equation}
For these final states, the SM backgrounds could be composed of several processes
\begin{eqnarray}
 {\rm B}_{4W, 1}:\mu^+\mu^-, VV & \to & W^{+}W^{+}W^{-}W^{-} \to \mu^{\pm} \mu^{\pm} \mathop{\nu_{\mu}}\limits^{(-)} \mathop{\nu_{\mu}}\limits^{(-)} j~j~j~j  \nonumber \; , \\
 {\rm B}_{4W,2}:~~~ \qquad VV             & \to & W^{\pm} W^{\pm} W^{\mp} Z \quad \to \mu^{\pm} \mu^{\pm} \mathop{\nu_{\mu}}\limits^{(-)} \mathop{\nu_{\mu}}\limits^{(-)} j~j~j~j \nonumber \; , \\
 {\rm B}_{4W,3}:~~~ \qquad VV             & \to & t_{(\to b~W^+ )}~\Bar{t}_{(\to \Bar{b}~W^-) }~W^{\pm} \to \mu^{\pm} \mu^{\pm} \mathop{\nu_{\mu}}\limits^{(-)} \mathop{\nu_{\mu}}\limits^{(-)} b~\Bar{b}~j~j \;.
\label{eqn:4w-bkg}
\end{eqnarray}
In ${\rm B}_{4W,1}$ case, from both $\mu^+\mu^-$ annihilation and VBF processes, the intermediate states are four $W$ bosons which are the same as the signal. The jets in the final states can also be produced from $Z$ boson decay. We thus consider this situation in the ${\rm B}_{4W,2}$ case where the intermediate states are composed of $W^{\pm} W^{\pm} W^{\mp} Z$. Due to the conservation of electric charge, this background is only achieved through the VBF processes. In the detector of collider, $b$ or $\Bar{b}$ quark is likely to be mistakenly identified as a light jet. Thus, the ${\rm B}_{4W,3}$ case is also considered and the intermediate states in this case are a pair of $t~\Bar{t}$ and $W^{\pm}$ with the $t (\overline{t})$ quark decaying to the $b (\overline{b})$ quark and $W^{\pm}$ boson. For the three $W$ bosons, the two with same-sign decay to the leptons ($W^{\pm} \to \mu^{\pm} \mathop{\nu_{\mu}}\limits^{(-)} $) and the third one hadronically decays ($W^{\mp} \to j~j$). The backgrounds discussed here will also be used in the analysis of $W^\pm W^\pm W^\mp Z$ final state in subsection~\ref{sec:3wz}.

\begin{figure}[t!]
\centering
\includegraphics[height=0.23\textheight]{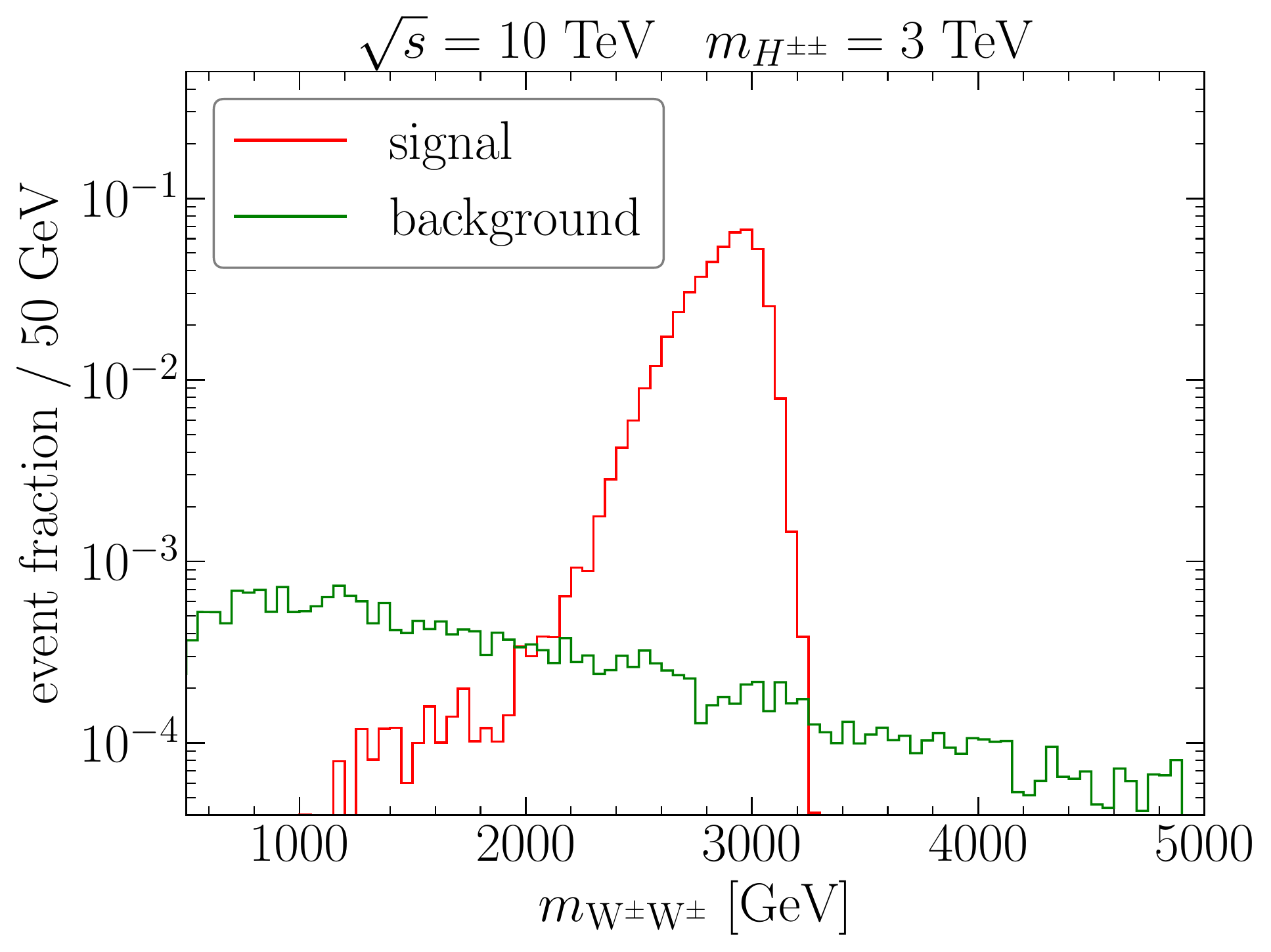}
\includegraphics[height=0.23\textheight]{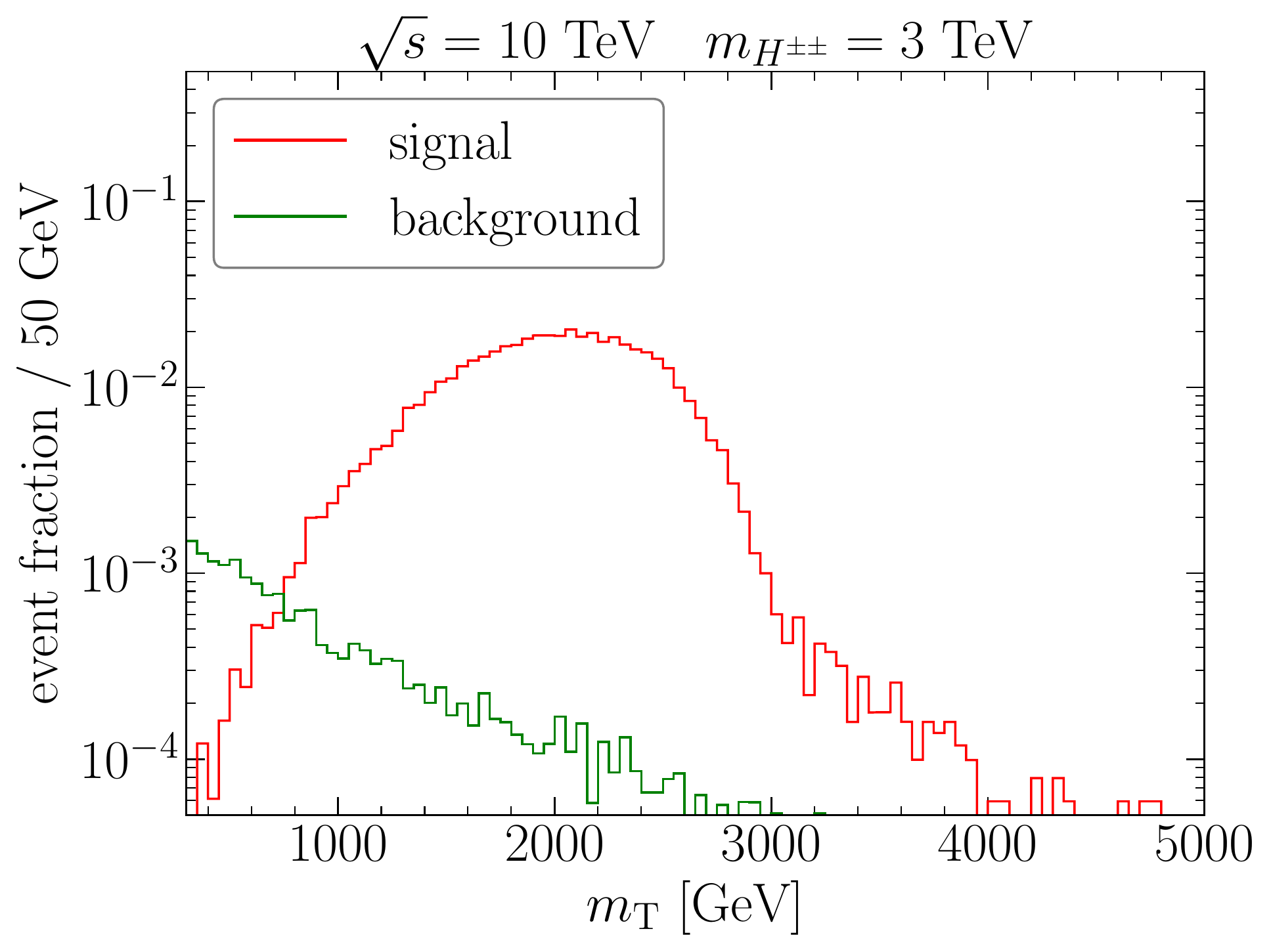}
\caption{The invariant mass of $W^\pm W^\pm$ (left) and the transverse mass $m_T$ (right) from $H^{\pm\pm}\to W^{\pm}W^{\pm}$ channel, with $\sqrt{s}=10$ TeV and $m_{H^{\pm\pm}}=3$ TeV.
}
\label{fig:recon_m_ww}
\end{figure}

We select the events containing at least two same-sign muons and two fat jets.
The fat jet is reconstructed via the ``Valencia'' algorithm with $R$ = 0.7 and is identified as $W$ boson with $65$ GeV $< M_J < 95$ GeV~\cite{Li:2022kkc}. We also employ the following basic cuts for the muons and missing neutrinos
\begin{eqnarray}
    p_T(\mu)> 50 ~{\rm GeV}~, \quad  \cancel{E}_T > 50 ~{\rm GeV} , \quad
    \left| \eta(\mu)\right| < 2.5~, \quad \Delta R_{\mu\mu} > 0.4~.
\label{eqn:3mu-basic-cut}
\end{eqnarray}
Then we reconstruct doubly charged Higgs with hadronic decay products and employ the judicious cuts of invariant mass:
\begin{equation}
    |m_{W^\pm W^\pm}-m_{H^{\pm\pm}}| < m_{H^{\pm\pm}}/5\,,
\label{eq:4w-mww-cut}
\end{equation}
where $m_{W^\pm W^\pm}$ is the invariant mass of the two same-sign $W$ bosons identified by fat jets. The invariant mass distributions of signal and background are shown in the left panel of Fig.~\ref{fig:recon_m_ww}. We can efficiently reduce the SM backgrounds by the above cut. On the other hand, due to the missing neutrinos in final states, one cannot directly reconstruct the other same-sign $W$ pair. We instead define the leptonic transverse mass~\cite{FileviezPerez:2008jbu}
\begin{equation}
    m_T = \sqrt{\left(\sqrt{m_{\mu\mu}^2+\left(\sum \Vec{p}_T(\mu)\right)^2}+\cancel{E}_T\right)^2-\left(\sum \Vec{p}_T(\mu)+\Vec{\cancel{p}}_T\right)^2}\; .
\label{eqn:4w-mT}
\end{equation}
The distribution of transverse mass is shown in the right panel of Fig.~\ref{fig:recon_m_ww} and there appears a mild cutoff around doubly charged Higgs mass. We apply the corresponding cut of $m_T$
\begin{equation}
   m_{H^{++}}/5 < m_T < m_{H^{++}} \; .
\label{eqn:4w-mT-cut}
\end{equation}

\begin{figure}[ht!]
\centering
\includegraphics[height=0.23\textheight]{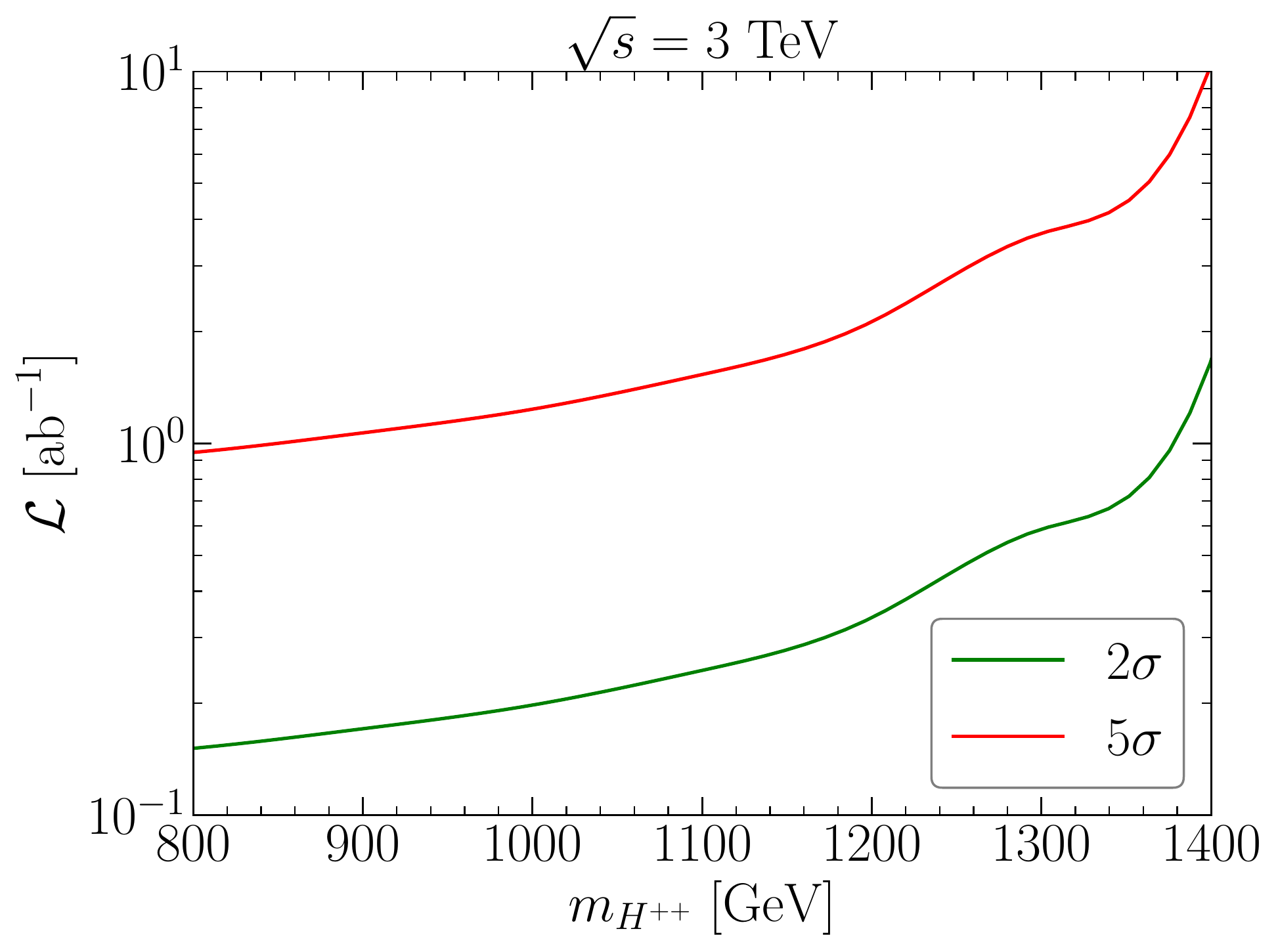}
\includegraphics[height=0.23\textheight]{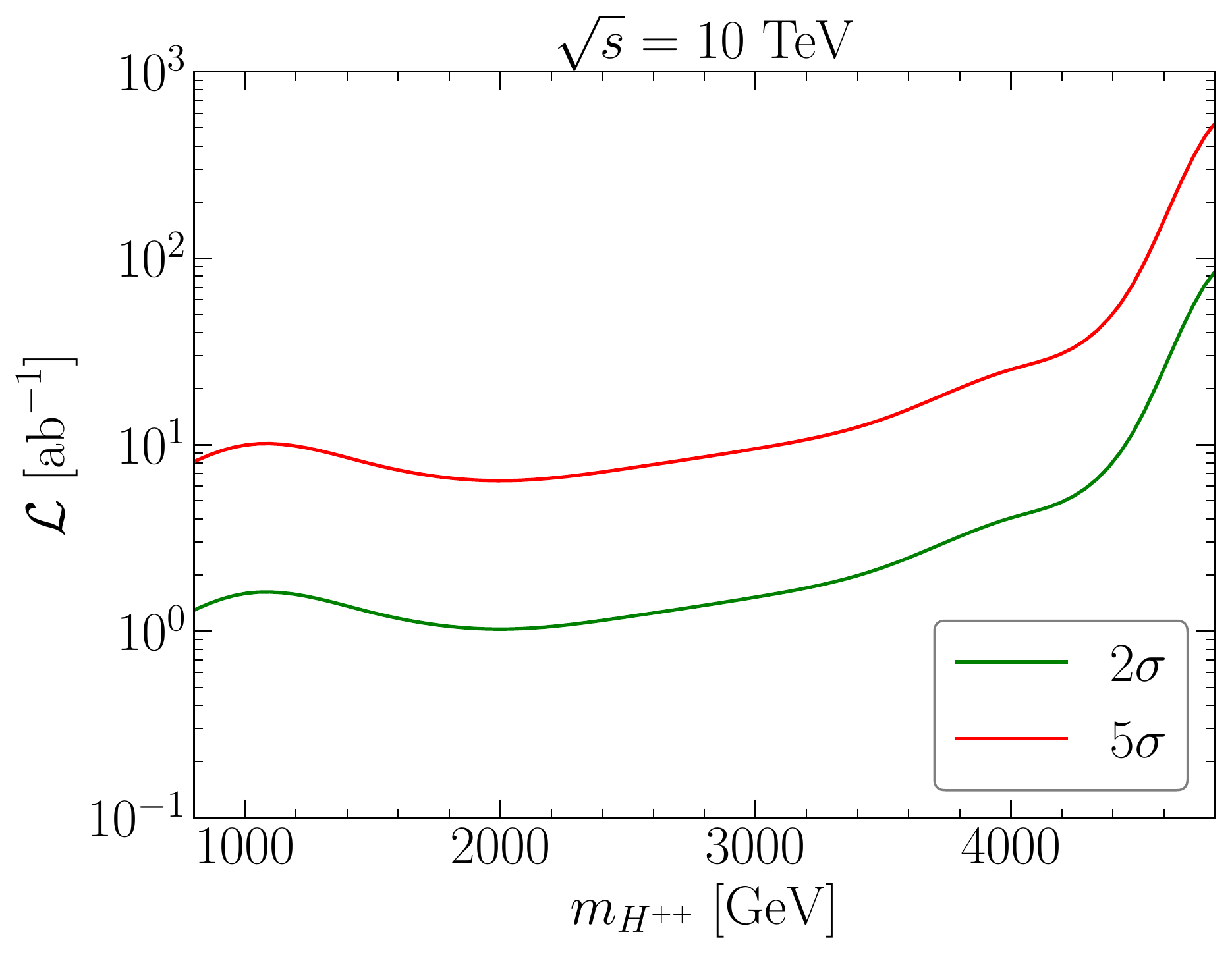}
\includegraphics[height=0.23\textheight]{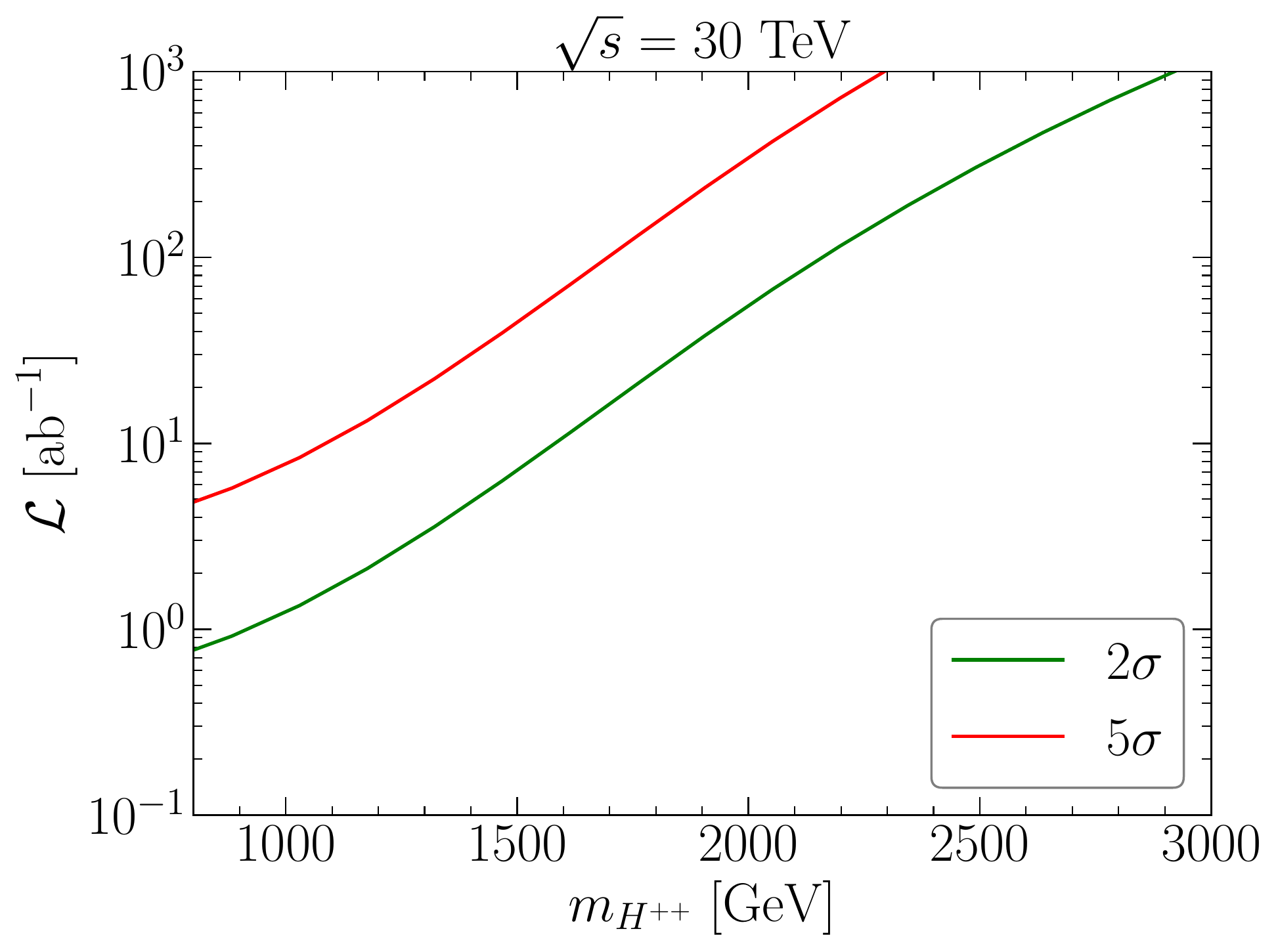}
\caption{The integrated luminosities for 2$\sigma$ (green) and 5$\sigma$ (red) significance versus $m_{H^{++}}$ for $H^{++}H^{--}\to W^+W^+W^-W^-$ channel at muon colliders with $\sqrt{s}=3$ TeV, 10 TeV and 30 TeV.}
\label{fig:4w_lumi}
\end{figure}

We also use the significance formula in Eq.~\eqref{eqn:significance} to perform the local significance analysis. The event numbers of the signal $N_{\rm S}$ and background $N_{\rm B}$ in this channel are
\begin{align}
N_{\rm S}&=(\sigma_{\rm S}^{\rm Ann} \epsilon_{\rm S}^{\rm Ann}+\sigma_{\rm S}^{\rm{VBF}} \epsilon_{\rm S}^{\rm{VBF}})\times {\rm BR}^2(H^{\pm\pm}\to W^{\pm} W^{\pm})  \times {\rm BR}^2(W^{\pm} \to \mu^{\pm} \mathop{\nu_{\mu}}\limits^{(-)}) \nonumber \\
&\quad  \times {\rm BR}^2(W^{\mp} \to q \overline{q}') \times \mathcal{L}\times 2\;, \nonumber \\
N_{\rm B}&=N_{{\rm B}_{4W,1}}\times 2+N_{{\rm B}_{4W,2}}+N_{{\rm B}_{4W,3}},
\end{align}
where the factor of 2 takes into account the charge conjugation of final states. $N_{{\rm B}_{4W,1}}$, $N_{{\rm B}_{4W,2}}$ and $N_{{\rm B}_{4W,3}}$ are the event numbers of background ${\rm B}_{4W,1}$, ${\rm B}_{4W,2}$ and ${\rm B}_{4W,3}$, respectively. They are given by
\begin{align}
N_{{\rm B}_{4W,1}}&=(\sigma_{{\rm B}_{4W,1}}^{\rm Ann} \epsilon_{{\rm B}_{4W,1}}^{\rm Ann}+\sigma_{{\rm B}_{4W,1}}^{\rm{VBF}} \epsilon_{{\rm B}_{4W,1}}^{\rm{VBF}})\times {\rm BR}^2(W^{\pm} \to \mu^{\pm} \mathop{\nu_{\mu}}\limits^{(-)}) \times {\rm BR}^2(W^{\pm} \to q \bar{q}') \times \mathcal{L}\;, \nonumber \\
N_{{\rm B}_{4W,2}}&=(\sigma_{{\rm B}_{4W,2}}^{\rm{VBF}} \epsilon_{{\rm B}_{4W,2}}^{\rm{VBF}})\times {\rm BR}^2(W^{\pm} \to \mu^{\pm} \mathop{\nu_{\mu}}\limits^{(-)}) \times {\rm BR}(W^{\pm} \to q \Bar{q}') \times {\rm BR}(Z \to q \Bar{q}) \times \mathcal{L}\;, \nonumber \\
N_{{\rm B}_{4W,3}}&=(\sigma_{{\rm B}_{4W,3}}^{\rm{VBF}} \epsilon_{{\rm B}_{4W,3}}^{\rm{VBF}})\times {\rm BR}^2(t \to b W^+) \times {\rm BR}^2(W^{\pm} \to \mu^{\pm} \mathop{\nu_{\mu}}\limits^{(-)}) \times {\rm BR}(W^{\pm} \to q \Bar{q}') \times \mathcal{L}\;, \nonumber \\
\label{eqn:4w-bkg-numbers}
\end{align}
where the decay branching fraction of $H^{++} \to W^+ W^+ $ is assumed to be $100\%$~\cite{FileviezPerez:2008jbu}.
The required luminosities for $2\sigma$ and $5\sigma$ significance at muon collider are shown in Fig.~\ref{fig:4w_lumi}, with $\sqrt{s}=$ 3, 10 and 30 TeV. It is almost impossible to reach $5\sigma$ significance with the integrated luminosity below 1 ${\rm ab}^{-1}$ for $\sqrt{s}=3$ TeV or 10 ${\rm ab}^{-1}$ for $\sqrt{s}=10$ TeV. For $\sqrt{s}=30$ TeV, the discovery significance can be reached for $m_{H^{\pm\pm}}$ lower than 1.7 TeV with $\mathcal{L}=90~{\rm ab}^{-1}$.

\subsection{$H^{\pm\pm}H^{\mp}~~\to \ell^\pm \ell^\pm \ell^\mp \nu$}

\begin{figure}[ht!]
\centering
\includegraphics[height=0.23\textheight]{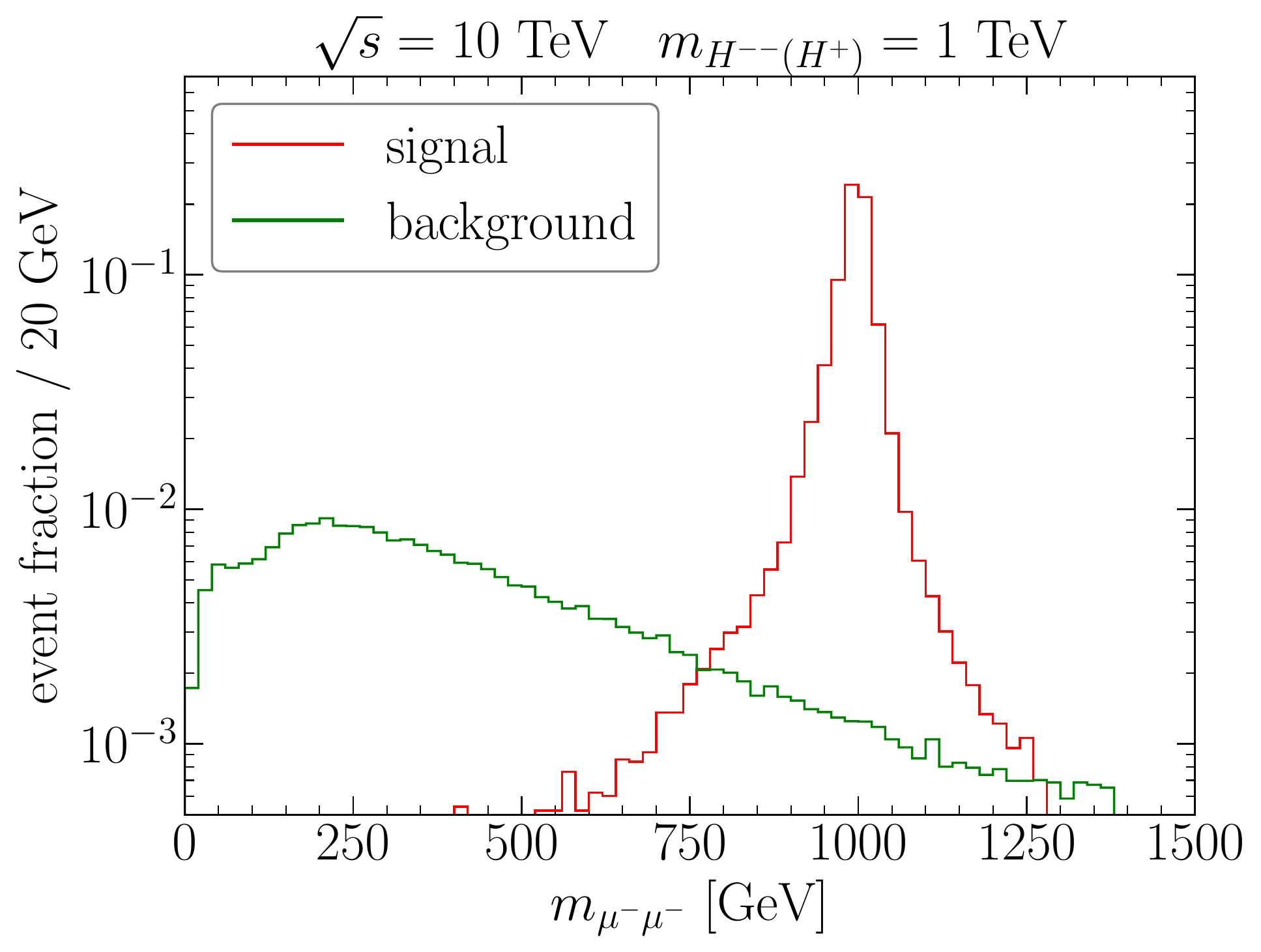}
\includegraphics[height=0.23\textheight]{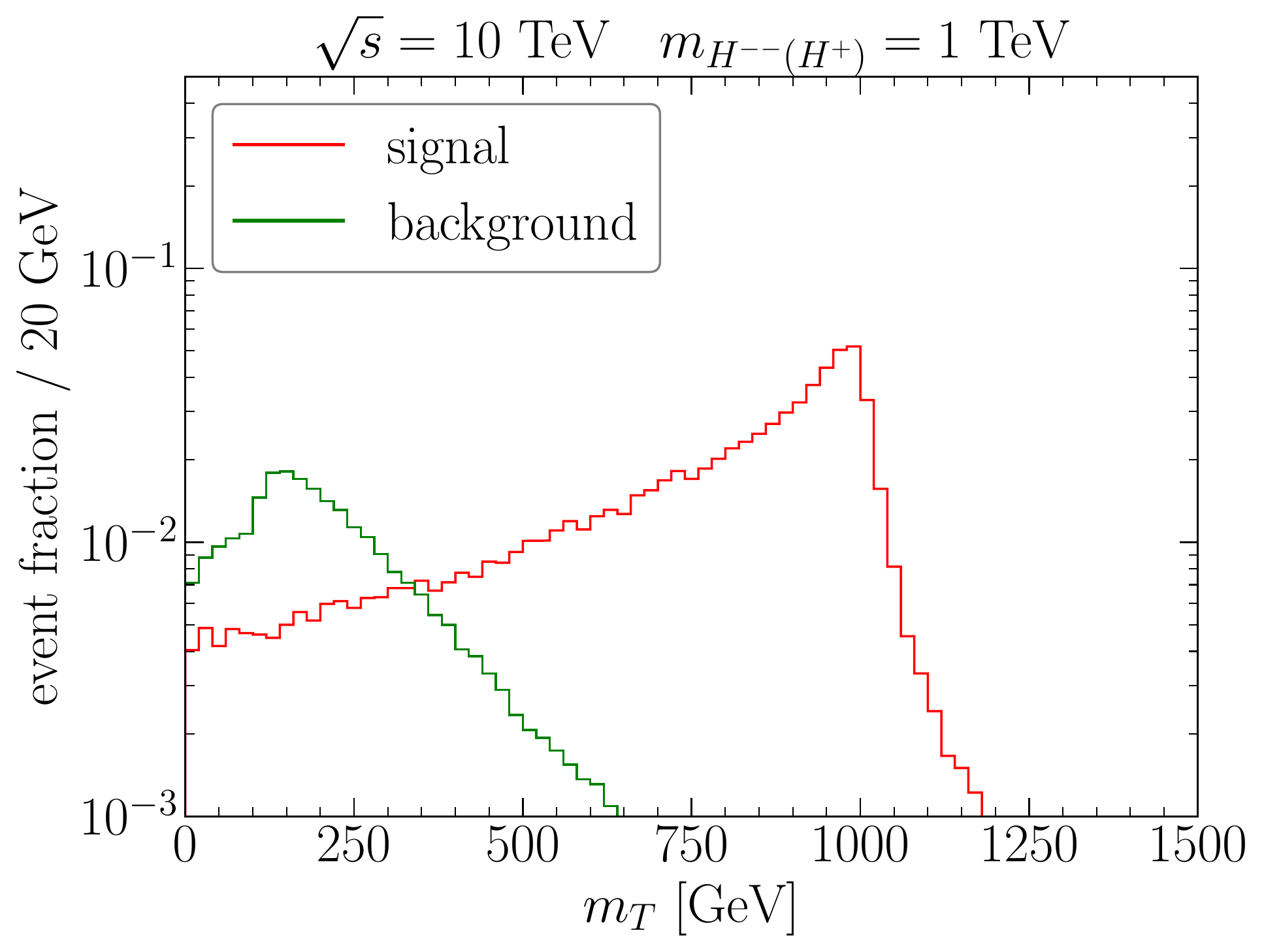}
\caption{The invariant mass of $\mu^{\pm}\mu^{\pm}$ from $H^{\pm\pm}\to \mu^{\pm}\mu^{\pm}$ and the transverse mass $m_T(\mu^{\mp}\nu)$ from $H^{\mp} \to \mu^{\mp}\nu$ with $\sqrt{s}=10$ TeV and $m_{H^{\pm\pm}(H^\mp)}=1$ TeV.}
\label{fig:SB-m-3mu}
\end{figure}

For the associated production of $H^{\pm\pm}H^{\mp}$ only induced by VBF processes at muon collider, we will start from the channel with leptonic Higgs decay and only consider the decay products of $\mu^{\pm}$ and light neutrinos. The decay modes of this channel are thus $H^{\pm\pm} \to \mu^{\pm} \mu^{\pm}$ and $H^{\mp} \to \mu^{\mp} \nu$. The signal is composed of a pair of same-sign muons, another opposite sign muon and missing neutrinos.
The main SM backgrounds include
\begin{eqnarray}
{\rm B}_{3\ell,1}: \quad VV &\to& \mu^{\pm} \mu^{\pm} \mu^{\mp} \mathop{\nu_{\mu}}\limits^{(-)}  \nonumber \; , \\
{\rm B}_{3\ell,2}: \quad VV &\to& W^{\pm} W^{\pm} W^{\mp} \to \mu^{\pm} \mu^{\pm} \mu^{\mp} \mathop{\nu_{\mu}}\limits^{(-)}\mathop{\nu_{\mu}}\limits^{(-)}\mathop{\nu_{\mu}}\limits^{(-)} \nonumber \; , \\
{\rm B}_{3\ell,3}: \quad VV &\to& Z Z W^{\pm} \to \mu^{\pm} \mu^{\pm} \mu^{\mp} \mathop{\nu_{\mu}}\limits^{(-)} ~\nu ~\overline{\nu}\;.
\label{eqn:3mu-SB}
\end{eqnarray}
In ${\rm B}_{3\ell,3}$ case, the final neutrino pair comes from one $Z$ boson decay.
We first employ the following basic cuts for the event selection
\begin{eqnarray}
p_T(\mu)> 50 ~{\rm GeV}~, \quad  \cancel{E}_T > 50 ~{\rm GeV}, \quad
\left| \eta(\mu)\right| < 2.5~, \quad \Delta R_{\mu\mu} > 0.4\;.
\label{eqn:3mu-basic-cut1}
\end{eqnarray}
Since the kinematics is so different between the signal and the backgrounds, we employ the judicious cuts to reduce the backgrounds:
\begin{itemize}
\item
To reconstruct doubly charged Higgs $H^{\pm\pm}$, we pair the final muons by taking advantage of the feature that they have same-sign of electric charge.
Then we take the invariant mass close to $m_{H^{\pm\pm}}$ with
\begin{align}
\left | m_{\mu^+\mu^+}-m_{H^{++}} \right| < m_{H^{++}}/5\;.
\end{align}
The reconstructed mass is shown in the left panel of Fig.~\ref{fig:SB-m-3mu}.
\item
For the singly charged Higgs $H^{\mp}$, we first define a transverse mass $m_T$ constructed by the opposite sign muon and missing transverse energy $\cancel{E}_T$~\cite{FileviezPerez:2008jbu}
\begin{equation}
m_T(\mu^{\mp}\nu) = \sqrt{(E_T(\mu)+\cancel{E}_T)^2-(\Vec{p}_T(\mu)+\Vec{\cancel{p}}_T)^2}\;.
\label{eqn:3mu-mT}
\end{equation}
This variable is displayed in the right panel of Fig.~\ref{fig:SB-m-3mu}. We then impose a cut further tightened up for heavier Higgs
\begin{equation}
m_T(\mu^{\mp}\nu) > 500~{\rm GeV}\;.
\end{equation}
\end{itemize}

After employing the above cuts, the SM backgrounds are significantly reduced.
The event numbers of the signal $N_S$ and backgrounds $N_B$ in this channel are
\begin{eqnarray}
N_S&=&\sigma_{S}^{\rm{VBF}} \cdot \epsilon_{S}^{\rm{VBF}}\times {\rm BR}(H^{\pm\pm}\to \mu^{\pm}\mu^{\pm}) \times {\rm BR}(H^{\mp}\to \mu^{\mp} \mathop{\nu}\limits^{(-)}) \times \mathcal{L} \;, \nonumber \\
N_B&=& N_{{\rm B}_{3\ell,1}} + N_{{\rm B}_{3\ell,2}} +N_{{\rm B}_{3\ell,3}}\;.
\end{eqnarray}
where the branching ratios of the charged Higgses are shown in Table~\ref{BR-Hpp} and Table~\ref{BR-Hp} for both NH and IH mass patterns.
The background event numbers $N_{{\rm B}_{3\ell,1}}$, $N_{{\rm B}_{3\ell,2}}$ and $N_{{\rm B}_{3\ell,3}}$ are given by
\begin{eqnarray}
N_{{\rm B}_{3\ell,1}} &=&   \sigma_{{\rm B}_{3\ell,1}}^{\rm{VBF}} \cdot \epsilon_{{\rm B}_{3\ell,1}}^{\rm{VBF}} \times \mathcal{L}  \;, \nonumber\\
N_{{\rm B}_{3\ell,2}} &=&   \sigma_{{\rm B}_{3\ell,2}}^{\rm{VBF}} \cdot \epsilon_{{\rm B}_{3\ell,2}}^{\rm{VBF}} \times {\rm BR}^2(W^{\pm}\to \mu^{\pm} \mathop{\nu_{\mu}}\limits^{(-)}) \times {\rm BR}(W^{\mp}\to \mu^{\mp} \mathop{\nu_{\mu}}\limits^{(-)}) \times \mathcal{L}  \;, \nonumber \\
N_{{\rm B}_{3\ell,3}} &=&   \sigma_{{\rm B}_{3\ell,3}}^{\rm{VBF}} \cdot \epsilon_{{\rm B}_{3\ell,3}}^{\rm{VBF}}  \times {\rm BR}(W^{\pm}\to \mu^{\pm} \mathop{\nu_{\mu}}\limits^{(-)}) \times {\rm BR}(Z \to \mu^{\pm} \mu^{\mp}) \times {\rm BR}(Z \to \nu \Bar{\nu})  \times \mathcal{L}  \;. \nonumber \\
\end{eqnarray}

\begin{figure}[ht!]
\centering
\includegraphics[height=0.23\textheight]{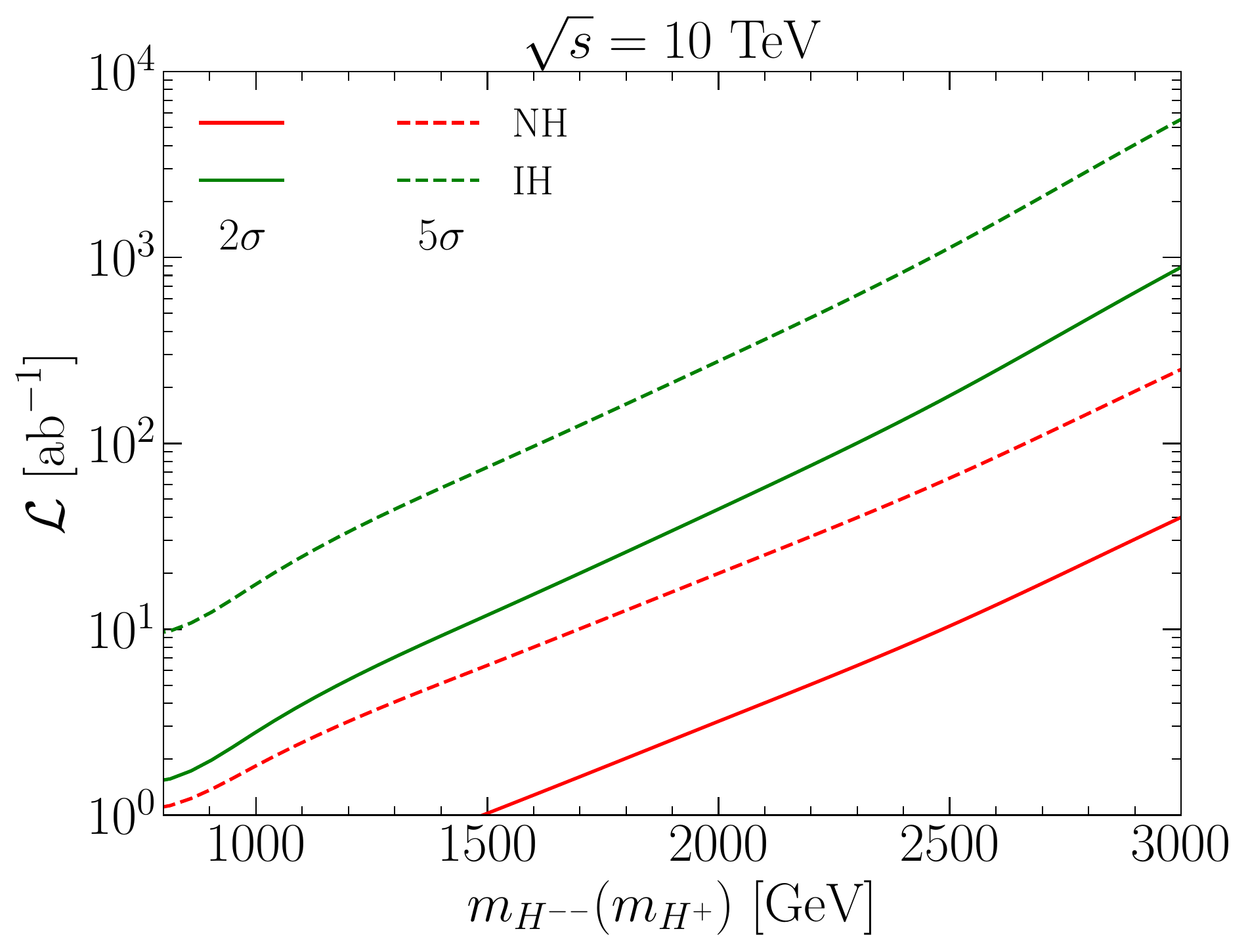}
\includegraphics[height=0.23\textheight]{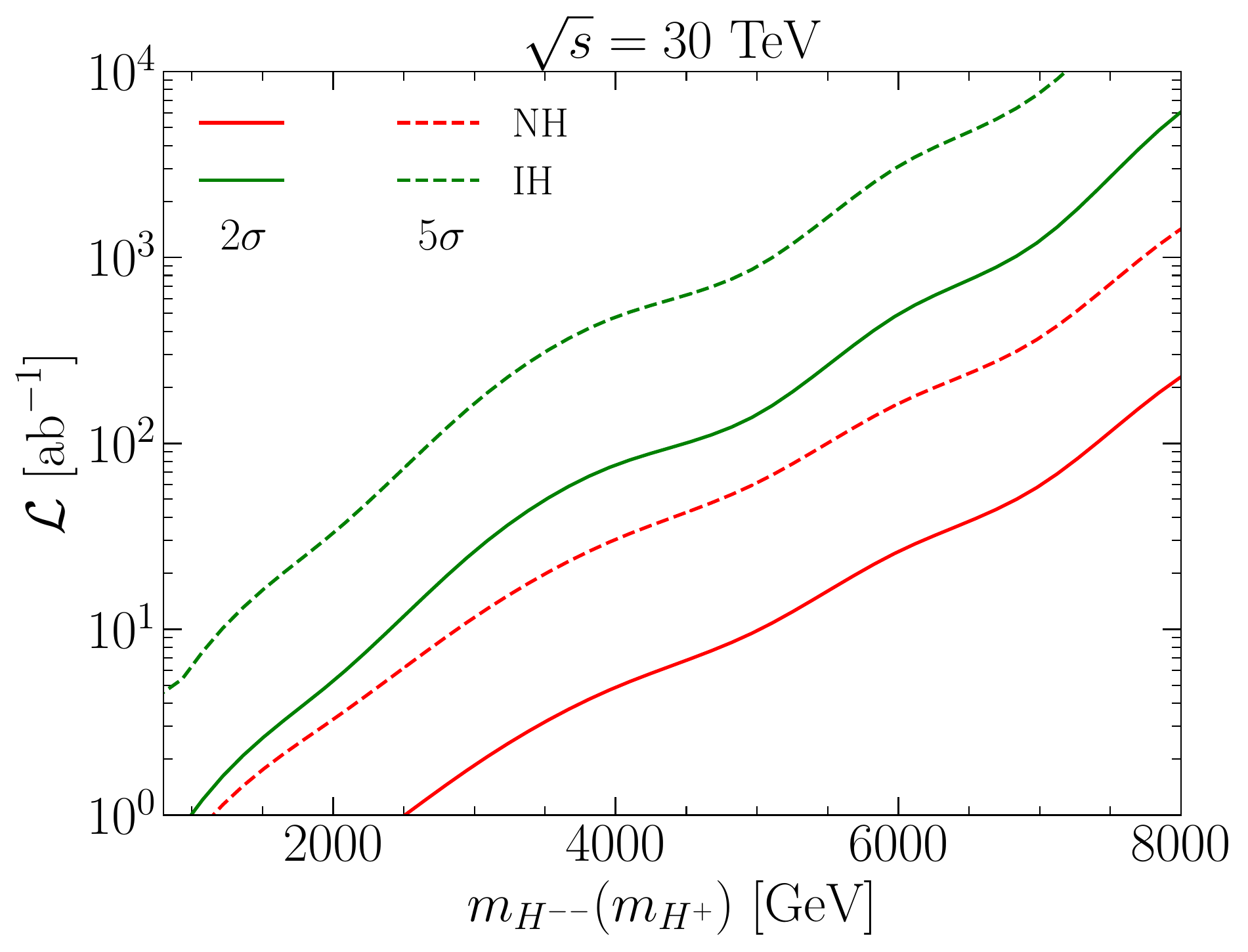}
\caption{The integrated luminosities for 2$\sigma$ (solid lines) and 5$\sigma$ (dashed lines) significance versus $m_{H^{\pm\pm}(H^\mp)}$ for $H^{\pm\pm}H^{\mp}\to \mu^\pm\mu^\pm\mu^+\nu$ at muon collider with $\sqrt{s}=10$ TeV (left) and 30 TeV (right) for NH (red) or IH (green).
}
\label{fig:3mu-lumi}
\end{figure}

\begin{figure}[ht!]
\centering
\includegraphics[height=0.23\textheight]{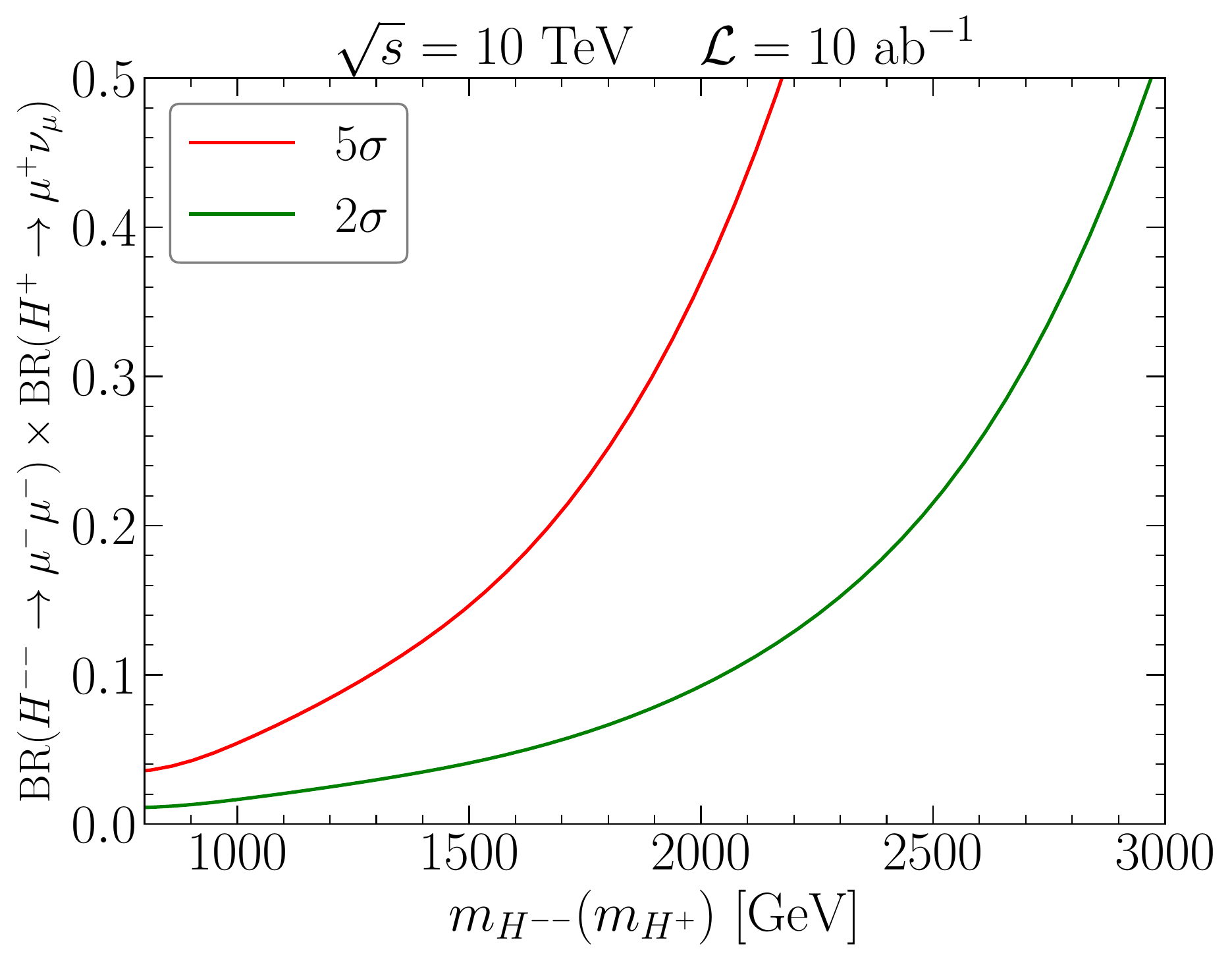}
\includegraphics[height=0.23\textheight]{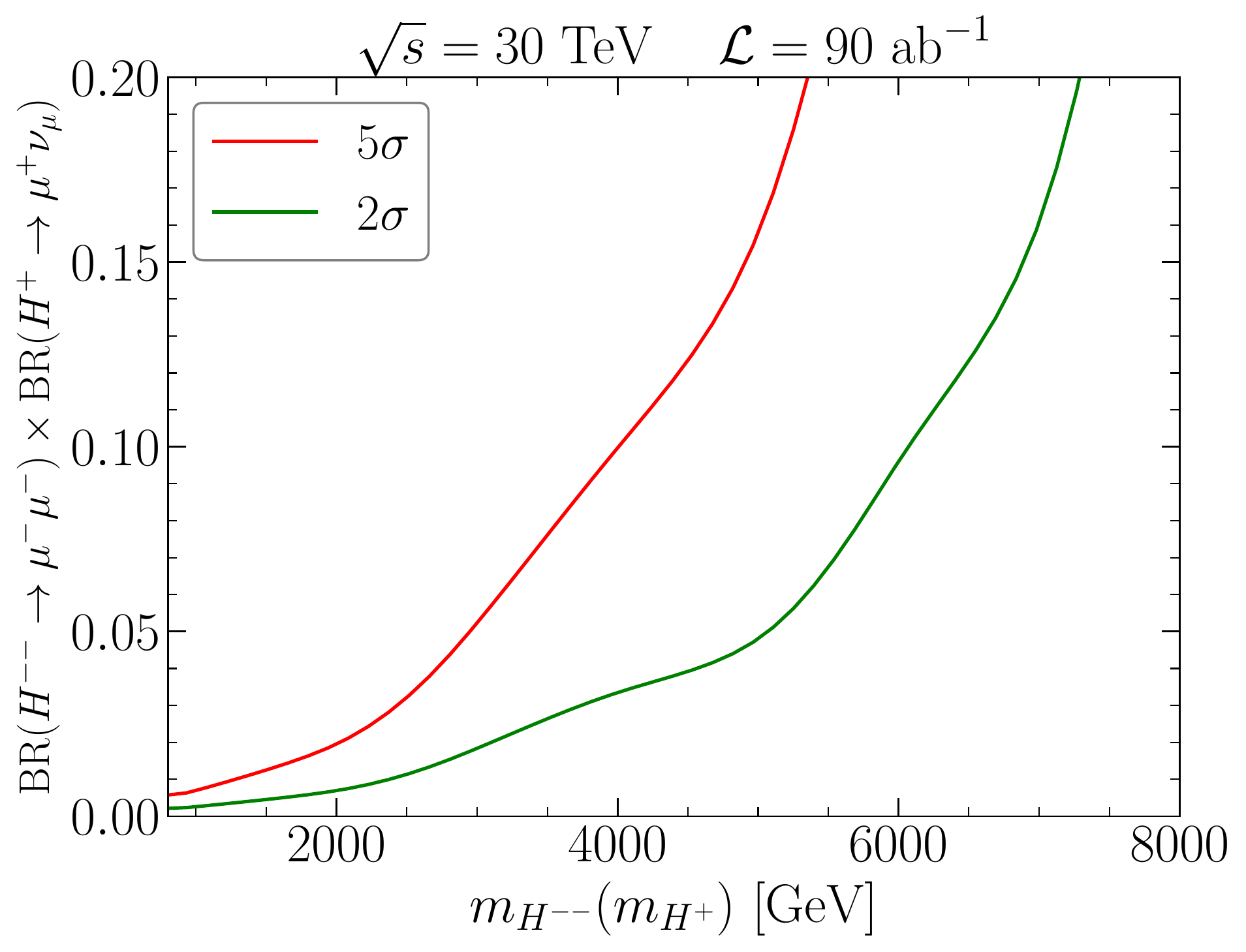}
\caption{The product of reachable branching ratios of $H^{\pm\pm}\to \mu^\pm\mu^\pm $ and $H^{\mp}\to \mu^\mp \nu$ corresponding to 2$\sigma$ (green) and 5$\sigma$ (red) significance versus $m_{H^{\pm\pm}(H^\mp)}$ for $H^{\pm\pm}H^{\mp}\to \mu^\pm\mu^\pm\mu^\mp\nu$ at muon collider with $\sqrt{s}=10$ TeV, $\mathcal{L}=10~{\rm ab}^{-1}$ (left) and $\sqrt{s}=30$ TeV, $\mathcal{L}=90~{\rm ab}^{-1}$ (right).
}
\label{fig:BR-3mu}
\end{figure}

In Fig.~\ref{fig:3mu-lumi}, we show the integrated luminosities for $2\sigma$ (solid lines) and $5\sigma$ (dashed lines) significance at muon collider with $\sqrt{s}=10$ TeV (left) and $\sqrt{s}=30$ TeV (right).
For $\sqrt{s}=10$ TeV with $\mathcal{L}=10~{\rm ab}^{-1}$, the charged Higgs as heavy as 1.7 TeV can be probed with $5 \sigma$ significance in NH.
For $\sqrt{s}=30$ TeV with $\mathcal{L}=90~{\rm ab}^{-1}$, $5 \sigma$ significance can be reached for the 5.5 (2.5) TeV charged Higgs in NH (IH).
The reachable branching ratio product of ${\rm BR}(H^{\pm\pm}\to \mu^{\pm}\mu^{\pm}) \times {\rm BR}(H^{\mp}\to \mu^{\mp}\nu)$ corresponding to 2$\sigma$ and 5$\sigma$ significance is given in Fig.~\ref{fig:BR-3mu}, with $\sqrt{s}=10$ TeV, $\mathcal{L}=10~{\rm ab}^{-1}$ (left) and $\sqrt{s}=30$ TeV, $\mathcal{L}=90~{\rm ab}^{-1}$ (right). For $m_{H^{\pm\pm}(H^\mp)}=2$ TeV at $\sqrt{s} = 10$ TeV and 5 TeV at $\sqrt{s} = 30$ TeV for illustration, one can see that the product BR($H^{++}\to\mu^+\mu^+$)$\times$BR($H^{-}\to\mu^-\nu$) as large as 36.4\% (9.2\%) and 15.8\% (4.8\%) can be reached for 5$\sigma$ (2$\sigma$) significance.

\subsection{$H^{\pm\pm}H^{\mp}~~\to W^\pm W^\pm W^\mp Z$}
\label{sec:3wz}

\begin{figure}[htb!]
\centering
\includegraphics[width=0.48\textwidth]{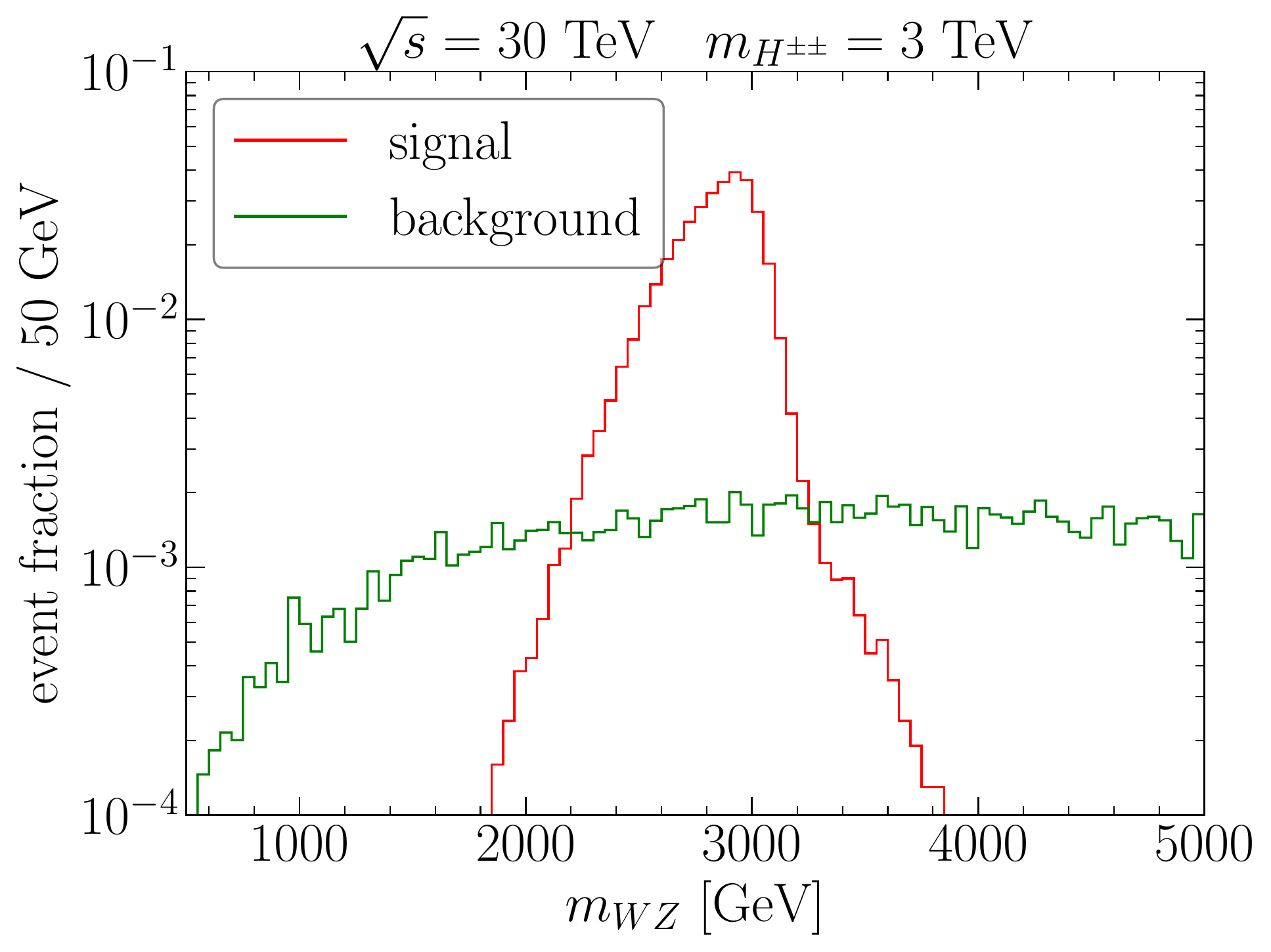}
\includegraphics[width=0.48\textwidth]{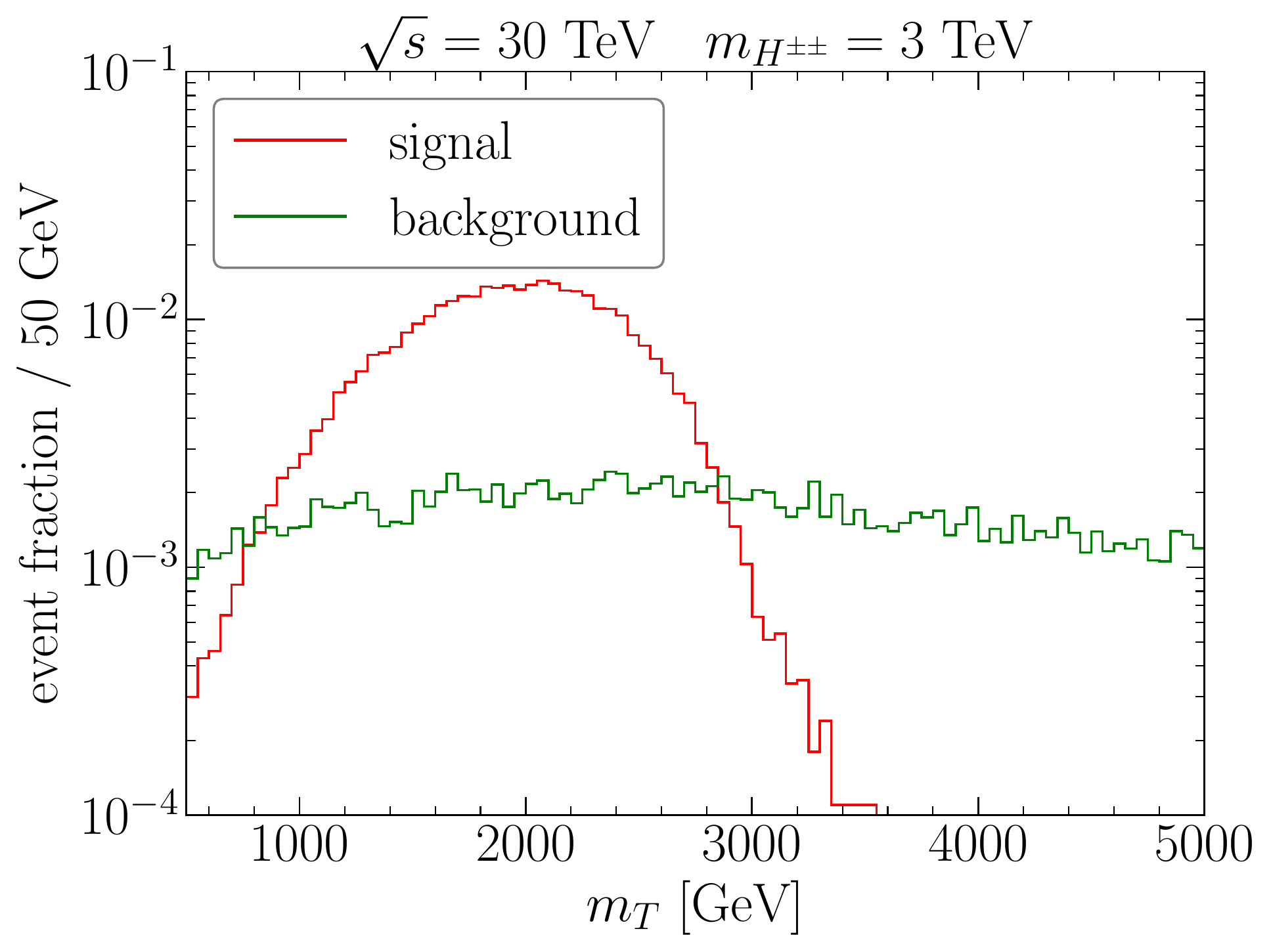}
\caption{The invariant mass of reconstructed $W$ and $Z$ bosons (left) and the transverse mass of the muons and missing energy from the same-sign $W$ bosons (right) in the signal $H^{\pm\pm}H^\mp\to WWWZ\to \mu~\mu~\nu_{\mu}\nu_{\mu}~JJ$ and SM backgrounds.
}
\label{fig:3w-mww}
\end{figure}

Finally, we consider the gauge boson decay modes in the associated production $H^{\pm\pm}H^{\mp}$ via only VBF processes at muon collider. For the decay channels of charged Higgses, we choose $H^{\pm\pm} \to W^{\pm} W^{\pm}$ and $H^{\mp} \to W^{\mp} Z$. The final states are the same as those in subsection~\ref{sec:4w} and they contain charged leptons, fat jets and missing neutrinos
\begin{equation}
VV \to H^{\pm\pm}H^{\mp} \to W^{\pm}W^{\pm}W^{\mp} Z \to  \mu^{\pm} \mu^{\pm} \mathop{\nu_{\mu}}\limits^{(-)} \mathop{\nu_{\mu}}\limits^{(-)} J~J\;.
\end{equation}
This channel is again only produced via VBF process and the two jets in the signal are from the decays of $Z$ boson and the opposite sign $W$. The SM backgrounds here completely agree with the Eq.~\eqref{eqn:4w-bkg}, i.e., ${\rm B}_{4W,1}$, ${\rm B}_{4W,2}$ and ${\rm B}_{4W,3}$.
We choose the following basic cuts
\begin{eqnarray}
p_T(\mu) > 50~\mathrm{GeV},~\left| \eta(\mu) \right| < 2.5,~
\Delta R_{\mu\mu} > 0.4~,~
\cancel{E}_T > 50~\mathrm{GeV}\;.
\end{eqnarray}
Then, we follow the same procedure of selections as subsection~\ref{sec:4w}.
For the two fat jets, the heavier one with $75$ GeV $< M_J <105$ GeV and the lighter one with $65$ GeV $< M_J <95$ GeV are identified as boosted $Z$ and $W$ boson, respectively.
The invariant mass of the reconstructed $W^{\mp}$ and $Z$ bosons is shown in the left panel of Fig.~\ref{fig:3w-mww} and we employ the same cut as that in Eq.~\eqref{eq:4w-mww-cut} for the invariant mass.
The right panel of Fig.~\ref{fig:3w-mww} shows the leptonic transverse mass Eq.~\eqref{eqn:4w-mT} and the cut in Eq.~\eqref{eqn:4w-mT-cut} is also applicable here.

The event number of the signal is
\begin{align}
N_{\rm S}&=\sigma_{\rm S}^{\rm{VBF}} \cdot \epsilon_{\rm S}^{\rm{VBF}} \times {\rm BR}(H^{\pm\pm}\to W^{\pm} W^{\pm}) \times {\rm BR}(H^{\mp}\to W^{\mp} Z)  \nonumber \\
&\quad \times {\rm BR}^2(W^{\pm} \to \mu^{\pm} \mathop{\nu_{\mu}}\limits^{(-)})  \times {\rm BR}(W^{\mp} \to q \Bar{q}') \times {\rm BR}(Z \to q \Bar{q})  \times \mathcal{L}\;,
\end{align}
where the decay branching fraction of $H^{\pm} \to W^{\pm} Z $ is taken to be $50\%$~\cite{FileviezPerez:2008jbu}. The $\sigma_S^{\rm VBF}$ is the cross section of the $H^{\pm\pm}H^{\mp}$ production from VBF processes.
The $\sigma_S^{\rm VBF}$ is quite small at low energies as we can see in Fig.~\ref{fig:xsection_hpphm}. It is thus difficult to produce more than one signal event for $\sqrt{s}=3$, 10 TeV. Although the signal event number can be enhanced at $\sqrt{s}=$ 30 TeV with $\mathcal{L}=90~{\rm ab}^{-1}$, there are still many backgrounds after applying the above cuts. We take $m_{H^{\pm\pm}}=m_{H^{\mp}}= 3$ TeV for illustration and summarize the results of the signal and backgrounds in Table~\ref{table:3w}.
One can see that the main background is from the VBF process of the case ${\rm B}_{4W,1}$ ($VV \to WWWW \to \mu~\mu~\nu_{\mu}\nu_{\mu}~ j~j~j~j~$).
Although the cuts are efficient for suppressing the backgrounds, the rates of backgrounds are still far larger than the signal.
As we have seen, it is not optimistic to probe the charged Higgs bosons at muon collider through this channel.

\begin{table}[htb!]
\centering
\begin{tabular}{|c|c|c|c|c|c|}
\hline
\hline
    $\sigma~({\rm pb})\times \epsilon $ & no cuts & basic &  $M_{W,Z}$ rec. & $m_{WZ} $ & $m_T$ \\
    $\times {\rm BRs}$& ($\sigma \times 100\% \times {\rm BRs}$) & cuts & $M_{W,Z} \pm 15~{\rm GeV}$ &  Eq.~(\ref{eq:4w-mww-cut}) & Eq.~(\ref{eqn:4w-mT-cut}) \\
    \hline
    S &  1.15$\times 10^{-7}$ & 8.72$\times 10^{-8}$& 5.04$\times 10^{-8}$ & 4.85$\times 10^{-8}$ & 4.80$\times 10^{-8}$ \\
    \hline
    ${\rm B}_{4W,1}$ &  2.60$\times 10^{-3}$ & 1.41$\times 10^{-3}$& 5.78$\times 10^{-4}$ & 1.16$\times 10^{-4}$ & 4.88$\times 10^{-5}$ \\
    ${\rm B}_{4W,2}$ &  4.70$\times 10^{-4}$ & 1.33$\times 10^{-4}$& 5.12$\times 10^{-5}$ & 1.01$\times 10^{-5}$ & 4.54$\times 10^{-6}$ \\
    ${\rm B}_{4W,3}$ &  1.78$\times 10^{-4}$ & 8.38$\times 10^{-6}$& 5.09$\times 10^{-8}$ & $-$ & $-$ \\
    \hline
\hline
\end{tabular}
\caption{The rates of $\sigma~({\rm pb})\times \epsilon \times {\rm BRs}$ for the signal $H^{\pm\pm}H^\mp\to WWWZ\to \mu~\mu~\nu_{\mu}\nu_{\mu}~JJ$ and SM backgrounds, before and after selection cuts. We choose $m_{H^{\pm\pm}}=m_{H^{\mp}}=3~{\rm TeV}$ at the muon collider with $\sqrt{s}=30~{\rm TeV}$.
}
\label{table:3w}
\end{table}

\section{Conclusions}
\label{sec:Concl}

In this work, we study the search potential of the heavy Higgs triplet in the Type II Seesaw mechanism at muon colliders with high collision energy and high luminosity. The latest neutrino oscillation data are taken into account for the impact of neutrino parameters on the leptonic decay modes of the heavy charged Higgs $(H^{\pm\pm},~H^{\pm})$ in the Type II Seesaw.
One can distinguish the two neutrino mass hierarchies by investigating the flavor structure in charged Higgs decay. The benchmark decay branching fractions are taken for both NH and IH. The decays of doubly charged Higgs moderately depend on the Majorana phase $\Phi_2$ in NH case, but strongly depend on the $\Phi_1$ phase in IH case. The leptonic decays of the singly charged Higgs are irrelevant to the Majorana phase.

The pair of the doubly charged Higgs ($H^{++}H^{--}$) is produced through $\mu^+\mu^-$ annihilation and VBF processes. For the leptonic decay channel $H^{++}H^{--}\to \mu^+\mu^+\mu^-\mu^-$, except for the
near threshold regime of mass, the doubly charged Higgs in NH mass pattern can be discovered with the c.m. energy $\sqrt{s}=3$ TeV, 10 TeV or 30 TeV and the optimistic integrated luminosities. In IH case, the 5$\sigma$ significance for $H^{\pm\pm}$ can be reached for $m_{H^{\pm\pm}}$ below 1 TeV, 3.5 TeV and 10 TeV given the optimistic integrated luminosity and $\sqrt{s}$ = 3, 10 and 30 TeV, respectively. We also obtain the reachable limits of doubly charged Higgs decay branching fraction with different collision energies. This collider measurement can provide complementary information for the neutrino properties. For the gauge boson channel $H^{++}H^{--}\to W^+W^+W^-W^-$, the 5$\sigma$ significance can only be reached for $m_{H^{\pm\pm}}$ below 1.7 TeV with $\sqrt{s}$ = 30 TeV and $\mathcal{L}=90~{\rm ab}^{-1}$.

The associated production of doubly and singly charged Higgs can only be induced by the VBF processes. In the leptonic decay channel $H^{\pm\pm}H^{\mp}\to \mu^{\pm}\mu^{\pm}\mu^{\mp} \nu$, the charged Higgs as heavy as 1.7 TeV can be probed for $5 \sigma$ significance in NH with $\sqrt{s}=10$ TeV and $\mathcal{L}=10~{\rm ab}^{-1}$.
For $\sqrt{s}=30$ TeV with $\mathcal{L}=90~{\rm ab}^{-1}$, $5 \sigma$ significance can be reached for the charged Higgs lighter than 5.5 (2.5) TeV in NH (IH). We then show the limits on the product of branching ratios ${\rm BR}(H^{\pm\pm}\to \mu^{\pm}\mu^{\pm}) \times {\rm BR}(H^{\mp}\to \mu^{\mp}\nu)$ with optimistic integrated luminosities. We also find that it is not optimistic to probe the triplet Higgs through the bosonic decay channel $H^{\pm\pm}H^{\mp}\to W^{\pm}W^{\pm}W^{\mp} Z$ at muon collider.

Besides the above LNV signatures, the lepton flavor violating processes $\mu^+\mu^-\to \ell_i^+\ell_j^-$ with $i\neq j$ may provide a clean and promising test of Yukawa couplings in the Type II Seesaw as there is no SM contribution in the amplitude level. We once studied the sensitivity of future $e^+e^-$ colliders to charged lepton flavor violation from bileptons in Refs.~\cite{Li:2018cod,Li:2019xvv}. The bileptons in our most general SM gauge invariant Lagrangian include the scalar triplet in the Type II Seesaw model. We leave the investigation of the sensitivity of muon colliders to lepton flavor violation in future studies.


\acknowledgments
T.L. is supported by the National Natural Science Foundation of China (Grants No. 11975129, 12035008) and ``the Fundamental Research Funds for the Central Universities'', Nankai University (Grant No. 63196013). C.Y.Y. is supported in part by the Grants No. NSFC-11975130, No. NSFC-12035008, No. NSFC-12047533, the Helmholtz-OCPC International Postdoctoral Exchange Fellowship Program, the National Key Research and Development Program of China under Grant No. 2017YFA0402200, the China Postdoctoral Science Foundation under Grant No. 2018M641621, and the Deutsche Forschungsgemeinschaft (DFG, German Research Foundation) under Germany's Excellence Strategy --- EXC 2121 ``Quantum Universe'' --- 390833306.

\bibliographystyle{JHEP}
\bibliography{refs}

\providecommand{\href}[2]{#2}\begingroup\raggedright\begin{thebibliography}{100}

\bibitem{Weinberg:1979sa}
S.~Weinberg, \emph{{Baryon and Lepton Nonconserving Processes}},
  \href{http://dx.doi.org/10.1103/PhysRevLett.43.1566}{\emph{Phys. Rev. Lett.}
  {\bf 43} (1979) 1566--1570}.

\bibitem{Ma:1998dn}
E.~Ma, \emph{{Pathways to naturally small neutrino masses}},
  \href{http://dx.doi.org/10.1103/PhysRevLett.81.1171}{\emph{Phys. Rev. Lett.}
  {\bf 81} (1998) 1171--1174}, [\href{http://arxiv.org/abs/hep-ph/9805219}{{\tt
  hep-ph/9805219}}].

\bibitem{Minkowski:1977sc}
P.~Minkowski, \emph{{$\mu \to e\gamma$ at a Rate of One Out of $10^{9}$ Muon
  Decays?}}, \href{http://dx.doi.org/10.1016/0370-2693(77)90435-X}{\emph{Phys.
  Lett. B} {\bf 67} (1977) 421--428}.

\bibitem{Yanagida:1979as}
T.~Yanagida, \emph{{Horizontal gauge symmetry and masses of neutrinos}},
  {\emph{Conf. Proc. C} {\bf 7902131} (1979) 95--99}.

\bibitem{GellMann:1980vs}
M.~Gell-Mann, P.~Ramond and R.~Slansky, \emph{{Complex Spinors and Unified
  Theories}}, {\emph{Conf. Proc. C} {\bf 790927} (1979) 315--321},
  [\href{http://arxiv.org/abs/1306.4669}{{\tt 1306.4669}}].

\bibitem{Glashow:1979nm}
S.~L. Glashow, \emph{{The Future of Elementary Particle Physics}},
  \href{http://dx.doi.org/10.1007/978-1-4684-7197-7_15}{\emph{NATO Sci. Ser. B}
  {\bf 61} (1980) 687}.

\bibitem{Mohapatra:1979ia}
R.~N. Mohapatra and G.~Senjanovic, \emph{{Neutrino Mass and Spontaneous Parity
  Nonconservation}},
  \href{http://dx.doi.org/10.1103/PhysRevLett.44.912}{\emph{Phys. Rev. Lett.}
  {\bf 44} (1980) 912}.

\bibitem{Shrock:1980ct}
R.~E. Shrock, \emph{{General Theory of Weak Leptonic and Semileptonic Decays.
  1. Leptonic Pseudoscalar Meson Decays, with Associated Tests For, and Bounds
  on, Neutrino Masses and Lepton Mixing}},
  \href{http://dx.doi.org/10.1103/PhysRevD.24.1232}{\emph{Phys. Rev. D} {\bf
  24} (1981) 1232}.

\bibitem{Schechter:1980gr}
J.~Schechter and J.~W.~F. Valle, \emph{{Neutrino Masses in SU(2) x U(1)
  Theories}}, \href{http://dx.doi.org/10.1103/PhysRevD.22.2227}{\emph{Phys.
  Rev. D} {\bf 22} (1980) 2227}.

\bibitem{Konetschny:1977bn}
W.~Konetschny and W.~Kummer, \emph{{Nonconservation of Total Lepton Number with
  Scalar Bosons}},
  \href{http://dx.doi.org/10.1016/0370-2693(77)90407-5}{\emph{Phys. Lett. B}
  {\bf 70} (1977) 433--435}.

\bibitem{Cheng:1980qt}
T.~P. Cheng and L.-F. Li, \emph{{Neutrino Masses, Mixings and Oscillations in
  SU(2) x U(1) Models of Electroweak Interactions}},
  \href{http://dx.doi.org/10.1103/PhysRevD.22.2860}{\emph{Phys. Rev. D} {\bf
  22} (1980) 2860}.

\bibitem{Lazarides:1980nt}
G.~Lazarides, Q.~Shafi and C.~Wetterich, \emph{{Proton Lifetime and Fermion
  Masses in an SO(10) Model}},
  \href{http://dx.doi.org/10.1016/0550-3213(81)90354-0}{\emph{Nucl. Phys. B}
  {\bf 181} (1981) 287--300}.

\bibitem{Mohapatra:1980yp}
R.~N. Mohapatra and G.~Senjanovic, \emph{{Neutrino Masses and Mixings in Gauge
  Models with Spontaneous Parity Violation}},
  \href{http://dx.doi.org/10.1103/PhysRevD.23.165}{\emph{Phys. Rev. D} {\bf 23}
  (1981) 165}.

\bibitem{Foot:1988aq}
R.~Foot, H.~Lew, X.~G. He and G.~C. Joshi, \emph{{Seesaw Neutrino Masses
  Induced by a Triplet of Leptons}},
  \href{http://dx.doi.org/10.1007/BF01415558}{\emph{Z. Phys. C} {\bf 44} (1989)
  441}.

\bibitem{FileviezPerez:2008jbu}
P.~Fileviez~Perez, T.~Han, G.-y. Huang, T.~Li and K.~Wang, \emph{{Neutrino
  Masses and the CERN LHC: Testing Type II Seesaw}},
  \href{http://dx.doi.org/10.1103/PhysRevD.78.015018}{\emph{Phys. Rev. D} {\bf
  78} (2008) 015018}, [\href{http://arxiv.org/abs/0805.3536}{{\tt 0805.3536}}].

\bibitem{Arhrib:2011uy}
A.~Arhrib, R.~Benbrik, M.~Chabab, G.~Moultaka, M.~C. Peyranere, L.~Rahili
  et~al., \emph{{The Higgs Potential in the Type II Seesaw Model}},
  \href{http://dx.doi.org/10.1103/PhysRevD.84.095005}{\emph{Phys. Rev. D} {\bf
  84} (2011) 095005}, [\href{http://arxiv.org/abs/1105.1925}{{\tt 1105.1925}}].

\bibitem{BhupalDev:2013xol}
P.~S. Bhupal~Dev, D.~K. Ghosh, N.~Okada and I.~Saha, \emph{{125 GeV Higgs Boson
  and the Type-II Seesaw Model}},
  \href{http://dx.doi.org/10.1007/JHEP03(2013)150}{\emph{JHEP} {\bf 03} (2013)
  150}, [\href{http://arxiv.org/abs/1301.3453}{{\tt 1301.3453}}].

\bibitem{Huitu:1996su}
K.~Huitu, J.~Maalampi, A.~Pietila and M.~Raidal, \emph{{Doubly charged Higgs at
  LHC}}, \href{http://dx.doi.org/10.1016/S0550-3213(97)87466-4}{\emph{Nucl.
  Phys. B} {\bf 487} (1997) 27--42},
  [\href{http://arxiv.org/abs/hep-ph/9606311}{{\tt hep-ph/9606311}}].

\bibitem{Gunion:1996pq}
J.~F. Gunion, C.~Loomis and K.~T. Pitts, \emph{{Searching for doubly charged
  Higgs bosons at future colliders}}, {\emph{eConf} {\bf C960625} (1996)
  LTH096}, [\href{http://arxiv.org/abs/hep-ph/9610237}{{\tt hep-ph/9610237}}].

\bibitem{Chun:2003ej}
E.~J. Chun, K.~Y. Lee and S.~C. Park, \emph{{Testing Higgs triplet model and
  neutrino mass patterns}},
  \href{http://dx.doi.org/10.1016/S0370-2693(03)00770-6}{\emph{Phys. Lett. B}
  {\bf 566} (2003) 142--151}, [\href{http://arxiv.org/abs/hep-ph/0304069}{{\tt
  hep-ph/0304069}}].

\bibitem{Han:2005nk}
T.~Han, H.~E. Logan, B.~Mukhopadhyaya and R.~Srikanth, \emph{{Neutrino masses
  and lepton-number violation in the littlest Higgs scenario}},
  \href{http://dx.doi.org/10.1103/PhysRevD.72.053007}{\emph{Phys. Rev. D} {\bf
  72} (2005) 053007}, [\href{http://arxiv.org/abs/hep-ph/0505260}{{\tt
  hep-ph/0505260}}].

\bibitem{Han:2007bk}
T.~Han, B.~Mukhopadhyaya, Z.~Si and K.~Wang, \emph{{Pair production of
  doubly-charged scalars: Neutrino mass constraints and signals at the LHC}},
  \href{http://dx.doi.org/10.1103/PhysRevD.76.075013}{\emph{Phys. Rev. D} {\bf
  76} (2007) 075013}, [\href{http://arxiv.org/abs/0706.0441}{{\tt 0706.0441}}].

\bibitem{Garayoa:2007fw}
J.~Garayoa and T.~Schwetz, \emph{{Neutrino mass hierarchy and Majorana CP
  phases within the Higgs triplet model at the LHC}},
  \href{http://dx.doi.org/10.1088/1126-6708/2008/03/009}{\emph{JHEP} {\bf 03}
  (2008) 009}, [\href{http://arxiv.org/abs/0712.1453}{{\tt 0712.1453}}].

\bibitem{Kadastik:2007yd}
M.~Kadastik, M.~Raidal and L.~Rebane, \emph{{Direct determination of neutrino
  mass parameters at future colliders}},
  \href{http://dx.doi.org/10.1103/PhysRevD.77.115023}{\emph{Phys. Rev. D} {\bf
  77} (2008) 115023}, [\href{http://arxiv.org/abs/0712.3912}{{\tt 0712.3912}}].

\bibitem{Akeroyd:2007zv}
A.~G. Akeroyd, M.~Aoki and H.~Sugiyama, \emph{{Probing Majorana Phases and
  Neutrino Mass Spectrum in the Higgs Triplet Model at the CERN LHC}},
  \href{http://dx.doi.org/10.1103/PhysRevD.77.075010}{\emph{Phys. Rev. D} {\bf
  77} (2008) 075010}, [\href{http://arxiv.org/abs/0712.4019}{{\tt 0712.4019}}].

\bibitem{Chao:2007mz}
W.~Chao, S.~Luo, Z.-z. Xing and S.~Zhou, \emph{{A Compromise between Neutrino
  Masses and Collider Signatures in the Type-II Seesaw Model}},
  \href{http://dx.doi.org/10.1103/PhysRevD.77.016001}{\emph{Phys. Rev. D} {\bf
  77} (2008) 016001}, [\href{http://arxiv.org/abs/0709.1069}{{\tt 0709.1069}}].

\bibitem{Chao:2008mq}
W.~Chao, Z.-G. Si, Z.-z. Xing and S.~Zhou, \emph{{Correlative signatures of
  heavy Majorana neutrinos and doubly-charged Higgs bosons at the Large Hadron
  Collider}},
  \href{http://dx.doi.org/10.1016/j.physletb.2008.08.003}{\emph{Phys. Lett. B}
  {\bf 666} (2008) 451--454}, [\href{http://arxiv.org/abs/0804.1265}{{\tt
  0804.1265}}].

\bibitem{delAguila:2008cj}
F.~del Aguila and J.~A. Aguilar-Saavedra, \emph{{Distinguishing seesaw models
  at LHC with multi-lepton signals}},
  \href{http://dx.doi.org/10.1016/j.nuclphysb.2008.12.029}{\emph{Nucl. Phys. B}
  {\bf 813} (2009) 22--90}, [\href{http://arxiv.org/abs/0808.2468}{{\tt
  0808.2468}}].

\bibitem{FileviezPerez:2008wbg}
P.~Fileviez~Perez, T.~Han, G.-Y. Huang, T.~Li and K.~Wang, \emph{{Testing a
  Neutrino Mass Generation Mechanism at the LHC}},
  \href{http://dx.doi.org/10.1103/PhysRevD.78.071301}{\emph{Phys. Rev. D} {\bf
  78} (2008) 071301}, [\href{http://arxiv.org/abs/0803.3450}{{\tt 0803.3450}}].

\bibitem{Akeroyd:2009hb}
A.~G. Akeroyd and C.-W. Chiang, \emph{{Doubly charged Higgs bosons and
  three-lepton signatures in the Higgs Triplet Model}},
  \href{http://dx.doi.org/10.1103/PhysRevD.80.113010}{\emph{Phys. Rev. D} {\bf
  80} (2009) 113010}, [\href{http://arxiv.org/abs/0909.4419}{{\tt 0909.4419}}].

\bibitem{Akeroyd:2010ip}
A.~G. Akeroyd, C.-W. Chiang and N.~Gaur, \emph{{Leptonic signatures of doubly
  charged Higgs boson production at the LHC}},
  \href{http://dx.doi.org/10.1007/JHEP11(2010)005}{\emph{JHEP} {\bf 11} (2010)
  005}, [\href{http://arxiv.org/abs/1009.2780}{{\tt 1009.2780}}].

\bibitem{Akeroyd:2011zza}
A.~G. Akeroyd and H.~Sugiyama, \emph{{Production of doubly charged scalars from
  the decay of singly charged scalars in the Higgs Triplet Model}},
  \href{http://dx.doi.org/10.1103/PhysRevD.84.035010}{\emph{Phys. Rev. D} {\bf
  84} (2011) 035010}, [\href{http://arxiv.org/abs/1105.2209}{{\tt 1105.2209}}].

\bibitem{Melfo:2011nx}
A.~Melfo, M.~Nemevsek, F.~Nesti, G.~Senjanovic and Y.~Zhang, \emph{{Type II
  Seesaw at LHC: The Roadmap}},
  \href{http://dx.doi.org/10.1103/PhysRevD.85.055018}{\emph{Phys. Rev. D} {\bf
  85} (2012) 055018}, [\href{http://arxiv.org/abs/1108.4416}{{\tt 1108.4416}}].

\bibitem{Aoki:2011pz}
M.~Aoki, S.~Kanemura and K.~Yagyu, \emph{{Testing the Higgs triplet model with
  the mass difference at the LHC}},
  \href{http://dx.doi.org/10.1103/PhysRevD.85.055007}{\emph{Phys. Rev. D} {\bf
  85} (2012) 055007}, [\href{http://arxiv.org/abs/1110.4625}{{\tt 1110.4625}}].

\bibitem{Chiang:2012dk}
C.-W. Chiang, T.~Nomura and K.~Tsumura, \emph{{Search for doubly charged Higgs
  bosons using the same-sign diboson mode at the LHC}},
  \href{http://dx.doi.org/10.1103/PhysRevD.85.095023}{\emph{Phys. Rev. D} {\bf
  85} (2012) 095023}, [\href{http://arxiv.org/abs/1202.2014}{{\tt 1202.2014}}].

\bibitem{Chun:2012zu}
E.~J. Chun and P.~Sharma, \emph{{Same-Sign Tetra-Leptons from Type II Seesaw}},
  \href{http://dx.doi.org/10.1007/JHEP08(2012)162}{\emph{JHEP} {\bf 08} (2012)
  162}, [\href{http://arxiv.org/abs/1206.6278}{{\tt 1206.6278}}].

\bibitem{Chun:2013vma}
E.~J. Chun and P.~Sharma, \emph{{Search for a doubly-charged boson in four
  lepton final states in type II seesaw}},
  \href{http://dx.doi.org/10.1016/j.physletb.2013.11.056}{\emph{Phys. Lett. B}
  {\bf 728} (2014) 256--261}, [\href{http://arxiv.org/abs/1309.6888}{{\tt
  1309.6888}}].

\bibitem{delAguila:2013mia}
F.~del \'Aguila and M.~Chala, \emph{{LHC bounds on Lepton Number Violation
  mediated by doubly and singly-charged scalars}},
  \href{http://dx.doi.org/10.1007/JHEP03(2014)027}{\emph{JHEP} {\bf 03} (2014)
  027}, [\href{http://arxiv.org/abs/1311.1510}{{\tt 1311.1510}}].

\bibitem{Kanemura:2013vxa}
S.~Kanemura, K.~Yagyu and H.~Yokoya, \emph{{First constraint on the mass of
  doubly-charged Higgs bosons in the same-sign diboson decay scenario at the
  LHC}}, \href{http://dx.doi.org/10.1016/j.physletb.2013.08.054}{\emph{Phys.
  Lett. B} {\bf 726} (2013) 316--319},
  [\href{http://arxiv.org/abs/1305.2383}{{\tt 1305.2383}}].

\bibitem{Kanemura:2014ipa}
S.~Kanemura, M.~Kikuchi, H.~Yokoya and K.~Yagyu, \emph{{LHC Run-I constraint on
  the mass of doubly charged Higgs bosons in the same-sign diboson decay
  scenario}}, \href{http://dx.doi.org/10.1093/ptep/ptv071}{\emph{PTEP} {\bf
  2015} (2015) 051B02}, [\href{http://arxiv.org/abs/1412.7603}{{\tt
  1412.7603}}].

\bibitem{Kang:2014lwn}
Z.~Kang, J.~Li, T.~Li, Y.~Liu and G.-Z. Ning, \emph{{Light Doubly Charged Higgs
  Boson via the $WW^*$ Channel at LHC}},
  \href{http://dx.doi.org/10.1140/epjc/s10052-015-3774-1}{\emph{Eur. Phys. J.
  C} {\bf 75} (2015) 574}, [\href{http://arxiv.org/abs/1404.5207}{{\tt
  1404.5207}}].

\bibitem{Godunov:2014waa}
S.~I. Godunov, M.~I. Vysotsky and E.~V. Zhemchugov, \emph{{Double Higgs
  production at LHC, see-saw type II and Georgi-Machacek model}},
  \href{http://dx.doi.org/10.1134/S1063776115030073}{\emph{J. Exp. Theor.
  Phys.} {\bf 120} (2015) 369--375},
  [\href{http://arxiv.org/abs/1408.0184}{{\tt 1408.0184}}].

\bibitem{Chen:2014qda}
C.-H. Chen and T.~Nomura, \emph{{Search for $\delta^{\pm\pm}$ with new decay
  patterns at the LHC}},
  \href{http://dx.doi.org/10.1103/PhysRevD.91.035023}{\emph{Phys. Rev. D} {\bf
  91} (2015) 035023}, [\href{http://arxiv.org/abs/1411.6412}{{\tt 1411.6412}}].

\bibitem{Han:2015hba}
Z.-L. Han, R.~Ding and Y.~Liao, \emph{{LHC Phenomenology of Type II Seesaw:
  Nondegenerate Case}},
  \href{http://dx.doi.org/10.1103/PhysRevD.91.093006}{\emph{Phys. Rev. D} {\bf
  91} (2015) 093006}, [\href{http://arxiv.org/abs/1502.05242}{{\tt
  1502.05242}}].

\bibitem{Han:2015sca}
Z.-L. Han, R.~Ding and Y.~Liao, \emph{{LHC phenomenology of the type II seesaw
  mechanism: Observability of neutral scalars in the nondegenerate case}},
  \href{http://dx.doi.org/10.1103/PhysRevD.92.033014}{\emph{Phys. Rev. D} {\bf
  92} (2015) 033014}, [\href{http://arxiv.org/abs/1506.08996}{{\tt
  1506.08996}}].

\bibitem{Bonilla:2015jdf}
C.~Bonilla, J.~C. Rom\~ao and J.~W.~F. Valle, \emph{{Electroweak breaking and
  neutrino mass: \textquoteleft{}invisible\textquoteright{} Higgs decays at the
  LHC (type II seesaw)}},
  \href{http://dx.doi.org/10.1088/1367-2630/18/3/033033}{\emph{New J. Phys.}
  {\bf 18} (2016) 033033}, [\href{http://arxiv.org/abs/1511.07351}{{\tt
  1511.07351}}].

\bibitem{Mitra:2016wpr}
M.~Mitra, S.~Niyogi and M.~Spannowsky, \emph{{Type-II Seesaw Model and
  Multilepton Signatures at Hadron Colliders}},
  \href{http://dx.doi.org/10.1103/PhysRevD.95.035042}{\emph{Phys. Rev. D} {\bf
  95} (2017) 035042}, [\href{http://arxiv.org/abs/1611.09594}{{\tt
  1611.09594}}].

\bibitem{Babu:2016rcr}
K.~S. Babu and S.~Jana, \emph{{Probing Doubly Charged Higgs Bosons at the LHC
  through Photon Initiated Processes}},
  \href{http://dx.doi.org/10.1103/PhysRevD.95.055020}{\emph{Phys. Rev. D} {\bf
  95} (2017) 055020}, [\href{http://arxiv.org/abs/1612.09224}{{\tt
  1612.09224}}].

\bibitem{Ghosh:2017pxl}
D.~K. Ghosh, N.~Ghosh, I.~Saha and A.~Shaw, \emph{{Revisiting the high-scale
  validity of the type II seesaw model with novel LHC signature}},
  \href{http://dx.doi.org/10.1103/PhysRevD.97.115022}{\emph{Phys. Rev. D} {\bf
  97} (2018) 115022}, [\href{http://arxiv.org/abs/1711.06062}{{\tt
  1711.06062}}].

\bibitem{Antusch:2018svb}
S.~Antusch, O.~Fischer, A.~Hammad and C.~Scherb, \emph{{Low scale type II
  seesaw: Present constraints and prospects for displaced vertex searches}},
  \href{http://dx.doi.org/10.1007/JHEP02(2019)157}{\emph{JHEP} {\bf 02} (2019)
  157}, [\href{http://arxiv.org/abs/1811.03476}{{\tt 1811.03476}}].

\bibitem{BhupalDev:2018tox}
P.~S. Bhupal~Dev and Y.~Zhang, \emph{{Displaced vertex signatures of doubly
  charged scalars in the type-II seesaw and its left-right extensions}},
  \href{http://dx.doi.org/10.1007/JHEP10(2018)199}{\emph{JHEP} {\bf 10} (2018)
  199}, [\href{http://arxiv.org/abs/1808.00943}{{\tt 1808.00943}}].

\bibitem{Primulando:2019evb}
R.~Primulando, J.~Julio and P.~Uttayarat, \emph{{Scalar phenomenology in
  type-II seesaw model}},
  \href{http://dx.doi.org/10.1007/JHEP08(2019)024}{\emph{JHEP} {\bf 08} (2019)
  024}, [\href{http://arxiv.org/abs/1903.02493}{{\tt 1903.02493}}].

\bibitem{deMelo:2019asm}
T.~B. de~Melo, F.~S. Queiroz and Y.~Villamizar, \emph{{Doubly Charged Scalar at
  the High-Luminosity and High-Energy LHC}},
  \href{http://dx.doi.org/10.1142/S0217751X19501574}{\emph{Int. J. Mod. Phys.
  A} {\bf 34} (2019) 1950157}, [\href{http://arxiv.org/abs/1909.07429}{{\tt
  1909.07429}}].

\bibitem{Chun:2019hce}
E.~J. Chun, S.~Khan, S.~Mandal, M.~Mitra and S.~Shil, \emph{{Same-sign
  tetralepton signature at the Large Hadron Collider and a future $pp$
  collider}}, \href{http://dx.doi.org/10.1103/PhysRevD.101.075008}{\emph{Phys.
  Rev. D} {\bf 101} (2020) 075008},
  [\href{http://arxiv.org/abs/1911.00971}{{\tt 1911.00971}}].

\bibitem{Padhan:2019jlc}
R.~Padhan, D.~Das, M.~Mitra and A.~Kumar~Nayak, \emph{{Probing doubly and
  singly charged Higgs bosons at the $pp$ collider HE-LHC}},
  \href{http://dx.doi.org/10.1103/PhysRevD.101.075050}{\emph{Phys. Rev. D} {\bf
  101} (2020) 075050}, [\href{http://arxiv.org/abs/1909.10495}{{\tt
  1909.10495}}].

\bibitem{Ashanujjaman:2021txz}
S.~Ashanujjaman and K.~Ghosh, \emph{{Revisiting type-II see-saw: present limits
  and future prospects at LHC}},
  \href{http://dx.doi.org/10.1007/JHEP03(2022)195}{\emph{JHEP} {\bf 03} (2022)
  195}, [\href{http://arxiv.org/abs/2108.10952}{{\tt 2108.10952}}].

\bibitem{Ghosh:2021khk}
P.~Ghosh, S.~Mahapatra, N.~Narendra and N.~Sahu, \emph{{TeV scale modified
  type-II seesaw mechanism and dark matter in a gauged U(1)B-L symmetric
  model}}, \href{http://dx.doi.org/10.1103/PhysRevD.106.015001}{\emph{Phys.
  Rev. D} {\bf 106} (2022) 015001},
  [\href{http://arxiv.org/abs/2107.11951}{{\tt 2107.11951}}].

\bibitem{Ashanujjaman:2022ofg}
S.~Ashanujjaman, K.~Ghosh and R.~Sahu, \emph{{Low-mass doubly-charged Higgs
  bosons at LHC}},  \href{http://arxiv.org/abs/2211.00632}{{\tt 2211.00632}}.

\bibitem{Butterworth:2022dkt}
J.~Butterworth, J.~Heeck, S.~H. Jeon, O.~Mattelaer and R.~Ruiz, \emph{{Testing
  the Scalar Triplet Solution to CDF's Fat $W$ Problem at the LHC}},
  \href{http://arxiv.org/abs/2210.13496}{{\tt 2210.13496}}.

\bibitem{Li:2018jns}
T.~Li, \emph{{Type II Seesaw and tau lepton at the HL-LHC, HE-LHC and FCC-hh}},
  \href{http://dx.doi.org/10.1007/JHEP09(2018)079}{\emph{JHEP} {\bf 09} (2018)
  079}, [\href{http://arxiv.org/abs/1802.00945}{{\tt 1802.00945}}].

\bibitem{Du:2018eaw}
Y.~Du, A.~Dunbrack, M.~J. Ramsey-Musolf and J.-H. Yu, \emph{{Type-II Seesaw
  Scalar Triplet Model at a 100 TeV $pp$ Collider: Discovery and Higgs Portal
  Coupling Determination}},
  \href{http://dx.doi.org/10.1007/JHEP01(2019)101}{\emph{JHEP} {\bf 01} (2019)
  101}, [\href{http://arxiv.org/abs/1810.09450}{{\tt 1810.09450}}].

\bibitem{Arhrib:2019ywg}
A.~Arhrib, K.~Cheung and C.-T. Lu, \emph{{Same-sign charged Higgs boson pair
  production in bosonic decay channels at the HL-LHC and HE-LHC}},
  \href{http://dx.doi.org/10.1103/PhysRevD.102.095026}{\emph{Phys. Rev. D} {\bf
  102} (2020) 095026}, [\href{http://arxiv.org/abs/1910.02571}{{\tt
  1910.02571}}].

\bibitem{Fuks:2019clu}
B.~Fuks, M.~Nemev\v{s}ek and R.~Ruiz, \emph{{Doubly Charged Higgs Boson
  Production at Hadron Colliders}},
  \href{http://dx.doi.org/10.1103/PhysRevD.101.075022}{\emph{Phys. Rev. D} {\bf
  101} (2020) 075022}, [\href{http://arxiv.org/abs/1912.08975}{{\tt
  1912.08975}}].

\bibitem{Aoki:2020til}
M.~Aoki, K.~Enomoto and S.~Kanemura, \emph{{Probing charged lepton number
  violation via $\ell^\pm \ell^{\prime \pm} W^\mp W^\mp$}},
  \href{http://dx.doi.org/10.1103/PhysRevD.101.115019}{\emph{Phys. Rev. D} {\bf
  101} (2020) 115019}, [\href{http://arxiv.org/abs/2002.12265}{{\tt
  2002.12265}}].

\bibitem{Rodejohann:2010jh}
W.~Rodejohann, \emph{{Inverse Neutrino-less Double Beta Decay Revisited:
  Neutrinos, Higgs Triplets and a Muon Collider}},
  \href{http://dx.doi.org/10.1103/PhysRevD.81.114001}{\emph{Phys. Rev. D} {\bf
  81} (2010) 114001}, [\href{http://arxiv.org/abs/1005.2854}{{\tt 1005.2854}}].

\bibitem{Shen:2015pih}
J.-F. Shen, Y.-P. Bi and Z.-X. Li, \emph{{Pair production of scalars at the ILC
  in the Higgs triplet model under the non-degenerate case}},
  \href{http://dx.doi.org/10.1209/0295-5075/112/31002}{\emph{EPL} {\bf 112}
  (2015) 31002}.

\bibitem{Blunier:2016peh}
S.~Blunier, G.~Cottin, M.~A. D\'\i{}az and B.~Koch, \emph{{Phenomenology of a
  Higgs triplet model at future $e^{+}e^{-}$ colliders}},
  \href{http://dx.doi.org/10.1103/PhysRevD.95.075038}{\emph{Phys. Rev. D} {\bf
  95} (2017) 075038}, [\href{http://arxiv.org/abs/1611.07896}{{\tt
  1611.07896}}].

\bibitem{Sui:2017qra}
Y.~Sui and Y.~Zhang, \emph{{Prospects of type-II seesaw models at future
  colliders in light of the DAMPE $e^+e^-$ excess}},
  \href{http://dx.doi.org/10.1103/PhysRevD.97.095002}{\emph{Phys. Rev. D} {\bf
  97} (2018) 095002}, [\href{http://arxiv.org/abs/1712.03642}{{\tt
  1712.03642}}].

\bibitem{Nomura:2017abh}
T.~Nomura, H.~Okada and H.~Yokoya, \emph{{Discriminating leptonic Yukawa
  interactions with doubly charged scalar at the ILC}},
  \href{http://dx.doi.org/10.1016/j.nuclphysb.2018.02.011}{\emph{Nucl. Phys. B}
  {\bf 929} (2018) 193--206}, [\href{http://arxiv.org/abs/1702.03396}{{\tt
  1702.03396}}].

\bibitem{Agrawal:2018pci}
P.~Agrawal, M.~Mitra, S.~Niyogi, S.~Shil and M.~Spannowsky, \emph{{Probing the
  Type-II Seesaw Mechanism through the Production of Higgs Bosons at a Lepton
  Collider}}, \href{http://dx.doi.org/10.1103/PhysRevD.98.015024}{\emph{Phys.
  Rev. D} {\bf 98} (2018) 015024}, [\href{http://arxiv.org/abs/1803.00677}{{\tt
  1803.00677}}].

\bibitem{Dev:2018sel}
P.~S.~B. Dev, M.~J. Ramsey-Musolf and Y.~Zhang, \emph{{Doubly-Charged Scalars
  in the Type-II Seesaw Mechanism: Fundamental Symmetry Tests and High-Energy
  Searches}}, \href{http://dx.doi.org/10.1103/PhysRevD.98.055013}{\emph{Phys.
  Rev. D} {\bf 98} (2018) 055013}, [\href{http://arxiv.org/abs/1806.08499}{{\tt
  1806.08499}}].

\bibitem{Rahili:2019ixf}
L.~Rahili, A.~Arhrib and R.~Benbrik, \emph{{Associated production of SM Higgs
  with a photon in type-II seesaw models at the ILC}},
  \href{http://dx.doi.org/10.1140/epjc/s10052-019-7471-3}{\emph{Eur. Phys. J.
  C} {\bf 79} (2019) 940}, [\href{http://arxiv.org/abs/1909.07793}{{\tt
  1909.07793}}].

\bibitem{Li:2019xvv}
T.~Li and M.~A. Schmidt, \emph{{Sensitivity of future lepton colliders and
  low-energy experiments to charged lepton flavor violation from bileptons}},
  \href{http://dx.doi.org/10.1103/PhysRevD.100.115007}{\emph{Phys. Rev. D} {\bf
  100} (2019) 115007}, [\href{http://arxiv.org/abs/1907.06963}{{\tt
  1907.06963}}].

\bibitem{Gluza:2020icp}
J.~Gluza, M.~Kordiaczy\'nska and T.~Srivastava, \emph{{Doubly Charged Higgs
  Bosons and Spontaneous Symmetry Breaking at eV and TeV Scales}},
  \href{http://dx.doi.org/10.3390/sym12010153}{\emph{Symmetry} {\bf 12} (2020)
  153}.

\bibitem{Costantini:2020stv}
A.~Costantini, F.~De~Lillo, F.~Maltoni, L.~Mantani, O.~Mattelaer, R.~Ruiz
  et~al., \emph{{Vector boson fusion at multi-TeV muon colliders}},
  \href{http://dx.doi.org/10.1007/JHEP09(2020)080}{\emph{JHEP} {\bf 09} (2020)
  080}, [\href{http://arxiv.org/abs/2005.10289}{{\tt 2005.10289}}].

\bibitem{Bandyopadhyay:2020mnp}
P.~Bandyopadhyay, A.~Karan and C.~Sen, \emph{{Discerning Signatures of Seesaw
  Models and Complementarity of Leptonic Colliders}},
  \href{http://arxiv.org/abs/2011.04191}{{\tt 2011.04191}}.

\bibitem{Li:2021lnz}
T.~Li, M.~A. Schmidt, C.-Y. Yao and M.~Yuan, \emph{{Charged lepton flavor
  violation in light of the muon magnetic moment anomaly and colliders}},
  \href{http://dx.doi.org/10.1140/epjc/s10052-021-09569-9}{\emph{Eur. Phys. J.
  C} {\bf 81} (2021) 811}, [\href{http://arxiv.org/abs/2104.04494}{{\tt
  2104.04494}}].

\bibitem{Bai:2021ony}
X.-H. Bai, Z.-L. Han, Y.~Jin, H.-L. Li and Z.-X. Meng, \emph{{Same-sign
  tetralepton signature in type-II seesaw at lepton colliders}},
  \href{http://dx.doi.org/10.1088/1674-1137/ac2ed1}{\emph{Chin. Phys. C} {\bf
  46} (2022) 012001}, [\href{http://arxiv.org/abs/2105.02474}{{\tt
  2105.02474}}].

\bibitem{Ashanujjaman:2022tdn}
S.~Ashanujjaman, K.~Ghosh and K.~Huitu, \emph{{Type-II see-saw: searching the
  LHC elusive low-mass triplet-like Higgses at $e^-e^+$ colliders}},
  \href{http://dx.doi.org/10.1103/PhysRevD.106.075028}{\emph{Phys. Rev. D} {\bf
  106} (2022) 075028}, [\href{http://arxiv.org/abs/2205.14983}{{\tt
  2205.14983}}].

\bibitem{Mandal:2022ysp}
S.~Mandal, O.~G. Miranda, G.~S. Garcia, J.~W.~F. Valle and X.-J. Xu,
  \emph{{High-energy colliders as a probe of neutrino properties}},
  \href{http://dx.doi.org/10.1016/j.physletb.2022.137110}{\emph{Phys. Lett. B}
  {\bf 829} (2022) 137110}, [\href{http://arxiv.org/abs/2202.04502}{{\tt
  2202.04502}}].

\bibitem{Mandal:2022zmy}
S.~Mandal, O.~G. Miranda, G.~Sanchez~Garcia, J.~W.~F. Valle and X.-J. Xu,
  \emph{{Toward deconstructing the simplest seesaw mechanism}},
  \href{http://dx.doi.org/10.1103/PhysRevD.105.095020}{\emph{Phys. Rev. D} {\bf
  105} (2022) 095020}, [\href{http://arxiv.org/abs/2203.06362}{{\tt
  2203.06362}}].

\bibitem{Dev:2019hev}
P.~S.~B. Dev, S.~Khan, M.~Mitra and S.~K. Rai, \emph{{Doubly-charged Higgs
  boson at a future electron-proton collider}},
  \href{http://dx.doi.org/10.1103/PhysRevD.99.115015}{\emph{Phys. Rev. D} {\bf
  99} (2019) 115015}, [\href{http://arxiv.org/abs/1903.01431}{{\tt
  1903.01431}}].

\bibitem{Yang:2021skb}
X.-H. Yang and Z.-J. Yang, \emph{{Doubly charged Higgs production at future
  $ep$ colliders}},
  \href{http://dx.doi.org/10.1088/1674-1137/ac581b}{\emph{Chin. Phys. C} {\bf
  46} (2022) 063107}, [\href{http://arxiv.org/abs/2103.11412}{{\tt
  2103.11412}}].

\bibitem{Deppisch:2015qwa}
F.~F. Deppisch, P.~S. Bhupal~Dev and A.~Pilaftsis, \emph{{Neutrinos and
  Collider Physics}},
  \href{http://dx.doi.org/10.1088/1367-2630/17/7/075019}{\emph{New J. Phys.}
  {\bf 17} (2015) 075019}, [\href{http://arxiv.org/abs/1502.06541}{{\tt
  1502.06541}}].

\bibitem{Cai:2017mow}
Y.~Cai, T.~Han, T.~Li and R.~Ruiz, \emph{{Lepton Number Violation: Seesaw
  Models and Their Collider Tests}},
  \href{http://dx.doi.org/10.3389/fphy.2018.00040}{\emph{Front. in Phys.} {\bf
  6} (2018) 40}, [\href{http://arxiv.org/abs/1711.02180}{{\tt 1711.02180}}].

\bibitem{ATLAS:2022pbd}
{\scshape ATLAS} collaboration, \emph{{Search for doubly charged Higgs boson
  production in multi-lepton final states using 139 fb$^{-1}$ of proton-proton
  collisions at $\sqrt{s}$ = 13 TeV with the ATLAS detector}},
  \href{http://arxiv.org/abs/2211.07505}{{\tt 2211.07505}}.

\bibitem{ATLAS:2021jol}
{\scshape ATLAS} collaboration, G.~Aad et~al., \emph{{Search for doubly and
  singly charged Higgs bosons decaying into vector bosons in multi-lepton final
  states with the ATLAS detector using proton-proton collisions at $
  \sqrt{\mathrm{s}} $ = 13 TeV}},
  \href{http://dx.doi.org/10.1007/JHEP06(2021)146}{\emph{JHEP} {\bf 06} (2021)
  146}, [\href{http://arxiv.org/abs/2101.11961}{{\tt 2101.11961}}].

\bibitem{MICE:2019jkl}
{\scshape MICE} collaboration, M.~Bogomilov et~al., \emph{{Demonstration of
  cooling by the Muon Ionization Cooling Experiment}},
  \href{http://dx.doi.org/10.1038/s41586-020-1958-9}{\emph{Nature} {\bf 578}
  (2020) 53--59}, [\href{http://arxiv.org/abs/1907.08562}{{\tt 1907.08562}}].

\bibitem{Delahaye:2019omf}
J.~P. Delahaye, M.~Diemoz, K.~Long, B.~Mansouli\'e, N.~Pastrone, L.~Rivkin
  et~al., \emph{{Muon Colliders}},  \href{http://arxiv.org/abs/1901.06150}{{\tt
  1901.06150}}.

\bibitem{Bartosik:2020xwr}
N.~Bartosik et~al., \emph{{Detector and Physics Performance at a Muon
  Collider}},
  \href{http://dx.doi.org/10.1088/1748-0221/15/05/P05001}{\emph{JINST} {\bf 15}
  (2020) P05001}, [\href{http://arxiv.org/abs/2001.04431}{{\tt 2001.04431}}].

\bibitem{Han:2020uid}
T.~Han, Y.~Ma and K.~Xie, \emph{{High energy leptonic collisions and
  electroweak parton distribution functions}},
  \href{http://dx.doi.org/10.1103/PhysRevD.103.L031301}{\emph{Phys. Rev. D}
  {\bf 103} (2021) L031301}, [\href{http://arxiv.org/abs/2007.14300}{{\tt
  2007.14300}}].

\bibitem{AlAli:2021let}
H.~Al~Ali et~al., \emph{{The muon Smasher\textquoteright{}s guide}},
  \href{http://dx.doi.org/10.1088/1361-6633/ac6678}{\emph{Rept. Prog. Phys.}
  {\bf 85} (2022) 084201}, [\href{http://arxiv.org/abs/2103.14043}{{\tt
  2103.14043}}].

\bibitem{Bose:2022obr}
T.~Bose et~al., \emph{{Report of the Topical Group on Physics Beyond the
  Standard Model at Energy Frontier for Snowmass 2021}},
  \href{http://arxiv.org/abs/2209.13128}{{\tt 2209.13128}}.

\bibitem{Maltoni:2022bqs}
F.~Maltoni et~al., \emph{{TF07 Snowmass Report: Theory of Collider Phenomena}},
   \href{http://arxiv.org/abs/2210.02591}{{\tt 2210.02591}}.

\bibitem{Narain:2022qud}
M.~Narain et~al., \emph{{The Future of US Particle Physics - The Snowmass 2021
  Energy Frontier Report}},  \href{http://arxiv.org/abs/2211.11084}{{\tt
  2211.11084}}.

\bibitem{Han:2021udl}
T.~Han, S.~Li, S.~Su, W.~Su and Y.~Wu, \emph{{Heavy Higgs bosons in 2HDM at a
  muon collider}},
  \href{http://dx.doi.org/10.1103/PhysRevD.104.055029}{\emph{Phys. Rev. D} {\bf
  104} (2021) 055029}, [\href{http://arxiv.org/abs/2102.08386}{{\tt
  2102.08386}}].

\bibitem{Ruiz:2021tdt}
R.~Ruiz, A.~Costantini, F.~Maltoni and O.~Mattelaer, \emph{{The Effective
  Vector Boson Approximation in high-energy muon collisions}},
  \href{http://dx.doi.org/10.1007/JHEP06(2022)114}{\emph{JHEP} {\bf 06} (2022)
  114}, [\href{http://arxiv.org/abs/2111.02442}{{\tt 2111.02442}}].

\bibitem{Abe:2011fz}
{\scshape Double Chooz} collaboration, Y.~Abe et~al., \emph{{Indication of
  Reactor $\bar{\nu}_e$ Disappearance in the Double Chooz Experiment}},
  \href{http://dx.doi.org/10.1103/PhysRevLett.108.131801}{\emph{Phys. Rev.
  Lett.} {\bf 108} (2012) 131801}, [\href{http://arxiv.org/abs/1112.6353}{{\tt
  1112.6353}}].

\bibitem{Ahn:2012nd}
{\scshape RENO} collaboration, J.~K. Ahn et~al., \emph{{Observation of Reactor
  Electron Antineutrino Disappearance in the RENO Experiment}},
  \href{http://dx.doi.org/10.1103/PhysRevLett.108.191802}{\emph{Phys. Rev.
  Lett.} {\bf 108} (2012) 191802}, [\href{http://arxiv.org/abs/1204.0626}{{\tt
  1204.0626}}].

\bibitem{An:2012eh}
{\scshape Daya Bay} collaboration, F.~P. An et~al., \emph{{Observation of
  electron-antineutrino disappearance at Daya Bay}},
  \href{http://dx.doi.org/10.1103/PhysRevLett.108.171803}{\emph{Phys. Rev.
  Lett.} {\bf 108} (2012) 171803}, [\href{http://arxiv.org/abs/1203.1669}{{\tt
  1203.1669}}].

\bibitem{Abe:2013hdq}
{\scshape T2K} collaboration, K.~Abe et~al., \emph{{Observation of Electron
  Neutrino Appearance in a Muon Neutrino Beam}},
  \href{http://dx.doi.org/10.1103/PhysRevLett.112.061802}{\emph{Phys. Rev.
  Lett.} {\bf 112} (2014) 061802}, [\href{http://arxiv.org/abs/1311.4750}{{\tt
  1311.4750}}].

\bibitem{Abe:2017uxa}
{\scshape T2K} collaboration, K.~Abe et~al., \emph{{Combined Analysis of
  Neutrino and Antineutrino Oscillations at T2K}},
  \href{http://dx.doi.org/10.1103/PhysRevLett.118.151801}{\emph{Phys. Rev.
  Lett.} {\bf 118} (2017) 151801}, [\href{http://arxiv.org/abs/1701.00432}{{\tt
  1701.00432}}].

\bibitem{Adamson:2016tbq}
{\scshape NOvA} collaboration, P.~Adamson et~al., \emph{{First measurement of
  electron neutrino appearance in NOvA}},
  \href{http://dx.doi.org/10.1103/PhysRevLett.116.151806}{\emph{Phys. Rev.
  Lett.} {\bf 116} (2016) 151806}, [\href{http://arxiv.org/abs/1601.05022}{{\tt
  1601.05022}}].

\bibitem{Hyper-KamiokandeProto-:2015xww}
{\scshape Hyper-Kamiokande Proto-} collaboration, K.~Abe et~al., \emph{{Physics
  potential of a long-baseline neutrino oscillation experiment using a J-PARC
  neutrino beam and Hyper-Kamiokande}},
  \href{http://dx.doi.org/10.1093/ptep/ptv061}{\emph{PTEP} {\bf 2015} (2015)
  053C02}, [\href{http://arxiv.org/abs/1502.05199}{{\tt 1502.05199}}].

\bibitem{JUNO:2015sjr}
{\scshape JUNO} collaboration, Z.~Djurcic et~al., \emph{{JUNO Conceptual Design
  Report}},  \href{http://arxiv.org/abs/1508.07166}{{\tt 1508.07166}}.

\bibitem{DUNE:2015lol}
{\scshape DUNE} collaboration, R.~Acciarri et~al., \emph{{Long-Baseline
  Neutrino Facility (LBNF) and Deep Underground Neutrino Experiment (DUNE)}:
  {Conceptual Design Report, Volume 2: The Physics Program for DUNE at LBNF}},
  \href{http://arxiv.org/abs/1512.06148}{{\tt 1512.06148}}.

\bibitem{Cheng:2022hbo}
Y.~Cheng, X.-G. He, F.~Huang, J.~Sun and Z.-P. Xing, \emph{{Electroweak
  precision tests for triplet scalars}},
  \href{http://arxiv.org/abs/2208.06760}{{\tt 2208.06760}}.

\bibitem{Muong-2:2021ovs}
{\scshape Muon g-2} collaboration, T.~Albahri et~al., \emph{{Magnetic-field
  measurement and analysis for the Muon $g-2$ Experiment at Fermilab}},
  \href{http://dx.doi.org/10.1103/PhysRevA.103.042208}{\emph{Phys. Rev. A} {\bf
  103} (2021) 042208}, [\href{http://arxiv.org/abs/2104.03201}{{\tt
  2104.03201}}].

\bibitem{Muong-2:2021vma}
{\scshape Muon g-2} collaboration, T.~Albahri et~al., \emph{{Measurement of the
  anomalous precession frequency of the muon in the Fermilab Muon $g-2$
  Experiment}},
  \href{http://dx.doi.org/10.1103/PhysRevD.103.072002}{\emph{Phys. Rev. D} {\bf
  103} (2021) 072002}, [\href{http://arxiv.org/abs/2104.03247}{{\tt
  2104.03247}}].

\bibitem{Muong-2:2021ojo}
{\scshape Muon g-2} collaboration, B.~Abi et~al., \emph{{Measurement of the
  Positive Muon Anomalous Magnetic Moment to 0.46 ppm}},
  \href{http://dx.doi.org/10.1103/PhysRevLett.126.141801}{\emph{Phys. Rev.
  Lett.} {\bf 126} (2021) 141801}, [\href{http://arxiv.org/abs/2104.03281}{{\tt
  2104.03281}}].

\bibitem{Esteban:2020cvm}
I.~Esteban, M.~C. Gonzalez-Garcia, M.~Maltoni, T.~Schwetz and A.~Zhou,
  \emph{{The fate of hints: updated global analysis of three-flavor neutrino
  oscillations}}, \href{http://dx.doi.org/10.1007/JHEP09(2020)178}{\emph{JHEP}
  {\bf 09} (2020) 178}, [\href{http://arxiv.org/abs/2007.14792}{{\tt
  2007.14792}}].

\bibitem{nufit2021}
{\emph{NuFIT 5.1, www.nu-fit.org} (2021) }.

\bibitem{Vagnozzi:2017ovm}
S.~Vagnozzi, E.~Giusarma, O.~Mena, K.~Freese, M.~Gerbino, S.~Ho et~al.,
  \emph{{Unveiling $\nu$ secrets with cosmological data: neutrino masses and
  mass hierarchy}},
  \href{http://dx.doi.org/10.1103/PhysRevD.96.123503}{\emph{Phys. Rev. D} {\bf
  96} (2017) 123503}, [\href{http://arxiv.org/abs/1701.08172}{{\tt
  1701.08172}}].

\bibitem{Planck:2018vyg}
{\scshape Planck} collaboration, N.~Aghanim et~al., \emph{{Planck 2018 results.
  VI. Cosmological parameters}},
  \href{http://dx.doi.org/10.1051/0004-6361/201833910}{\emph{Astron.
  Astrophys.} {\bf 641} (2020) A6},
  [\href{http://arxiv.org/abs/1807.06209}{{\tt 1807.06209}}].

\bibitem{Alwall:2014hca}
J.~Alwall, R.~Frederix, S.~Frixione, V.~Hirschi, F.~Maltoni, O.~Mattelaer
  et~al., \emph{{The automated computation of tree-level and next-to-leading
  order differential cross sections, and their matching to parton shower
  simulations}}, \href{http://dx.doi.org/10.1007/JHEP07(2014)079}{\emph{JHEP}
  {\bf 07} (2014) 079}, [\href{http://arxiv.org/abs/1405.0301}{{\tt
  1405.0301}}].

\bibitem{Artoisenet:2012st}
P.~Artoisenet, R.~Frederix, O.~Mattelaer and R.~Rietkerk, \emph{{Automatic
  spin-entangled decays of heavy resonances in Monte Carlo simulations}},
  \href{http://dx.doi.org/10.1007/JHEP03(2013)015}{\emph{JHEP} {\bf 03} (2013)
  015}, [\href{http://arxiv.org/abs/1212.3460}{{\tt 1212.3460}}].

\bibitem{Sjostrand:2014zea}
T.~Sj\"ostrand, S.~Ask, J.~R. Christiansen, R.~Corke, N.~Desai, P.~Ilten
  et~al., \emph{{An introduction to PYTHIA 8.2}},
  \href{http://dx.doi.org/10.1016/j.cpc.2015.01.024}{\emph{Comput. Phys.
  Commun.} {\bf 191} (2015) 159--177},
  [\href{http://arxiv.org/abs/1410.3012}{{\tt 1410.3012}}].

\bibitem{deFavereau:2013fsa}
{\scshape DELPHES 3} collaboration, J.~de~Favereau, C.~Delaere, P.~Demin,
  A.~Giammanco, V.~Lema\^\i{}tre, A.~Mertens et~al., \emph{{DELPHES 3, A
  modular framework for fast simulation of a generic collider experiment}},
  \href{http://dx.doi.org/10.1007/JHEP02(2014)057}{\emph{JHEP} {\bf 02} (2014)
  057}, [\href{http://arxiv.org/abs/1307.6346}{{\tt 1307.6346}}].

\bibitem{Li:2022kkc}
T.~Li, H.~Qin, C.-Y. Yao and M.~Yuan, \emph{{Probing heavy triplet leptons of
  the type-III seesaw mechanism at future muon colliders}},
  \href{http://dx.doi.org/10.1103/PhysRevD.106.035021}{\emph{Phys. Rev. D} {\bf
  106} (2022) 035021}, [\href{http://arxiv.org/abs/2205.04214}{{\tt
  2205.04214}}].

\bibitem{Li:2018cod}
T.~Li and M.~A. Schmidt, \emph{{Sensitivity of future lepton colliders to the
  search for charged lepton flavor violation}},
  \href{http://dx.doi.org/10.1103/PhysRevD.99.055038}{\emph{Phys. Rev. D} {\bf
  99} (2019) 055038}, [\href{http://arxiv.org/abs/1809.07924}{{\tt
  1809.07924}}].

\end{thebibliography}\endgroup

\end{document}